\documentclass[doctor,final,11pt]{iscs-thesis}
\usepackage{amsmath, amsfonts, amssymb}
\usepackage{fontspec}
\usepackage{xeCJK}
\setCJKsansfont{HaranoAjiGothic-Regular.otf}[
  BoldFont=HaranoAjiGothic-Bold.otf
]
\usepackage{graphicx}
\usepackage{hyperref}
\hypersetup{hidelinks}
\usepackage{caption}
\usepackage[subrefformat=parens]{subcaption}
\usepackage{enumerate}
\usepackage{ascmac}
\usepackage{tabularx}
\usepackage{array}
\usepackage{booktabs}
\usepackage{bm}
\usepackage{microtype}

\def\BibTeX{{\rm B\kern-.05em{\sc i\kern-.025em b}\kern-.08em
    T\kern-.1667em\lower.7ex\hbox{E}\kern-.125emX}}
\newcommand{\ctext}[1]{\raise0.2ex\hbox{\textcircled{\scriptsize{#1}}}}

\etitle{AI-Guided Learning:\\
Research on Knowledge and Skill Acquisition Support Methods\\
Using Deep Learning Audio\mbox{-}Video Processing Techniques}
\jtitle{深層学習による音声動画処理技術を用いた\\
知識および技能獲得支援手法の研究}
\eauthor{Kazuki Kawamura}
\jauthor{河村和紀}
\esupervisor{Jun Rekimoto}
\jsupervisor{暦本純一}
\supervisortitle{Professor}
\date{September 19, 2025}

\begin{document}

\thispagestyle{empty}
\setlength{\parindent}{0pt}
\vspace*{3.85cm}

\noindent\makebox[\textwidth][c]{%
\begin{minipage}{150mm}
\raggedleft
{\sffamily\bfseries\fontsize{19.5}{25.5}\selectfont
AI-Guided Learning:\\
Research on Knowledge and Skill Acquisition\\
Support Methods Using Deep Learning\\
Audio\mbox{-}Video Processing Techniques\par}

\vspace{0.43cm}

{\rmfamily\fontsize{17.28}{21.5}\selectfont
深層学習による音声動画処理技術を用いた\\[0.08cm]
知識および技能獲得支援手法の研究\par}
\end{minipage}}

\vfill

\noindent\makebox[\textwidth][c]{%
\begin{minipage}{150mm}
\begin{minipage}{142.5mm}
\raggedleft
{\sffamily\mdseries\fontsize{23.3}{28}\selectfont Kazuki Kawamura\par}
\vspace{0.34cm}
{\rmfamily\fontsize{17.28}{21.5}\selectfont 河村 和紀\par}
\end{minipage}
\hfill
\end{minipage}}

\vspace{0.56cm}
\noindent\makebox[\textwidth][c]{\rule{150mm}{0.9pt}}
\vspace{0.03cm}

\noindent\makebox[\textwidth][c]{%
\begin{minipage}{150mm}
\begin{minipage}[b]{125.3mm}
\raggedleft
{\rmfamily\fontsize{12}{14.4}\selectfont
A Doctoral Dissertation, 2025\\
Doctor of Interdisciplinary Informatics\\
Degree awarded September 19, 2025\par}
\vspace{3.93pt}
{\rmfamily\fontsize{12}{14.4}\selectfont
Applied Computer Science Course\\
Graduate School of Interdisciplinary Information Studies\\
The University of Tokyo\par}
\end{minipage}
\hspace{4.2mm}
\begin{minipage}[b]{12.5mm}
\raggedright
\includegraphics[trim=156.75bp 54bp 160.5bp 55.5bp,clip,width=12.5mm]{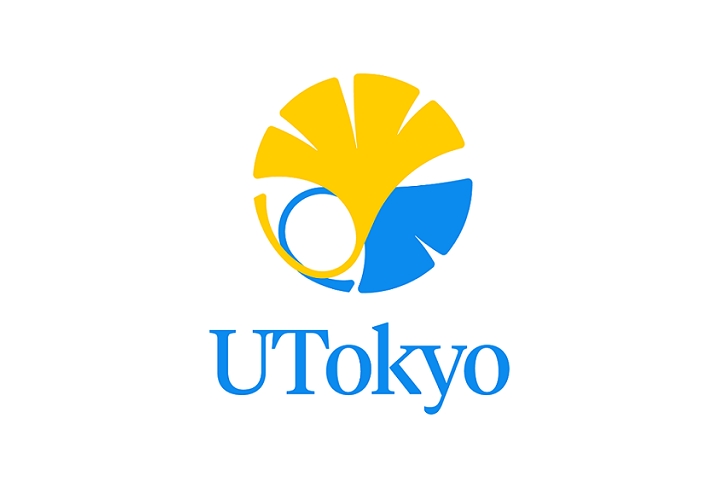}
\end{minipage}
\hfill
\end{minipage}}

\vspace*{2.586cm}
\newpage
\begin{eabstract}
The advancement of digital technology is driving the diversification of learning media. With the widespread adoption of the internet and the advancement of streaming technology, audio and video content have become established as primary learning tools alongside traditional text-based materials. In particular, the widespread adoption of mobile devices such as smartphones and tablets has enabled learners to access this content anytime and anywhere, leading to a significant recognition of its educational value. Audio and video media offer unique advantages that are difficult to achieve through text-based learning. For example, lecture videos integrate verbal explanations with visual demonstrations, educational podcasts enable learning while commuting or working, and videos capturing experts' performances provide learners with opportunities to observe and emulate advanced skills. In this way, audio and video content alleviate temporal and spatial constraints on learning, enabling richer and more diverse learning experiences.

However, two fundamental challenges specific to audio-visual learning exist: (1) the time inefficiency of needing to consume long-form content sequentially from beginning to end, and (2) the absence of scalable feedback mechanisms for skill acquisition through imitation. To improve both time efficiency and learning outcomes in audio-visual learning, this research proposes an integrated learning support framework that provides necessary support at each stage as learners progress through three interconnected phases: Consume (efficient content consumption), Understand (deep content comprehension), and Imitate (skill acquisition). To realize this framework, we leverage deep learning (AI) technologies that excel in automatically learning patterns from large-scale data and extracting high-level features from audio, video, and text data. To address the complete learning cycle from content consumption to practical skill acquisition, we developed three deep learning systems: a phoneme-level speed optimization system for audio content, a multimodal video summarization system that preserves visual and auditory information, and an automated feedback system for imitation-based skill learning.

The first study, ``AIxSpeed," addresses the listening inefficiency of long-form audio content where speech clarity varies significantly across segments. When podcasts and audiobooks lasting several hours are played at constant speed, time is wasted on easily comprehensible sections while important content in difficult-to-hear sections is missed due to the same playback speed. Based on the hypothesis that speech recognition model confidence correlates with human listening difficulty, we developed a system that dynamically adjusts playback speed at the phoneme level using only acoustic features without requiring semantic analysis. Technical evaluation produced average playback factors of 1.30x for LibriSpeech and 1.29x for UME-ERJ. In a blind evaluation with 50 participants, the resulting speech received higher mean opinion scores than constant-speed playback matched to the same average speeds.

The second study, ``FastPerson," tackles the challenge of efficient learning from lecture videos where critical information is distributed across multiple modalities—spoken explanations, visual slides, and demonstrations. Existing summarization methods primarily generate text transcripts, losing rich multimodal nature and breaking continuity with the original content. Our system analyzes both visual and audio streams to generate condensed videos, using voice cloning technology to maintain the original speaker's voice and preserve continuity with the original lecture. Furthermore, to enable efficient access to necessary information according to one's comprehension level, we implemented an interface that allows learners to seamlessly switch between summary and full versions on a chapter-by-chapter basis. In experiments with 40 participants watching educational videos, FastPerson achieved an average 53\% reduction in viewing time, with no statistically significant difference in quiz scores compared with normal playback; 78\% of participants rated the switching functionality as useful or very useful.

The third study, ``Profy," develops a feedback system for the problem that learners cannot objectively evaluate how much their performance deviates from expert models and where the differences lie in skill acquisition through imitation. Taking second-language pronunciation as a concrete example with applicability to various skill domains, we constructed a deep learning model that directly learns what constitutes good and poor performance from unannotated audio data. The system highlights waveform regions emphasized by the classifier and shows model-derived acoustic distances from native-speaker distributions in latent space. In an experiment with 10 Japanese learners of English evaluated by five American raters, Profy showed a larger observed improvement in pronunciation intelligibility than elicited imitation; unlike elicited imitation, its pre- and post-practice confidence intervals did not overlap.

These three systems demonstrate how AI can support each stage of learning. AIxSpeed enables efficient information acquisition in the consumption stage by adapting to an ASR-based estimate of speech-recognition difficulty. FastPerson reduces learning time while presenting multimodal information, with no statistically significant quiz-score difference from normal playback. Profy provides model-derived feedback in the imitation stage. Together, these systems support time-efficient consumption, content access, and repeated skill practice.

In conclusion, this research has established a technical foundation for audio-visual learning support in the digital education era. Through the development of systems addressing specific challenges at each stage of the learning cycle—from efficient information discovery and understanding to practical skill acquisition through imitation—we demonstrated a 53\% reduction in viewing time and observed improvements in pronunciation intelligibility. These technologies alleviate traditional temporal constraints and comprehension difficulties in learning from audio-visual content, enabling more efficient and effective learning. As a result, audio-visual learning becomes a more practical and attractive option for learners, promoting diversification of learning methods. While current limitations include the need for long-term effect validation and extension beyond audio-visual domains, this work demonstrates new application possibilities for AI technologies in audio-visual learning support.
\end{eabstract}
\pagenumbering{roman}
\makeabstract

\begin{acknowledge}
I would like to express my deepest gratitude to \textbf{Professor Jun Rekimoto}, my principal supervisor at the Interfaculty Initiative in Information Studies, The University of Tokyo. Over the past five years\,---\,including my tenure as a student researcher at the Kyoto Lab of Sony Computer Science Laboratories (Sony CSL Kyoto)\,---\,his insightful guidance and unfailing support have been indispensable to both my academic progress and personal growth.

I am also sincerely grateful to the following members of the Rekimoto Laboratory\,---\,\textbf{Associate Professor Yoshio Ishiguro}, \textbf{Project Associate Professor Yuta Itoh} (both at the Interfaculty Initiative in Information Studies, The University of Tokyo), and \textbf{Associate Professor Hiromi Nakamura} (Faculty of Informatics, Tokyo City University)\,---\,whose constructive discussions and expert advice during our regular meetings have greatly enriched my research.

My academic journey has been shaped by several distinguished mentors. I wish to thank \textbf{Professor Kuniaki Uehara}, Osaka Gakuin University \emph{and Professor Emeritus, Kobe University}, and \textbf{Professor Takashi Matsubara}, Hokkaido University, for providing me with a solid foundation in artificial intelligence during my undergraduate studies, as well as \textbf{Professor Akihiro Yamamoto}, Kyoto University, for guiding me in research methodology throughout my master's program.

I am deeply appreciative of my fellow doctoral candidates, senior colleagues, and all members of the Rekimoto Laboratory. Their camaraderie and collaborative spirit have fostered a stimulating and supportive research environment. I also wish to acknowledge the anonymous reviewers of my conference papers and journal articles; their insightful comments and constructive feedback have substantially improved the quality of my publications.

My heartfelt thanks extend to the team at Sony CSL Kyoto. \textbf{Dr.\,Lana Sinapayen} and \textbf{Dr.\,Yuichiro Takeuchi} offered valuable discussions during my tenure as a student researcher. I am likewise grateful for the administrative support, patent assistance, and thoughtful conversations provided by \textbf{General Manager Kojiro Kashiwa}, and \textbf{Ms.~Mari Mitani}.

I would also like to express my sincere appreciation to \textbf{Mr.~Ryo Mukaiyama}, General Manager, and \textbf{Mr.~Kei Tateno}, Manager, at Sony Group Corporation. Their encouragement and willingness to allow me to pursue this doctoral degree while continuing my professional duties have been instrumental to my academic journey.

Finally, I wish to express my profound gratitude to my family. Their unwavering encouragement and understanding have been the cornerstone of this endeavor, giving me the strength and resolve to complete my doctoral research.
\end{acknowledge}

\frontmatter
\tableofcontents
\listoffigures
\listoftables
\mainmatter

\chapter{Introduction}

\begin{quote}\itshape
    ``The public is indoctrinated to believe that skills are valuable and reliable only if they are the result of formal schooling.''
    \begin{flushright}
    --- Ivan Illich, \textit{Deschooling Society}, 1971
    \end{flushright}
\end{quote}

\section{Motivation and Historical Context: The Challenge of Efficient Learning from Audio-Visual Media}
\label{sec:intro_motivation_historical_context}

\begin{figure}[t]
    \centering
    \includegraphics[width=0.8\textwidth]{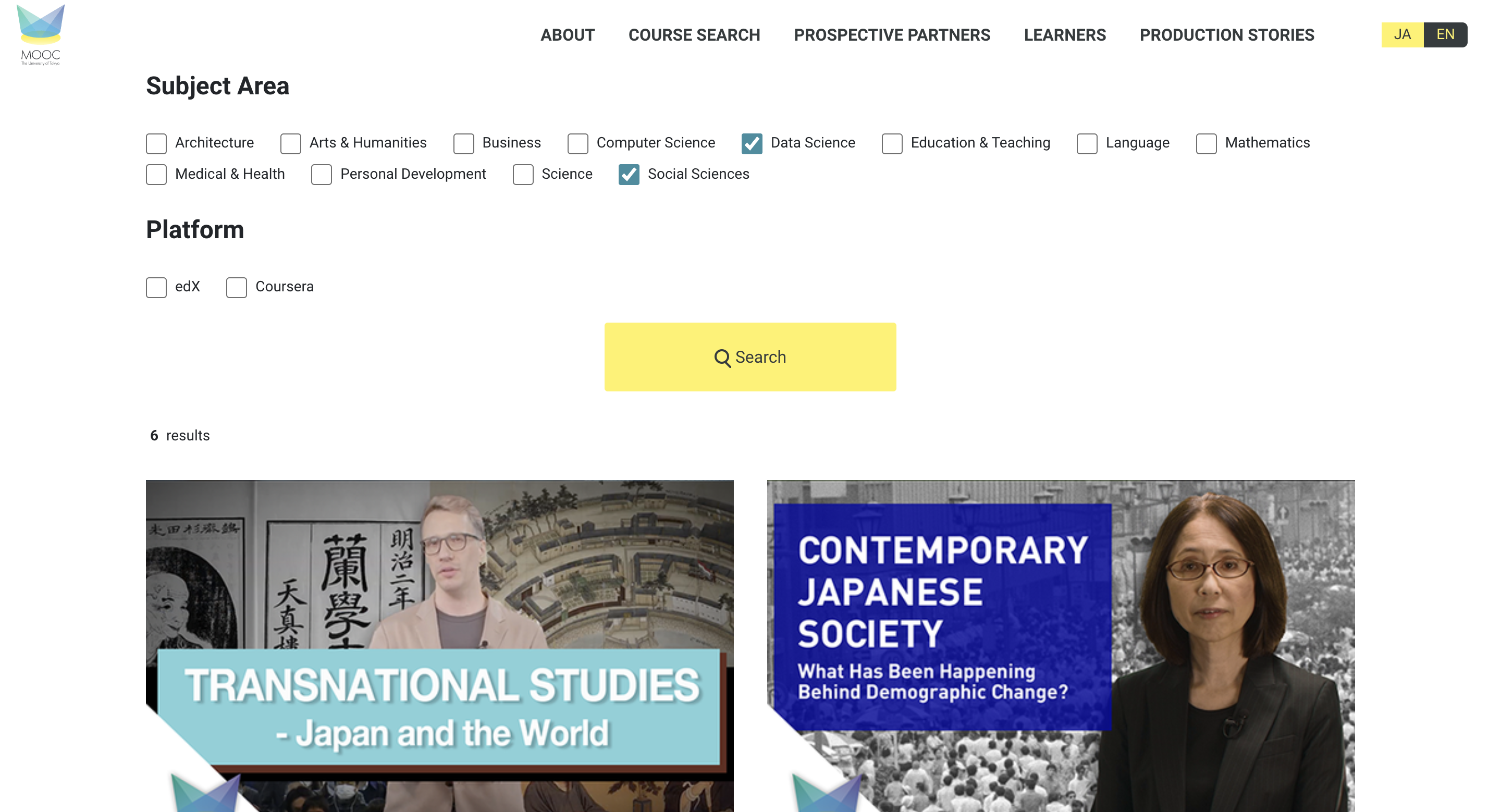} 
    \caption[An example of a Massive Open Online Course (MOOC) interface]{An example of a Massive Open Online Course (MOOC) interface, typically featuring a video lecture, supplementary materials, and interactive elements. This illustrates the shift towards accessible online learning environments.}
    \label{fig:mooc_example}
\end{figure}

The proliferation of the internet and digital technologies has fundamentally transformed the educational paradigm. Learning that was once constrained by physical classrooms and fixed curricula is now possible anytime, anywhere. The emergence of MOOCs (see Figure~\ref{fig:mooc_example}), educational video platforms, podcasts, and audiobooks has democratized access to expertise worldwide~\cite{Dillahunt2014, Martin2020}, enabling learning that transcends geographical, economic, and temporal constraints.

This transformation deeply resonates with the vision Ivan Illich proposed in ``Deschooling Society"~\cite{Illich1971}. Illich criticized the monopoly of institutionalized schooling and advocated for building ``learning webs" where individuals could autonomously learn from diverse resources. Modern online learning environments are realizing parts of this ideal. Learners can now freely choose expert lectures, demonstrations, and experiences according to their interests and pace, beyond the boundaries of formal educational institutions.

The most notable change brought by this democratization of learning is the emergence of audio and video as primary educational media. Unlike traditional text-based materials, audio-visual content provides rich learning experiences including actual expert performances, nuanced explanations, and visual demonstrations. As multimedia learning theory suggests~\cite{mayer2020multimedia,mayer2003social}, lecture recordings, educational podcasts, tutorial videos, and audiobooks go beyond mere information transmission to provide learners with a sense of ``being with the expert."

Audio content in particular has created new learning opportunities through its portability. Listening to podcasts during commutes, enjoying audiobooks while exercising, playing lecture recordings while doing housework—these all embody the ``learning anytime, anywhere" world that Illich envisioned. Similarly, video content enables visual explanations of complex concepts, practical demonstrations, and virtual laboratory experiences from remote locations, aligning with principles of effective educational videos~\cite{brame2016effective}.

\begin{figure}[t]
    \centering
    \includegraphics[width=0.9\linewidth]{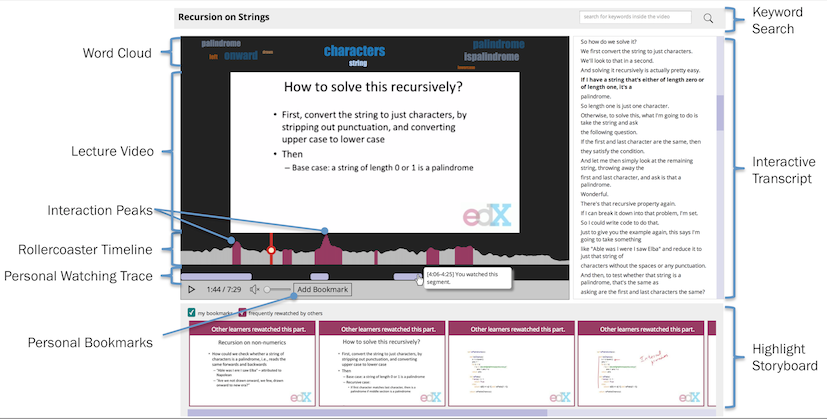}
    \caption[Information-density variations inferred from collective
             click-frequency traces]{Information-density variations inferred from collective
             click-frequency traces in an online lecture video
             (reconstructed from the LectureScape interface~\cite{kim2014understanding}).}
    \label{fig:linear_consumption_challenge}
\end{figure}

However, the existence of these abundant audio-visual resources also creates new challenges. First, due to the linear and time-bound nature of these media, learners must consume lengthy content to obtain necessary information. A large-scale analysis of 6.9 million MOOC video-viewing sessions found that the fraction viewed decreased as videos became longer and that median engagement was less than half of the duration for videos longer than nine minutes~\cite{Guo2014} (see Figure~\ref{fig:linear_consumption_challenge}). This problem becomes even more pronounced with podcasts and audiobooks.

\begin{figure}[t]
  \centering
  \includegraphics[width=0.8\textwidth]{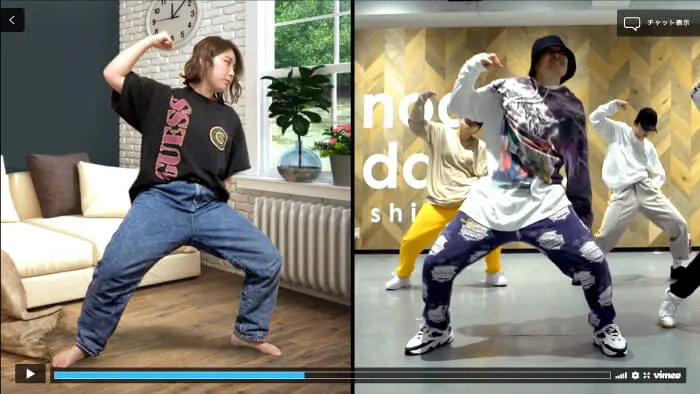}
  \caption[An example of imitation learning in an online dance class]{An example of imitation learning in an online dance class. Learners attempt to replicate the instructor's postures and movements, often without immediate, granular feedback on their performance, highlighting a key challenge in online skill acquisition.}
  \label{fig:online_dance_class}
\end{figure}

Second, observational learning—watching and imitating expert performances—is what audio-visual media excels at, yet it also reveals its greatest limitation. Learners can repeatedly listen to native speaker pronunciation, observe skilled musicians' performances, and study professional athletes' movements (see Figure~\ref{fig:online_dance_class}). For instance, in an online dance class, learners try to mimic the instructor's movements. However, objectively evaluating how closely their own attempts approach the expert model remains extremely difficult. This ``lack of feedback" remains an unresolved challenge in computer-assisted learning systems~\cite{neri2002pedagogy}.

This research proposes solutions to these audio-visual media-specific challenges using deep learning technologies. Our goal is to realize autonomous and effective learning with modern technology. Specifically, we build an integrated AI-guided system that can efficiently extract knowledge from lengthy content (Consume and Understand) and support practice of observed skills through model-derived feedback (Imitate).

Through this, we aim to maximize the potential of rich educational media like audio and video while overcoming their inherent constraints, realizing a truly ``effective learning anytime, anywhere" environment. We hope this will contribute not merely as a technical solution but to building a new educational ecosystem where learners can proactively design and optimize their own learning.

\section{The Dual Challenge: Inefficient Content Consumption and Lack of Objective Feedback}
\label{sec:intro_dual_challenge}

The challenges facing modern learners are not merely about access to information, but about how to efficiently process that information and transform it into practical skills. This research identifies two interrelated challenges in learning with audio and video content.

\subsection{Challenge 1: Time-Inefficient Consumption of Long-Form Audio-Visual Content}
\label{subsec:intro_challenge_time_efficiency}

The first challenge is the time inefficiency in consuming long-form audio-visual content. A typical university lecture lasts 60-90 minutes, podcast episodes range from 30-180 minutes, and audiobooks span several hours to tens of hours. However, as recent learning analytics research reveals~\cite{kim2014understanding}, the truly valuable information for learners often constitutes only a portion of the entire content.

In audio content particularly, information density is not constant. Some segments, such as introductions, repetitions, filler words, or explanations of familiar concepts, can be understood at high speed without hindering comprehension. Conversely, sections with technical terminology, complex concepts, or unclear speaker pronunciation require listening at standard speed or slower. Current uniform playback speed adjustments cannot accommodate this variable information density, as studies on adaptive playback have shown~\cite{pastore2015using}.

Video content presents an even more complex challenge as it requires processing visual information in addition to audio. According to cognitive load theory~\cite{sweller2019cognitive}, learners must integrate information from multiple modalities—speaker explanations, text and diagrams on slides, demonstrations—while efficiently finding and understanding necessary information.

\subsection{Challenge 2: Absence of Scalable, Objective Feedback for Imitation-Based Learning}
\label{subsec:intro_challenge_feedback}

The second challenge is the lack of objective, scalable feedback in imitation-based learning. Many practical skills (language pronunciation, musical instrument performance, sports techniques) are acquired by observing and imitating expert performances. However, it is extremely difficult for learners to objectively evaluate the differences between their own performance and the expert model.

In traditional face-to-face education, teachers directly observe learner performances and provide specific feedback. However, in online learning environments, providing such personalized feedback at scale is impractical from cost and logistics perspectives~\cite{kasch2021educational}. As a result, many learners must rely on self-assessment, which tends to be subjective and inaccurate.

Moreover, many existing automated feedback systems are based on predefined error patterns and cannot adapt to individual learners' specific difficulties or progress~\cite{witt2000phone}. These systems typically provide only binary judgments of ``correct" or ``incorrect," lacking detailed information about where and how differences occur and how to improve.

These two challenges are not independent but closely related in an effective learning process. Learners must first efficiently consume content to acquire knowledge (Challenge 1), then require appropriate feedback when transforming that acquired knowledge into practical skills (Challenge 2). This research aims to solve these challenges in an integrated manner.

\section{The Bottleneck of Long-Form Media in Digital Learning}
\label{sec:intro_bottleneck_long_form_media}

\subsection{Quantifying the Time Cost of Audio-Visual Learning}
\label{subsec:intro_time_cost}

Quantitatively examining the time cost of audio-visual learning reveals its inefficiency. For example, a typical university online course includes 2-3 hours of video lectures per week over 10-15 weeks, totaling 20-45 hours of content. In a large-scale MOOC log analysis, median engagement time was at most six minutes regardless of video length, and the fraction viewed decreased for longer videos~\cite{Guo2014}.

In the realm of podcasts and audiobooks, the situation is even more pronounced. Popular educational podcasts average 60-90 minutes per episode, with many series containing over 100 episodes. Audiobooks average 8-12 hours for non-fiction titles, with more comprehensive works exceeding 20 hours. For learners seeking to leverage these resources for professional development or lifelong learning, the time investment can become prohibitive.

The linear nature of audio content in particular presents unique challenges. Unlike text, audio cannot be easily scanned or jumped to relevant sections. Learners must listen through irrelevant content to reach specific information of interest. This ``forced linear consumption" creates a significant time burden, especially for learners searching for specific topics or concepts, as noted in studies on video navigation~\cite{kim2014understanding}.

\subsection{The Variable Nature of Content Complexity and Clarity}
\label{subsec:intro_content_variability}

The complexity and clarity of audio-visual content varies significantly even within the same content. Acoustic analysis reveals the following variations in educational audio content:
(1) Speech rate variation: ranging from 100 to over 200 words per minute even for the same speaker
(2) Pronunciation clarity: varying due to technical terms, foreign words, or speaker fatigue  
(3) Background noise: signal-to-noise ratio (SNR) changes depending on recording environment
(4) Information density: varies greatly between sections such as concept explanations, examples, and summaries

Current common approaches of uniform playback speed changes have fundamental limitations for these variations. Research shows that uniform high-speed playback exceeding 1.5x significantly reduces comprehension of complex content and increases cognitive load~\cite{Murphy2022LectureSpeed, pastore2015using}. However, a recent meta-analysis shows that, for clear and simple material, comprehension remains largely intact up to roughly 1.5x speed~\cite{tharumalingam2025increasing}.

\subsection{Limitations of Current Navigation and Summarization Approaches}
\label{subsec:intro_current_limitations}

Current navigation and summarization approaches for audio-visual content have several fundamental limitations. While transcription through automatic speech recognition (ASR) has become widely available, it provides only a partial solution.

First, for audio content (podcasts, audiobooks), transcription loses prosodic features of speech (emphasis, emotion, irony, etc.). These features play important roles in conveying speaker intent and importance, especially in educational content. Furthermore, transcriptions of long audio become massive texts themselves, creating new information overload problems.

Second, for video content, many existing summarization technologies rely only on speech transcripts and do not adequately utilize visual information (slides, diagrams, demonstrations, body language). Recent work on abstractive video lecture summarization highlights this limitation~\cite{benedetto2024abstractive}. This loses the richness of video content's multimodal nature and can generate incomplete or misleading summaries.

Third, many current summarization systems present summary results as new text, which breaks continuity with the original content and can reduce learner trust and engagement. In particular, the original speaker's voice and speaking style play important roles in educational content, and preserving these is important from a learning effectiveness perspective, as research on multimodal feedback suggests~\cite{mayer2003social}.

\section{The Feedback Deficit in Imitation-Based Skill Acquisition}
\label{sec:intro_feedback_deficit}

\subsection{The Critical Role of Feedback in Skill Learning}
\label{subsec:intro_feedback_importance}

In practical skill acquisition, feedback is an essential element. Cognitive science research shows that skill learning progresses through cycles of observation, practice, feedback, and correction~\cite{bandura1977social}. However, in online learning environments, the ``feedback" stage of this cycle is notably lacking.

For imitation-based skills (language pronunciation, musical instrument performance, physical movements), learners need the following specific information:
(1) Which parts of their performance deviate from the expert model
(2) The degree of that deviation
(3) Which direction corrections should take
In traditional face-to-face instruction, skilled teachers provide this information immediately, but achieving this in large-scale online environments has been difficult with current technology, as motor learning research demonstrates~\cite{sullivan2008motor}.

\subsection{The Scalability Challenge of Human Feedback}
\label{subsec:intro_scalability_challenge}

While feedback from human experts is high quality, it faces fundamental scalability problems. For example, if 1000 learners each practice pronunciation 5 times per week, providing 2 minutes of feedback per practice would require approximately 167 hours of expert time per week. This is equivalent to about 4 full-time teachers. On large-scale platforms like MOOCs where enrollment can reach tens of thousands, this problem becomes even more severe~\cite{kasch2021educational}.

Furthermore, in asynchronous online learning environments, feedback delay is also an important issue. Learning science research shows that immediate feedback is more effective for learning than delayed feedback~\cite{sullivan2008motor}. However, in systems dependent on human experts, feedback delays are unavoidable.

\subsection{Limitations of Existing Automated Feedback Systems}
\label{subsec:intro_existing_limitations}

Existing automated feedback systems are mainly classified into two approaches: rule-based systems and supervised learning systems. Both have fundamental limitations in providing the flexible and adaptive feedback needed in online learning contexts.

Rule-based systems rely on databases containing predefined error patterns and their correction methods. While these systems function for anticipated errors, they cannot handle individual learners' unique error patterns or errors not anticipated during system design. Recent systematic reviews of CAPT systems confirm these limitations~\cite{ec2022systematic}. Additionally, adapting to new languages, dialects, or skill domains requires complete redesign.

Supervised learning systems require large amounts of labeled data (pairs of correct and incorrect examples). Collecting and annotating such data is time-consuming and costly, and particularly impractical when needing to cover diverse error patterns and learner backgrounds. Furthermore, these systems tend to have low generalization ability for patterns not included in the training data.

Due to these limitations, many current automated feedback systems are limited to providing binary ``correct/incorrect" judgments or limited quantitative scores. They cannot provide the detailed, actionable feedback that learners truly need regarding ``where," ``how much," and ``how" their performance differs. Recent advances in ASR-based pronunciation training still struggle with this granularity~\cite{ngo2024effectiveness}.

\begin{figure}[t]
  \centering
  \includegraphics[width=0.8\linewidth]{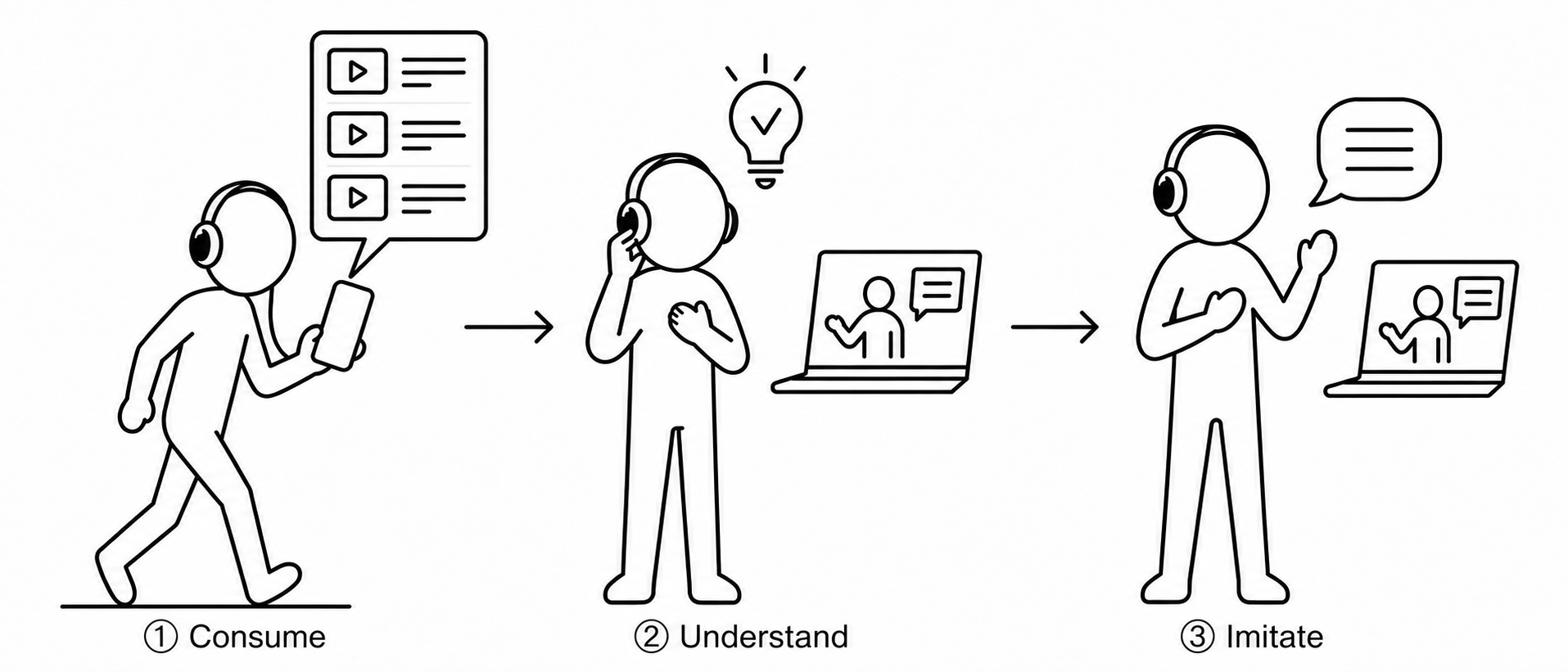}
  \caption[The iterative Consume–Understand–Imitate learning loop]{
    The iterative \textit{Consume–Understand–Imitate} learning loop.
    Each stage is later augmented by a dedicated AI subsystem
    (\emph{AIxSpeed}, \emph{FastPerson}, and \emph{Profy}; see \S\ref{subsec:intro_proposed_systems}).}
  \label{fig:cui_pipeline}
\end{figure}

\section{Towards an Integrated Solution: The Consume-Understand-Imitate Learning Pipeline and AI-Guided Learning}
\label{sec:intro_integrated_solution}

\subsection{The Natural Flow of Skill-Based Learning}
\label{subsec:intro_learning_flow}

For the purposes of this dissertation, we organize practical skill learning as a staged process. Informed by multimedia learning research~\cite{mayer2020multimedia}, we conceptualize learning with audio and video as three interrelated stages: Consume, Understand, and Imitate. This model is an organizing framework proposed in this dissertation rather than an empirically established sequence. Figure~\ref{fig:cui_pipeline} visualises the cyclical relationship of these three stages.

In the Consume stage, learners explore and identify relevant information from large amounts of available content. The main challenge at this stage is time efficiency. They need to quickly find parts relevant to their learning goals from lengthy podcasts, lecture recordings, and educational videos.

In the Understand stage, learners deeply process identified information and build conceptual understanding. The challenge here is to efficiently grasp core concepts while dealing with content complexity and variations in information density. Speech clarity, speaking rate, and use of technical terms affect the speed and depth of understanding, as cognitive load theory predicts~\cite{sweller2019cognitive}.

In the Imitate stage, learners put observed knowledge and skills into practice. Most important at this stage is informative feedback on their own performance. Learners need specific information to recognize differences from expert models and correct them, aligning with Bandura's observational learning theory~\cite{bandura1997self}.

\subsection{AI-Guided Learning: A Framework for Integrated Support}
\label{subsec:intro_ai_guided_learning}

This research proposes ``AI-Guided Learning" as a new paradigm for comprehensively supporting these learning stages. AI-Guided Learning is an integrated approach that leverages deep learning technologies to provide adaptive support at each learning stage, enabling learners to efficiently and effectively acquire knowledge and skills from audio-visual content.

The core of AI-Guided Learning is that AI functions as a ``guide" for learners. While traditional AI education systems focused on automating specific tasks, AI-Guided Learning aims to reduce the cognitive load required at each stage and maximize learning effectiveness while maintaining learner autonomy. This aligns with recent perspectives on AI in education that emphasize augmentation rather than replacement~\cite{crompton2023artificial}.

AI-Guided Learning is based on three principles:
(1) Adaptivity: Dynamic adjustment according to learners' individual needs, comprehension speed, and skill levels
(2) Integration: Seamless coordination and mutual support between Consume-Understand-Imitate stages
(3) Transparency: Presentation of AI's judgments and recommendations in a form learners can understand

\subsection{The Need for Integrated Technological Support}
\label{subsec:intro_integrated_support}

The three stages of Consume-Understand-Imitate are not linear but iterative and cyclical. AI-Guided Learning recognizes and leverages this cyclical nature. For example, learners who receive feedback in the imitation stage may return to specific content segments to deepen understanding (revisiting the Understand stage) or search for new expert examples (revisiting the Consume stage).

Many current educational technologies are ``point" solutions focused on one of these stages. Video summarization tools, playback speed control apps, pronunciation checkers are each useful, but lack an integrated approach supporting the entire learning process. AI-Guided Learning integrates these individual functions and provides a comprehensive framework that offers appropriate tools and support according to learners' situations.

\subsection{AI-Driven Subsystems for AI-Guided Learning}
\label{subsec:intro_proposed_systems}

To realize AI-Guided Learning, this research develops three AI-driven subsystems. These systems are designed to support specific stages of the Consume-Understand-Imitate pipeline while providing an integrated AI-Guided Learning experience as a whole.

AIxSpeed handles the efficiency of Consume-Understand stages in AI-Guided Learning. Through phoneme-level acoustic feature analysis, it adjusts playback speed while using speech-recognition performance as a proxy for intelligibility.

FastPerson revolutionizes the Consume-Understand stages for long-form video content in AI-Guided Learning. By integrating multimodal information and providing summaries in the original speaker's voice using recent voice cloning advances~\cite{xtts2024}, it guides learners to efficiently grasp core content while also providing options to access details as needed.

Profy guides learners to objectively evaluate and improve their own skills in the Imitate stage of AI-Guided Learning. By analyzing and visualizing differences from expert performances in detail using self-supervised learning techniques~\cite{kim2022automatic}, it enables precise skill acquisition that was traditionally difficult without human instructors.

Through integration of these systems, AI-Guided Learning provides learners not merely a collection of individual tools but a coherent learning experience. Following AI guidance, learners can realize a complete learning cycle: efficiently finding necessary information from large amounts of content (Consume), deeply understanding it (Understand), and mastering it as practical skills (Imitate).

AI-Guided Learning aims for essential transformation of learning beyond technological innovation. It establishes a new learning paradigm where learners can progress efficiently and effectively toward their goals without getting lost in the vast sea of digital resources.

\section{Research Objectives and Questions}
\label{sec:intro_research_objectives_questions}

\subsection{Overall Research Objective}
\label{subsec:intro_overall_objective}

The overarching objective of this research is to solve three fundamental challenges in learning with audio and video under the AI-Guided Learning framework:
(1) Time inefficiency in listening to long-form audio content
(2) Difficulty in integrated summarization of visual and audio information in long-form video content
(3) Difficulty in identifying specific locations of errors in imitation learning

To achieve this objective, we set the following central research question:

\begin{quotation}
\noindent
\textbf{Central Research Question (CRQ):}
To what extent can adaptive support at each stage of consuming, understanding, and imitating audio-visual content through deep learning technologies achieve reductions in learning time and improvements in learning outcomes?
\end{quotation}

\subsection{Specific Research Questions}
\label{subsec:intro_specific_questions}

From this central question, we set three specific research questions:

\begin{itemize}
    \item \textbf{RQ1: Balancing Listening Time Reduction and Comprehension Maintenance in Audio Content}
    
    \textbf{Underlying Challenge:} Long-form audio content such as podcasts, audiobooks, and interviews is typically played at constant speed, despite variations in content importance and speaker clarity across segments. This results in wasting time on easily understandable parts and missing content in difficult-to-understand parts.
    
    \textbf{Research Question:} By automatically determining the listening difficulty of each part of audio content and optimizing playback speed for each segment, to what extent can we reduce listening time while maintaining comprehension? How much does such variable-speed playback improve listenability compared to constant high-speed playback?

    \item \textbf{RQ2: Efficient Summarization Integrating Visual and Audio Information in Lecture Videos}
    
    \textbf{Underlying Challenge:} Lecture videos contain multiple information channels including speaker explanations (audio), slides and board writing (visual), and gestures, but conventional summarization methods cannot handle these integratively. Additionally, when summary results are presented only as text, the original lecture context and speaker characteristics are lost, disrupting the continuity of the learning experience.
    
    \textbf{Research Question:} By integratively analyzing visual and audio information in lecture videos and providing summaries in the original video format, to what extent can we reduce viewing time? What impact does an interface allowing switching between summary and original videos have on learner comprehension?

    \item \textbf{RQ3: Model-Derived Localization of Performance Differences in Imitation Learning}
    
    \textbf{Underlying Challenge:} In imitation-based skill acquisition such as pronunciation learning, learners cannot objectively understand which parts of their performance deviate from the target. Conventional systems either provide overall scores or detect only specific error patterns, and cannot point out problem areas specific to individual learners in detail.
    
    \textbf{Research Question:} Is it possible to highlight model-derived regions associated with performance differences using small amounts of learner data and large amounts of expert data? What impact does such localized feedback have on learners' pronunciation improvement compared with an elicited-imitation baseline?
\end{itemize}

For these research questions, we will conduct the following evaluations:
For RQ1, we will measure subjective evaluation of listenability and actual time reduction rate;
For RQ2, we will conduct viewing time reduction rate and comprehension evaluation through quizzes;
For RQ3, we will verify the improvement in pronunciation evaluation by experts.

Through these studies, we evaluate time efficiency, perceived audio quality, quiz-measured comprehension, and pronunciation intelligibility within the scope of the three prototypes.

\section{Significance and Contributions of the Research}
\label{sec:intro_significance_contributions}

\subsection{Technical Innovations}
\label{subsec:intro_technical_innovations}

This research has the following technical novelties in developing audio and video processing technologies specifically for educational applications:

\textbf{1. Phoneme-level playback speed optimization using ASR confidence (AIxSpeed):}
Unlike conventional uniform speed changes, this method uses speech recognition model's recognition probability as a proxy for ``listenability" and dynamically controls playback speed for each phoneme. A key feature is determining optimal speed for each part based solely on acoustic features without requiring semantic understanding. This technology can be extended to other audio processing applications such as hearing aids that automatically slow down difficult-to-hear parts for the hearing impaired.

\textbf{2. Video-to-video summarization preserving visual and auditory information (FastPerson):}
While most existing video summarization technologies are limited to audio transcription or text output, this method integratively analyzes visual information (slides, board writing, gestures) and audio information, reconstructing summarized content as video with the original speaker's voice. It is particularly unique in allowing learners to select viewing granularity according to their understanding through an interface that switches between summary and original videos by chapter.

\textbf{3. Model-derived localization and visualization of pronunciation differences with small data (Profy):}
While conventional CAPT systems require detailed annotated data and comparison with specific speakers, this method learns pronunciation quality in a data-driven manner from small amounts of unannotated learner speech and large amounts of native speech. It is novel in providing more localized information than an overall score by highlighting waveform regions emphasized by the classifier and visualizing distances in latent space. This framework, inspired by recent self-supervised learning advances~\cite{gui2023survey}, is also expected to be applicable to imitation learning tasks beyond pronunciation.

\subsection{Validated Educational Impact}
\label{subsec:intro_educational_impact}

The principal empirical results reported for the three systems are as follows:

\textbf{Adaptive playback with improved listenability (AIxSpeed):}
Blind comparison experiments confirmed higher mean opinion scores than constant-speed playback matched to AIxSpeed's average playback factors (1.30x for LibriSpeech and 1.29x for UME-ERJ). The UME-ERJ result also suggests that AIxSpeed can improve the listenability of non-native-speaker utterances for native listeners.

\textbf{Halving viewing time without a significant quiz-score difference (FastPerson):}
In lecture-video viewing experiments with 40 participants, an average 53\% reduction in viewing time was achieved, and quiz-measured comprehension showed no statistically significant difference from conventional normal playback. Among the FastPerson users, 78\% rated the chapter-switching function as useful or very useful.

\textbf{Improved pronunciation clarity and helpful feedback features (Profy):}
In experiments with Japanese learners of English, American raters judged post-practice speech to be more intelligible, and all participants rated the pronunciation-score display as helpful.

\subsection{Contribution to AI-Guided Learning Paradigm}
\label{subsec:intro_broader_implications}

The most important contribution of this research is demonstrating a new learning support paradigm called AI-Guided Learning, beyond the three individual technologies. We showed specific ways AI can augment learner capabilities at each stage of Consume (efficient browsing), Understand (deep comprehension), and Imitate (skill acquisition).

These technologies, under the common vision of ``augmenting human learning and comprehension abilities with AI," build a comprehensive support ecosystem from video viewing to language use by covering different learning phases. In the future, this framework is expected to be extended to other learning domains and support for acquiring more advanced cognitive and physical skills, as envisioned in recent discussions of AI's future in education~\cite{zhang2023generative}.

\section{Dissertation Structure}
\label{sec:intro_dissertation_structure}

This dissertation is structured as follows to systematically present the concept, technical implementation, empirical evaluation, and educational implications of AI-Guided Learning:

\textbf{Chapter 2: Related Work} comprehensively reviews existing research in three areas to clarify the academic positioning of this research: (1) adaptive audio/video playback technologies, (2) educational content summarization technologies, (3) feedback systems in imitation learning. We clarify the technical limitations of each area and the novelty of our approach.

\textbf{Chapter 3: AIxSpeed—Adaptive High-Speed Audio Playback} details the phoneme-level playback speed control system based on ASR confidence. We present the system architecture, speed optimization algorithm, and results of evaluation experiments on listenability.

\textbf{Chapter 4: FastPerson—Multimodal Video Summarization} explains the design and implementation of a video summarization system integrating visual and audio information. We report on chapter-based summary generation methods, re-narration through voice synthesis, and evaluation results from viewing experiments with 40 participants.

\textbf{Chapter 5: Profy—Automated Feedback for Pronunciation Learning} presents a system that learns a pronunciation-proficiency classifier from small data and visualizes classifier-emphasized regions and model-derived distances. We explain the learning method, feedback generation, and evaluation with Japanese learners of English.

\textbf{Chapter 6: Integration and Discussion} discusses integratively how the three systems support the Consume-Understand-Imitate learning cycle. We consider design principles, implementation challenges, and ethical considerations of AI-Guided Learning.

\textbf{Chapter 7: Conclusion and Future Directions} summarizes the academic and practical contributions of this research and provides perspectives on future development possibilities of the AI-Guided Learning paradigm. We discuss challenges and opportunities for extension to other learning domains, integration with more advanced AI technologies, and large-scale deployment in educational settings.

\section{Concluding Remarks}
\label{sec:intro_concluding_remarks}

The proliferation of online education has dramatically improved access to learning resources. However, as this research has revealed, the true challenges lie beyond access. How to efficiently consume long-form audio-visual content, how to effectively understand complex audiovisual information, and how to reliably acquire observed skills—for these challenges, this research has presented an integrated solution called AI-Guided Learning.

The three systems—AIxSpeed, FastPerson, and Profy—each developed as independent research results, create greater educational value when integrated. These are not merely time-saving tools but ``learning augmentation technologies" designed to help learners maximize their capabilities.

The vision of AI-Guided Learning presented in this dissertation suggests a new relationship between technology and education. AI becomes a partner that augments the capabilities of teachers and learners rather than replacing them, enabling more effective learning. We hope this research will be a step toward realizing a more equitable and efficient learning environment where all learners can receive high-quality education.

\clearpage

\chapter{Related Work}
\label{ch:relatedwork}

\section{Introduction}
\label{sec:relwork_intro}

The vision of AI-Guided Learning emerges from decades of research in human-computer interaction, educational technology, and artificial intelligence. This chapter traces the intellectual lineage of our work, from early pioneers who imagined computers as learning companions to contemporary systems that leverage deep learning for personalized education. We organize this review around the three pillars of our Consume-Understand-Imitate framework, examining how each component has evolved and identifying opportunities for integration.

Our analysis reveals that while significant progress has been made in individual components—video navigation interfaces, content summarization, and automated feedback systems—the field lacks a unified approach that seamlessly connects these elements. This gap motivates our integrated AI-Guided Learning framework, which we position as a natural evolution of interactive learning systems that respects learner agency while providing intelligent support.

\section{Foundations of Interactive Learning Systems}
\label{sec:relwork_foundations}

\subsection{Historical Evolution: From PLATO to AI-Guided Learning}
\label{subsec:relwork_historical}

\begin{figure}[t]
  \centering
  \includegraphics[width=0.8\linewidth]{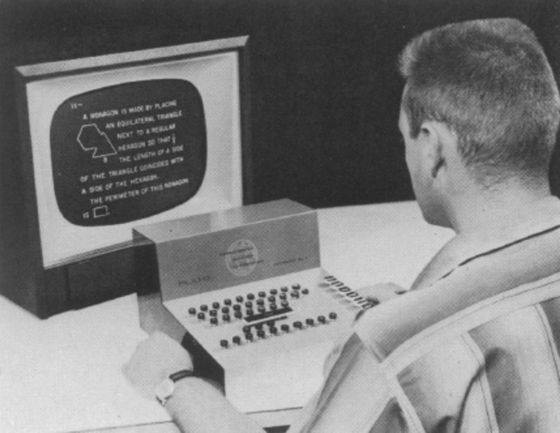}
  \caption[PLATO terminal.]{%
    A PLATO terminal in use (c.\,1972).  
    The networked PLATO system featured touch-screen plasma displays,
    real-time online forums, and thousands of interactive lessons—
    showcasing large-scale computer-based learning decades before the internet.}
  \label{fig:plato-terminal}
\end{figure}

The history of computer-assisted learning begins with PLATO (Programmed Logic for Automatic Teaching Operations), developed at the University of Illinois from 1960 to 2006~\cite{Bitzer1961PLATO, Dear2017PLATO}. PLATO pioneered numerous innovations that remain relevant today: the first flat-panel display (1964), touch-sensitive screens, real-time collaborative editing, and multimedia integration with random-access audio devices~\cite{Woolley1994PLATO}. With over 12,000 hours of courseware spanning subjects from language learning to flight simulation, PLATO demonstrated the potential for scalable, interactive education decades before the internet (see Figure \ref{fig:plato-terminal}).

PLATO's influence extends beyond technical innovation. It introduced the concept of learner-paced instruction, where students could progress through material at their own speed—a principle central to our AIxSpeed system. The platform's Notes system, arguably the first online community platform, fostered peer learning and collaboration, presaging modern MOOCs' discussion forums. Most remarkably, PLATO's foreign language courses incorporated speech synthesis and primitive speech recognition as early as the 1970s, anticipating our Profy system by nearly half a century~\cite{Sherwood1979ComputerSpeaks}.

Concurrent with PLATO's development, Ivan Sutherland's Sketchpad (1963) introduced direct manipulation interfaces~\cite{Sutherland1963Sketchpad}, fundamentally changing how humans interact with computers. This paradigm, later formalized by Ben Shneiderman~\cite{Shneiderman1983Direct, Shneiderman1997Direct}, established principles that remain foundational: continuous representation of objects, physical actions instead of command syntax, and immediate visible feedback. These principles deeply influence our interface design, particularly in FastPerson's timeline manipulation and Profy's waveform visualization.

The constructionist learning movement, led by Seymour Papert at MIT, brought a philosophical framework that complements these technical innovations~\cite{Papert1980Mindstorms}. Papert's Logo programming language embodied the principle that children learn best by constructing their own knowledge through creation and experimentation. This philosophy evolved through Mitchel Resnick's work on Scratch~\cite{Resnick2009Scratch}, which democratized programming for millions of children worldwide. The constructionist emphasis on learning through making directly informs our Imitate component, where learners improve through iterative practice with immediate feedback.

HyperCard (1987-2004) represents another pivotal moment in educational technology history~\cite{Goodman1987HyperCard}. By enabling non-programmers to create interactive multimedia applications, HyperCard democratized content creation and inspired a generation of educational software. Its card-and-stack metaphor influenced the design of presentation software and web navigation, while its emphasis on user authorship prefigures contemporary discussions about learner agency in AI systems.

\subsection{Theoretical Frameworks for Technology-Enhanced Learning}
\label{subsec:relwork_theory}

The theoretical foundations of AI-Guided Learning draw from multiple disciplines. Vygotsky's Zone of Proximal Development (ZPD) provides a framework for understanding when and how to provide assistance~\cite{vygotsky1978mind}. The ZPD represents the gap between what learners can accomplish independently and what they can achieve with guidance. It motivates future learner-adaptive assistance; the current AIxSpeed prototype adapts to acoustic properties of the content rather than estimating an individual learner's capability.

Cognitive Load Theory, developed by John Sweller, offers principles for managing the mental effort required for learning~\cite{sweller1988cognitive, sweller2011cognitive}. The theory distinguishes between intrinsic load (inherent task complexity), extraneous load (poor instructional design), and germane load (beneficial schema construction). Recent work extends this framework to AI-assisted learning environments~\cite{feng2025aial}, suggesting that intelligent systems can reduce extraneous load while maintaining appropriate intrinsic challenge.

Richard Mayer's Cognitive Theory of Multimedia Learning provides twelve principles for effective educational design~\cite{mayer2020multimedia, mayer2005cambridge}. The modality principle (people learn better from graphics and narration than from graphics and text) directly influences FastPerson's design choice to maintain audio narration rather than converting to text. The personalization principle (people learn better from conversational style) guides our decision to preserve the original speaker's voice through synthesis rather than using generic TTS~\cite{mayer2003social}.

Scaffolding theory, rooted in Vygotsky's work and elaborated by Wood, Bruner, and Ross~\cite{wood1976role}, describes how support should fade as learner competence increases. Recent research on adaptive scaffolding in digital environments demonstrates that dynamic adjustment based on real-time performance can improve learning outcomes~\cite{Azevedo2004AdaptiveScaffolding}. This principle suggests a future extension for Profy; the current prototype presents the same feedback types rather than fading support as competence increases.

Ericsson's deliberate practice framework provides essential insights for skill-based learning systems~\cite{ericsson1993role}. Deliberate practice requires four conditions: (1) well-defined specific goals, (2) focused attention, (3) immediate informative feedback, and (4) repetition with refinement. Traditional online learning often fails to provide immediate feedback, the third crucial component. Within the current systems, AIxSpeed provides automatic acoustic pacing, FastPerson provides direct access to summarized and original segments, and Profy provides immediate model-derived pronunciation feedback; AIxSpeed does not directly measure or report learner comprehension.

Bandura's social cognitive theory, particularly the concept of observational learning~\cite{bandura1997self}, informs our Imitate component. The theory's four constituent processes—attention, retention, reproduction, and motivation—offer a lens for interpreting the systems, but the current prototypes do not directly evaluate all four processes or implement a shared gamification mechanism.

The expertise reversal effect in skill acquisition suggests that optimal instruction changes as learners progress~\cite{kalyuga2003expertise}. Detailed guidance beneficial for novices can become redundant or detrimental for advanced learners. This principle motivates a future adaptive-feedback mechanism; the current Profy prototype does not vary feedback by proficiency level or transition learners to self-assessment prompts.

\subsection{The Evolution from Direct Manipulation to Intelligent Interfaces}
\label{subsec:relwork_evolution}

The transition from direct manipulation to intelligent interfaces represents a fundamental shift in HCI philosophy. Shneiderman's original vision emphasized user control and predictability~\cite{Shneiderman1997Direct}. However, as Pattie Maes argued in her debate with Shneiderman~\cite{Shneiderman1997DebateMaes}, certain tasks benefit from intelligent agents that act on users' behalf. This tension between direct manipulation and delegation remains central to AI interface design~\cite{Shneiderman2022HumanCenteredAI}.

Recent work proposes hybrid approaches that preserve user agency while leveraging AI capabilities. Horvitz's principles of mixed-initiative interaction~\cite{Horvitz1999Principles} suggest that systems should: (1) develop significant value-added automation, (2) consider uncertainty about user goals, (3) provide mechanisms for efficient agent-user collaboration, and (4) minimize the cost of poor guesses. These principles inform our design decisions throughout the Consume-Understand-Imitate pipeline.

The emergence of ``unremarkable AI" represents another important trend. As Yang et al. observe~\cite{Yang2020UnremarkableAI}, some successful AI systems fade into the background, supporting users without demanding attention. This philosophy motivates our design goal of minimizing disruption while AIxSpeed adjusts speed, FastPerson presents summaries, and Profy presents practice feedback; the degree of unobtrusiveness was not directly evaluated.

\section{The CONSUME Component: Efficient Access to Audio-Visual Content}
\label{sec:relwork_view}

\subsection{Video Navigation and Temporal Media Manipulation}
\label{subsec:relwork_video_nav}

The challenge of navigating long-form video content has inspired numerous HCI innovations. Pongnumkul et al.'s Content-Aware Dynamic Timeline~\cite{Pongnumkul2010Timeline} pioneered the decoupling of content structure from temporal representation, allowing users to navigate based on semantic importance rather than chronological time. This work introduced the concept of ``temporal remapping," where the timeline dynamically adjusts to show more detail in information-rich segments—a principle that influences FastPerson's chapter-based navigation.

\begin{figure}[t]
  \centering
  \includegraphics[width=0.8\linewidth]{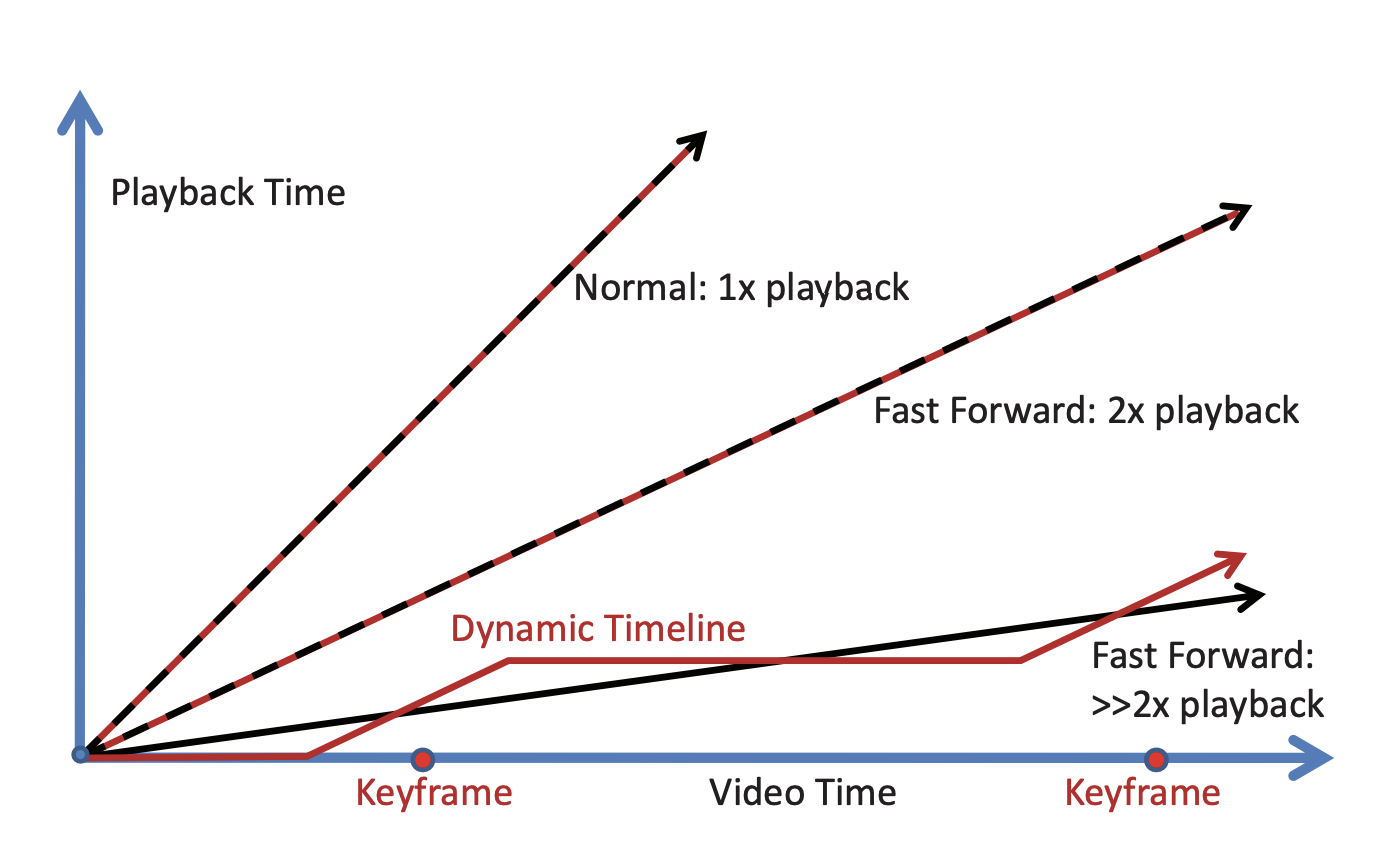}
  \caption[Content-aware dynamic timeline concept.]{%
    The dynamic timeline approach decouples video speed from playback speed, enabling intelligent content selection at high browsing speeds. 
    At normal speed (1x), playback time matches video time. At 2x speed, playback time is halved. 
    Beyond 8x speed, instead of uniform sampling, the system selects key clips to play at 2x, 
    followed by jumps to maintain desired average speed while preserving content intelligibility. 
    Adapted from Pongnumkul et al.~\cite{Pongnumkul2010Timeline}.}
  \label{fig:dynamic-timeline-concept}
\end{figure}

The core innovation of this dynamic timeline approach lies in decoupling video speed from playback speed (see Figure~\ref{fig:dynamic-timeline-concept}). Traditional timelines display sequences of unrelated frames during high-speed browsing, losing content coherence. However, the dynamic system selectively displays important segments through content-awareness, providing intelligible abstraction while maintaining average playback speed.

\begin{figure}[t]
  \centering
  \includegraphics[width=1.0\linewidth]{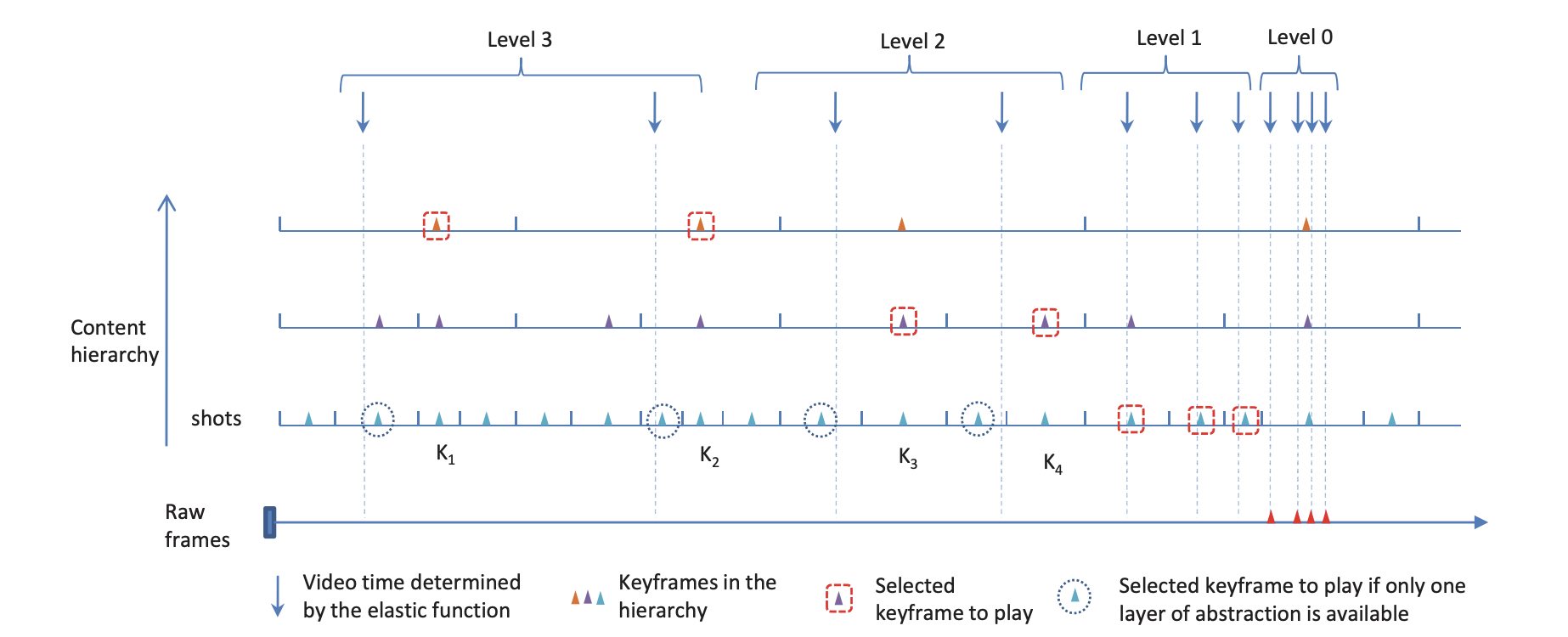}
  \caption[Hierarchical content analysis for dynamic skim generation.]{%
    Dynamic skim generation using hierarchical content analysis. 
    The system builds a hierarchy from raw frames (Level 0) through shots (Level 1) to higher-level clusters. 
    Based on desired video speed, key-clips are selected from appropriate hierarchy levels: 
    higher levels for fast browsing (showing only most important content), 
    lower levels for detailed navigation (showing more comprehensive coverage). 
    This approach ensures that different speeds reveal complementary views of the same content. 
    Adapted from Pongnumkul et al.~\cite{Pongnumkul2010Timeline}.}
  \label{fig:hierarchical-skim-generation}
\end{figure}

The system's hierarchical content analysis (see Figure~\ref{fig:hierarchical-skim-generation}) is particularly innovative. By segmenting raw frames into meaningful units through shot boundary detection and further clustering based on similarity, multiple levels of abstraction are constructed. During high-speed browsing, important clips are selected from higher hierarchy levels, while low-speed navigation provides more detailed content from lower levels. This adaptive approach ensures that browsing the same content at different speeds reveals complementary perspectives, directly influencing FastPerson's chapter structure-based navigation.

Dragicevic et al.'s work on video Browse by direct manipulation~\cite{Dragicevic2008VideoManipulation} demonstrated how physical metaphors could make video navigation more intuitive. Their technique of dragging objects within video frames to navigate temporally inspired subsequent research on object-based video Browse. The Swift system~\cite{Justin2012} addressed latency in online video scrubbing, achieving sub-100ms response times that maintain the illusion of direct manipulation even over network connections.

Recent advances in reactive video systems represent a paradigm shift. Reactive Video~\cite{Clarke2020ReactiveVideo} demonstrates how video playback can adapt in real-time to user behavior. Their system synchronizes instructional videos with learner movements, pausing when users fall behind and accelerating during mastered sections. This adaptive approach directly inspired our variable-speed playback in AIxSpeed, though we focus on comprehension rather than physical synchronization.

SmartSkip~\cite{Drucker2002SmartSkip} introduced content-based navigation for consumer video, using audio-visual analysis to identify segment boundaries. This early work presaged modern approaches using deep learning for scene segmentation. Recent systems like SoftVideo~\cite{Saelyne2022SoftVideo} leverage collective viewing patterns to identify important segments in tutorial videos, demonstrating how aggregate user behavior can improve individual navigation experiences.

The concept of egocentric elastic timelines, introduced by Higuchi et al.~\cite{Higuchi2017}, adaptively changes playback speed based on computer-vision-derived egocentric cues. This approach inspired AIxSpeed's philosophy of maintaining constant information flow rather than constant temporal speed. In their user study, EgoScanning reduced total event-finding time from 500.17 seconds to 350.95 seconds (about 30\%) relative to the baseline and increased average scanning speed. These results motivated our more sophisticated ASR-based approach.

\subsection{Variable-Speed Playback and Cognitive Processing}
\label{subsec:relwork_speed}

The relationship between playback speed and comprehension has been studied extensively. Arons' SpeechSkimmer~\cite{Barry1997} pioneered time-compressed audio, demonstrating that speech remains intelligible at up to 2.5x speed with pitch correction. However, recent meta-analyses reveal a more nuanced picture~\cite{tharumalingam2025increasing}: while comprehension remains stable up to 1.5x for clear content, it degrades rapidly beyond this threshold for complex material.

Chen et al.'s study of video playback speed in online lectures~\cite{chen2024effect} found that learner preferences vary significantly based on prior knowledge, content difficulty, and individual differences. Their work identified the ``comprehension cliff"—the speed at which understanding suddenly collapses—as highly variable across individuals and content types. This finding motivates AIxSpeed's content-aware approach rather than uniform acceleration.

Research on time-compressed audio in language learning reveals additional complexities. Playback speed and language proficiency both affect listening comprehension among multilingual English and EFL learners, and lower-proficiency learners recall less time-compressed speech~\cite{zong2024playback,conrad1989timecompressed}. Pastore's work on cognitive load in accelerated media~\cite{pastore2015using} demonstrates that time compression increases extraneous cognitive load, suggesting the need for compensatory strategies like our confidence-based speed modulation.

Psychoacoustic research on time-compressed speech perception reveals nuanced limits. Gabay et al.~\cite{gabay2017perceptual} demonstrated that gradual difficulty progression enhances learning outcomes compared to constant compression levels. Their protocols show that adaptive training, where compression gradually increases, leads to better long-term comprehension than immediate exposure to high compression rates. This finding directly informs AIxSpeed's gradual speed adjustment strategy.

The relationship between cognitive load and speech intelligibility involves multiple individual difference factors. Zekveld et al.'s pupillometry studies~\cite{zekveld2011cognitive} show that listening effort systematically increases with degraded speech quality, with marked individual differences based on age and hearing status. For AI-Guided Learning systems, this research suggests implementing adaptive algorithms that monitor cognitive load indicators and adjust speech clarity, presentation speed, and multimodal redundancy accordingly.

Janse~\cite{janse2004word} found that naturally produced fast speech was processed more slowly and was less preferred than linearly time-compressed speech, even when the naturally produced speech was intelligible. This distinction is crucial for AIxSpeed, which must balance compression efficiency with naturalness to maintain both comprehension and listening comfort.

\subsubsection{Critical Gaps in Current Variable-Speed Systems}
\label{subsubsec:relwork_speed_gaps}

Our comprehensive survey of variable-speed playback tools reveals three persistent gaps that hinder both learning efficiency and deployment.

\begin{table}[t]
\centering
\caption{Comparison of variable-speed playback approaches and their limitations (revised)}
\label{tab:speed_playback_comparison}
\scriptsize
\begingroup
\setlength{\tabcolsep}{4pt}
\renewcommand{\arraystretch}{1.12}
\microtypesetup{protrusion=false}
\sloppy
\begin{tabularx}{\linewidth}{
  >{\raggedright\arraybackslash\hspace{0pt}}p{2.2cm}%
  >{\raggedright\arraybackslash\hspace{0pt}}p{2.6cm}%
  >{\raggedright\arraybackslash\hspace{0pt}}p{2.2cm}%
  >{\raggedright\arraybackslash\hspace{0pt}}p{1.7cm}%
  >{\raggedright\arraybackslash\hspace{0pt}}X%
}
\toprule
\textbf{Approach} & \textbf{Speed Control} & \textbf{Granularity} & \textbf{Processing} & \textbf{Key Limitation} \\
\midrule
Uniform speedup
& Constant (global time-scale)
& Entire audio
& None
& Ignores content difficulty / listener variation. \\

SpeechSkimmer~\cite{Barry1997}
& Time-compressed speech + pause shortening; interactive detail control
& \textbf{Multi-level} (pause-based segments; post-pause chunks; sentence/topic-level)
& \textbf{Real-time}
& No automatic difficulty awareness; relies on acoustic heuristics. {\small\cite{Barry1997}} \\

Elastic timeline (EgoScanning)~\cite{Higuchi2017}
& Adaptive playback based on \textbf{egocentric cues} (hands\slash motion\slash people)
& Frame--segment (highlighted regions slowed; others fast-forwarded)
& CV cue extraction typically \textbf{precomputed/batch}; not purely real-time in practice
& Tailored to first-person videos; audio handling not central; preprocessing cost. {\small\cite{Higuchi2017}} \\

Dynamic timeline~\cite{Pongnumkul2010Timeline}
& Content-aware temporal remapping (hierarchical \textbf{shot/cluster} selection)
& \textbf{Shot/cluster} (not only “scene”)
& \textbf{Offline} (hierarchy building)
& \textbf{Video-only}; audio track not processed in the original paper. {\small\cite{Pongnumkul2010Timeline}} \\

\textbf{AIxSpeed (Ours)}
& ASR confidence–driven rate control
& \textbf{Subword\slash word\slash phoneme} (depends on aligner/ASR outputs)
& \textbf{Near real-time} (streaming ASR inference)
& ASR confidence is an imperfect proxy; noise/domain shifts; fine-grained rate changes may affect prosody; multilingual performance depends on ASR. \\
\bottomrule
\end{tabularx}
\endgroup
\end{table}

First, \emph{most production players lack fine-grained difficulty awareness}. Uniform speed control squanders time on trivial segments and overwhelms listeners on dense segments.

Second, \emph{semantic or vision-heavy processing introduces deployment barriers}. Methods that rely on transcripts, NLP, or GPU-based computer-vision cues incur preprocessing latency and complicate multilingual real-time use.

Third, \emph{coarse granularity yields sub-optimal speed profiles}. Utterance- or segment-level adjustment overlooks difficulty variations within words, despite phonetic evidence that comprehension load fluctuates at the phoneme level.

AIxSpeed addresses all three gaps by driving phoneme-level speed modulation solely from ASR confidence scores, requiring no transcripts or offline preprocessing.

\subsection{Self-Supervised Speech Representation Learning}
\label{subsec:relwork_self_supervised_speech}

The emergence of self-supervised learning in speech processing has revolutionized how we approach audio understanding with limited labeled data. Wav2Vec2~\cite{baevski2020wav2vec} demonstrates that contrastive learning in latent space can achieve state-of-the-art results with as little as 10 minutes of labeled data, achieving 4.8/8.2 WER on LibriSpeech benchmark. This approach is particularly suitable for educational applications where transcribed speech data is limited, aligning with AI-Guided Learning principles by reducing data requirements for personalized speech-based educational tools.

Recent advances in robust multilingual ASR, exemplified by OpenAI's Whisper~\cite{radford2023}, demonstrate multilingual transcription and translation using large-scale weak supervision. Such models suggest future multilingual extensions, but the AIxSpeed and Profy prototypes evaluated in this thesis use English speech and do not establish performance across diverse languages.

Forced alignment technologies, particularly Montreal Forced Aligner (MFA)~\cite{mcauliffe2017_interspeech}, remain crucial for phoneme-level processing. Recent benchmarking studies~\cite{rousso2024tradition} demonstrate that MFA outperforms modern ASR systems like WhisperX and MMS in alignment accuracy, validating our choice for phoneme-level speed optimization in AIxSpeed.

\subsection{Content-Aware Interfaces and Adaptive Systems}
\label{subsec:relwork_content_aware}

Content-aware interfaces adapt their behavior based on the material being presented. LectureScape~\cite{kim2014understanding} uses collective viewing patterns to create visual summaries of lecture videos, revealing ``hot spots" where many students re-watch content. This crowdsourced approach to identifying difficult concepts inspired FastPerson's potential for collaborative annotation, though our current implementation focuses on automatic analysis.

ChronoViz~\cite{Fouse2011ChronoViz} demonstrates how multiple data streams can be synchronized for analysis, supporting researchers in exploring connections between video, sensor data, and annotations. While designed for research rather than learning, its principles of multimodal synchronization inform our approach to integrating visual and audio streams in FastPerson.

The Video Digests project~\cite{pavel2014} automatically creates readable summaries of instructional videos by combining transcript analysis with visual change detection. Their finding that users prefer summaries maintaining the original narrative structure influenced FastPerson's design decision to preserve chronological flow while enabling non-linear access.

Recent work on semantic video navigation~\cite{lei2021moment} uses deep learning to enable natural language queries over video content. Users can search for concepts rather than keywords, finding relevant segments even when specific terms aren't mentioned. This capability would naturally extend FastPerson, allowing learners to formulate queries like ``the part where she explains why the algorithm fails" rather than searching for specific words.

\subsection{Large-Scale MOOC Analytics and Collective Intelligence}
\label{subsec:relwork_mooc_analytics}

Large-scale analysis of MOOC viewing behavior provides unprecedented insights into learning patterns. Zhou et al.~\cite{zhou2025crowd} analyzed extensive data from XuetangX platform, revealing that the relationship between effort and performance differs from traditional assumptions. Their finding that ``certificate buddies" increase completion rates threefold highlights the importance of social factors in online learning, informing FastPerson's potential for collaborative features.

Systematic reviews of video interaction patterns~\cite{yurum2024predictive} identify pause, play, rewind, and fast-forward as key behavioral indicators. These interaction patterns serve as implicit feedback about content difficulty and learner engagement. Zhang et al.'s collective attention model~\cite{zhang2022collective} demonstrates how network analysis through an ecosystem lens can reveal hidden behavioral patterns, informing our real-time adaptation strategies in AI-Guided Learning systems.

The concept of using collective viewing patterns for difficulty estimation extends beyond simple interaction counting. When many learners independently pause or replay specific segments, aggregated behaviors can provide signals about content difficulty~\cite{kim2014understanding}. Such signals could complement a future version of AIxSpeed; the current model adjusts speed from acoustic features and does not use collective viewing data.

\section{The UNDERSTAND Component: Supporting Deep Comprehension}
\label{sec:relwork_understand}

\subsection{Multimodal Information Processing and Integration}
\label{subsec:relwork_multimodal}

Understanding complex content requires integrating information across multiple modalities. Early work by Moreno and Mayer on multimedia learning~\cite{Moreno2000Multimedia} established that combining visual and auditory channels can enhance learning when designed properly. However, poor integration can cause the ``split-attention effect," where learners waste cognitive resources switching between sources.

Recent advances in multimodal fusion leverage deep learning to better integrate diverse information streams. Hori et al.'s attention-based fusion for video description~\cite{hori2017attention} demonstrates how neural networks can learn which modality provides the most relevant information at each moment. This dynamic weighting inspired FastPerson's approach to balancing slide content, speaker video, and audio narration.

The rise of multimodal large language models (MLLMs) opens new possibilities. GPT-4V's ability to understand both images and text enables more sophisticated content analysis~\cite{OpenAI2023GPT4V}. However, educational applications require special consideration: maintaining pedagogical coherence, preserving instructor intent, and avoiding oversimplification. Our work demonstrates how these powerful models can be grounded in educational theory.

Cross-modal learning analytics represent an emerging field. By combining clickstream data, eye tracking, and physiological sensors, researchers can model student attention and engagement more accurately~\cite{Mu2020MultimodalAnalytics}. While our current systems don't incorporate such sensors, the framework could naturally extend to include these richer data sources for more adaptive support.

\subsection{Summarization Technologies for Educational Content}
\label{subsec:relwork_summarization}

Video summarization research spans multiple approaches: keyframe extraction, highlight detection, and narrative summarization~\cite{truong2007video}. Educational videos pose unique challenges: preserving pedagogical structure, maintaining conceptual dependencies, and avoiding oversimplification. Generic summarization techniques often fail to respect these constraints.

Work on abstractive summarization of educational content shows promise. Benedetto et al.'s system for lecture video summarization~\cite{benedetto2024abstractive} uses transformer models to generate concise textual summaries while preserving key concepts and their relationships. However, text-only output loses the multimodal richness that makes video compelling for learning.

The How2 dataset and associated challenges have driven progress in instructional video understanding~\cite{Palaskar2019How2Summ}. Systems trained on this data can identify task structure, extract steps, and generate summaries. However, most focus on procedural content (cooking, crafts) rather than conceptual learning, limiting their applicability to academic lectures.

Recent work on personalized summarization~\cite{yoshitaka2012personalized} adapts output based on user knowledge and goals. A summary for an expert should differ from one for a novice—a principle FastPerson could incorporate by modeling user expertise. This personalization extends beyond content selection to presentation style, terminology choice, and detail level.

\subsubsection{Critical Gaps in Video Summarization and Access for Learning}
\label{subsubsec:relwork_fastperson_gaps}

Our analysis of existing video navigation and summarization technologies highlights several persistent challenges when applied to educational content, which FastPerson aims to mitigate (see Table~\ref{tab:fastperson_comparison}).

\begin{table}[t]
\centering
\caption{Comparison of video summarization and navigation approaches for learning content and their limitations (revised)}
\label{tab:fastperson_comparison}
\scriptsize
\begingroup
\setlength{\tabcolsep}{4pt}
\renewcommand{\arraystretch}{1.12}
\microtypesetup{protrusion=false}
\sloppy
\begin{tabularx}{\linewidth}{
  >{\raggedright\arraybackslash\hspace{0pt}}p{2.3cm}%
  >{\raggedright\arraybackslash\hspace{0pt}}p{3.0cm}%
  >{\raggedright\arraybackslash\hspace{0pt}}p{2.6cm}%
  >{\raggedright\arraybackslash\hspace{0pt}}X%
}
\toprule
\textbf{Approach} & \textbf{Summarization Output / Navigation Aid} & \textbf{Multimodal Preservation} & \textbf{Key Limitation} \\
\midrule
Manual Scrubbing / Skipping
& Full video (user-controlled)
& Full (original A/V)
& Time-consuming; depends on user to locate key segments; little guidance. \\

Keyframe Extraction
& Static image sequence (visual skim)
& Visual only; \textbf{loss of audio/prosody and motion dynamics}
& Insufficient for understanding complex spoken content. {\small\cite{truong2007video}} \\

Transcript-based (e.g., Video Digests~\cite{pavel2014}; Abstractive lecture summaries~\cite{benedetto2024abstractive})
& Text-centric summaries (often with thumbnails/segments)
& Audio $\rightarrow$ text (prosody lost); limited visual dynamics
& Loses multimodal richness; risk of reduced engagement vs watching. {\small\cite{pavel2014,benedetto2024abstractive}} \\

Dynamic Timelines~\cite{Pongnumkul2010Timeline}
& Content-aware \textbf{visual} skim via resampled clips; timeline supports fast browsing
& \textbf{Primarily visual}; \textbf{audio not handled} in the original implementation
& Designed for rapid browsing, not structured pedagogical summarization; offline preprocessing required. {\small\cite{Pongnumkul2010Timeline}} \\

Elastic Timelines (EgoScanning)~\cite{Higuchi2017}
& Non-linear time-compressed playback highlighting egocentric events
& Visual emphasis; audio handling \textbf{not central/unspecified}
& Requires per-frame CV cues; tailored to first-person videos; focus on speed, not structured summary. {\small\cite{Higuchi2017}} \\

\textbf{FastPerson (Ours)}
& Chapter-based video-to-video condensed summary
& Full visual stream (edited); synthesized speaker identity for narration
& Requires clear chapter/topic structure; synthesis quality depends on TTS/model and source audio. \\
\bottomrule
\end{tabularx}
\endgroup
\end{table}

Three primary gaps emerge from this comparison:

First, \emph{loss of multimodal richness}: Many summarization techniques reduce video content to text or keyframes. This approach discards the pedagogical benefits of the speaker's original voice, intonation, non-verbal cues, and the synchronized interplay between visual demonstrations and spoken explanations, which are crucial for engagement and comprehension in learning contexts.

Second, \emph{inefficient navigation and cognitive overhead}: While some tools offer improved navigation, they often lack structured summarization, forcing users to still invest significant time in finding and integrating information. Manually scrubbing or relying on simple timeline markers in long educational videos remains a cognitively demanding task.

Third, \emph{preservation of instructional integrity}: Automated summarization must not only shorten content but also preserve its pedagogical structure, conceptual dependencies, and narrative flow. Generic summarization may fail to identify and retain key instructional moments or may present information in a disjointed manner.

FastPerson is designed to address these gaps by providing a chapter-based video-to-video summarization that retains the original visual stream and the speaker's identity through voice synthesis. This approach aims to significantly reduce viewing time while preserving the rich, multimodal learning experience and instructional coherence of the original content.

\subsection{Cognitive Load Management in Digital Learning}
\label{subsec:relwork_cognitive_load}

Managing cognitive load in digital environments requires careful attention to information presentation. The expertise reversal effect~\cite{kalyuga2003expertise} shows that instructional techniques helping novices can hinder experts, suggesting the need for adaptive approaches. This finding motivates future proficiency-aware extensions; the current prototypes do not automatically estimate user proficiency to adjust their support.

Research on multimedia learning reveals specific techniques for load reduction. The modality principle suggests presenting words as narration rather than on-screen text when accompanied by graphics~\cite{Low2008Modality}. The redundancy principle warns against presenting identical information in multiple forms. These principles guide FastPerson's decision to maintain audio narration while selectively displaying visual elements.

Attention management in video learning presents unique challenges. Mind-wandering increases with video length, particularly in self-paced environments~\cite{Risko2012Wandering}. Techniques such as interpolated testing, perspective changes, and social presence may support engagement. Whether AIxSpeed's variable pacing reduces mind-wandering was not tested and remains an empirical question.

Recent work on cognitive load measurement using physiological signals~\cite{beauchemin2024enhancing} suggests possibilities for real-time adaptation. Eye tracking can detect when learners struggle with specific content, while EEG signals indicate cognitive overload. Though our current systems use content-based rather than learner-based adaptation, future versions could incorporate such biometric feedback.

Mind-wandering during video-based learning represents a significant challenge for maintaining engagement. Conrad and Newman's EEG research~\cite{conrad2021measuring} reports frequent mind-wandering during video lectures and a negative relationship with learning outcomes. This finding motivates testing whether dynamic pacing affects attention; the current AIxSpeed evaluation did not measure mind-wandering.

Individual differences in working memory capacity (WMC) moderate the effectiveness of multimedia learning interventions. Martini et al.~\cite{martini2020wmc} discovered that high-WMC individuals benefit from post-learning rest periods, while low-WMC individuals show opposite patterns. This interaction effect suggests that AI-Guided Learning systems must adapt not only to content difficulty but also to learner cognitive profiles.

Metacognitive awareness and self-regulation skills vary substantially among learners, creating both challenges and opportunities. Research on metacognitive interventions reports an average impact equivalent to seven additional months of progress~\cite{evidenceforlearning2021metacognition}, suggesting potential for personalized metacognitive support. FastPerson supports self-directed chapter navigation, but the current AIxSpeed and Profy prototypes do not implement learner-facing confidence indicators or progressive feedback fading.

\section{The IMITATE Component: Feedback for Skill Acquisition}
\label{sec:relwork_imitate}

\subsection{Computer-Assisted Pronunciation Training: Evolution and Current State}
\label{subsec:relwork_capt}

Second language acquisition theory provides crucial foundations for pronunciation training systems. Recent large-scale empirical studies have refined our understanding of critical periods in L2 acquisition. Hartshorne et al.~\cite{hartshorne2018critical} analyzed data from 669,498 participants, discovering that the critical period for syntax learning closes at 17.4 years—much later than previously thought. However, Singleton et al.~\cite{singleton2021cph} demonstrated that phonological acquisition becomes significantly more difficult after age 9, suggesting different critical periods for different linguistic subsystems.

These findings have profound implications for AI-guided pronunciation training. Systems must incorporate age-sensitive algorithms that adjust expectations and feedback mechanisms based on learner age. For adult learners past the phonological critical period, focusing on intelligibility rather than native-like accuracy may be more appropriate, while younger learners can still achieve near-native pronunciation with proper guidance.

L1 transfer mechanisms represent another crucial consideration. Flege et al.~\cite{flege1997experience} found that L2 English vowel perception and production varied with both language experience and learners' L1 background. This evidence motivates future L1-sensitive feedback. The current Profy prototype was trained and evaluated with Japanese learners of English but does not personalize feedback by L1.

The evolution of CAPT systems mirrors broader trends in AI and HCI. Early systems like PLATO's language courses used pattern matching and rule-based feedback~\cite{alpert1970advances}. Despite technical limitations, they demonstrated the potential for automated pronunciation assessment, inspiring decades of subsequent research.

Modern CAPT leverages advanced ASR and machine learning. Commercial systems like ELSA Speak~\cite{ravenor2024adequacy} and Duolingo~\cite{Duolingo2024} reach millions of users, providing personalized pronunciation practice at scale. Academic research has evaluated CAPT effectiveness: a systematic review maps findings across CAPT designs, and a meta-analysis of 31 controlled studies reports an overall medium effect on L2 pronunciation learning~\cite{amrate2024capt,almusharraf2024capt}.

Recent innovations focus on visual feedback. PTeacher~\cite{bu2021pteacher} uses exaggerated audio-visual corrective feedback to help learners perceive subtle pronunciation differences. By amplifying errors in both audio playback and animated lip movements, the system makes problems more noticeable. User studies showed significant improvement over traditional minimal-pair training.

The L2-ARCTIC corpus~\cite{zhao2018l2} and similar datasets enable data-driven approaches to pronunciation assessment. By training on diverse non-native speech, systems can recognize patterns specific to different L1 backgrounds. This speaker adaptation represents a significant advance over one-size-fits-all approaches, though it requires substantial data collection.

\subsection{Visualization Techniques for Performance Feedback}
\label{subsec:relwork_visualization}

Effective visualization can make abstract pronunciation concepts concrete. The Vowel Game~\cite{Paganus2006VowelGame} maps tongue position to a 2D space, allowing learners to see their vowel production in real-time. This spatial representation leverages human visual processing capabilities, making articulatory concepts intuitive.

Spectrograms, traditionally used in speech analysis, are increasingly employed for pedagogical purposes. By showing frequency patterns over time, they reveal aspects of speech invisible to naive listeners. However, interpreting spectrograms requires training, limiting their accessibility. Profy's waveform-based approach provides a simpler alternative while still conveying timing and intensity information.

Research on visual feedback effectiveness reveals important design principles~\cite{Hardison2004VisualFeedback}. Feedback should be immediate, interpretable without extensive training, and directly actionable. Abstract representations (numerical scores, complex graphs) prove less effective than concrete visualizations showing what to change. This finding influenced Profy's design, which highlights specific segments requiring attention.

Gamification elements can enhance engagement with pronunciation practice. Duolingo's streak system and achievement badges can encourage daily practice~\cite{shortt2023gamification}. However, excessive gamification can distract from learning goals. Progress indicators are a possible future extension for Profy; the current prototype does not implement game mechanics or a longitudinal progress system.

\subsection{Multidimensional Assessment Frameworks}
\label{subsec:relwork_assessment_frameworks}

The evolution of pronunciation assessment has moved beyond simple accuracy scores to multidimensional frameworks. Thomson's meta-analysis~\cite{thomson2018measurement} establishes that intelligibility, comprehensibility, and accentedness represent three partially independent constructs. Intelligibility measures actual understanding, comprehensibility captures perceived difficulty, and accentedness reflects deviation from native norms. This tripartite framework aligns with communicative language teaching goals, suggesting that perfect native-like pronunciation may be less important than clear communication.

Recent advances in multi-task pronunciation assessment demonstrate the feasibility of comprehensive evaluation across multiple dimensions. The MultiPA system~\cite{chen2024multipa} reports correlations exceeding 0.6 with human ratings on three dimensions, using transformer-based architectures to capture segmental and suprasegmental features. This work suggests a future extension for Profy; the current prototype does not separately score intelligibility, comprehensibility, and accentedness.

Long-term retention and transfer effects remain understudied in CAPT research. While most studies report immediate improvements, few track whether gains persist beyond the training period or transfer to spontaneous speech~\cite{saito2019effects}. The current Profy study used short, controlled sentence practice and did not measure retention or spontaneous-speech transfer; these require future evaluation.

\subsection{Embodied Interaction and Kinesthetic Learning}
\label{subsec:relwork_embodied}

Embodied cognition theory suggests that physical experience shapes abstract thinking~\cite{wilson2002six}. This principle extends to language learning: pronunciation involves motor skills that benefit from physical practice. Gesture-based interfaces can leverage this connection, linking movement to sound production.

Tangible interfaces for pronunciation training remain underexplored. Ishii's vision of Tangible Bits~\cite{Ishii1997TangibleBits} suggests possibilities: physical objects whose manipulation controls speech parameters. Imagine squeeze toys that teach stress patterns or elastic bands demonstrating vowel length. While our current work remains screen-based, future extensions could incorporate tangible elements.

Motion capture technology enables detailed analysis of articulatory movements. Systems tracking lip, jaw, and tongue position can provide precise feedback about articulation. Though currently limited to laboratory settings, advancing sensor technology may soon enable home-based motion analysis, extending Profy's capabilities beyond audio to visual speech features.

The relationship between musical training and pronunciation improvement suggests shared cognitive mechanisms~\cite{zhang2024embodied}. Both involve precise motor control, auditory discrimination, and rhythmic timing. This connection suggests a possible future extension; the current Profy prototype does not include music-based exercises.

\section{AI Integration Across the Learning Pipeline}
\label{sec:relwork_ai_integration}

\subsection{Large Language Models in Educational Interfaces}
\label{subsec:relwork_llms}

The emergence of LLMs has transformed educational technology. ChatGPT's ability to explain concepts, answer questions, and provide personalized feedback offers unprecedented support for learners~\cite{kasneci2023chatgpt}. However, challenges remain: hallucination, bias, and the risk of bypassing productive struggle in learning.

Research on LLMs for educational content generation shows promise and pitfalls. Wang et al. demonstrate how GPT-4 can create high-quality practice problems, but generated content requires careful validation. The VIVID system~\cite{Choi2024VIVID} converts lecture videos into interactive dialogues, allowing students to ``converse" with content. Early studies show improved engagement, though learning outcomes remain unclear.

Prompt engineering for educational applications requires special consideration. Effective prompts must maintain appropriate difficulty, encourage deep thinking, and avoid simply providing answers. Recent work on ``pedagogical prompting"~\cite{xiao2025pedagogical} develops principles for educationally effective LLM interactions, emphasizing scaffolding over direct instruction.

The question of appropriate reliance on AI assistance remains contentious. When does AI support become a crutch that prevents skill development? Research on ``AI-induced skill atrophy" in professional contexts offers cautionary lessons for education. Our systems address this by focusing on augmentation rather than replacement, maintaining learner agency throughout.

Beyond text-based applications, transformer architectures are revolutionizing multimodal educational content analysis. Vision Transformers (ViViT)~\cite{arnab2021vivit} enable sophisticated video understanding by extracting spatiotemporal tokens from educational videos. This technology allows automatic analysis of student engagement in video-based learning, identifying moments of confusion through facial expressions and body language. For AI-Guided Learning systems, these capabilities suggest future extensions where learner state is continuously monitored and support is provided proactively.

Recent educational applications of BERT and GPT models demonstrate both promise and limitations. Pilicita and Barra~\cite{pilicita2025llms} found high classification accuracy when evaluating BERT-, RoBERTa-, and GPT-based models on student comments indicating anxiety, depression, or neutral affect. Separately, a randomized trial found that teachers using ChatGPT spent 31\% less time on lesson and resource preparation, while curriculum-design research emphasizes educator review of generated outputs~\cite{roy2024lesson,jorgens2024curriculum}.

The convergence of diffusion models and educational content generation opens unprecedented opportunities. While current applications focus on visual content creation, the potential for generating personalized pronunciation examples, creating learner-specific practice materials, and synthesizing culturally appropriate educational scenarios remains largely unexplored~\cite{wang2024diffusion}. These generative capabilities could extend AI-Guided Learning systems to provide truly individualized learning experiences.

\subsection{Adaptive Learning Systems and Personalization}
\label{subsec:relwork_adaptive}

Adaptive learning systems adjust instruction based on learner characteristics and performance. Classic Intelligent Tutoring Systems (ITS) like the LISP Tutor~\cite{Anderson1985LISP} demonstrated the potential for individualized instruction at scale. Modern systems leverage machine learning for more sophisticated adaptation.

Bloom's ``2 sigma problem"~\cite{bloom19842} showed that one-on-one tutoring produces learning gains two standard deviations above classroom instruction. AI systems aim to approximate this personalized attention at scale. Recent meta-analyses confirm that well-designed adaptive systems can achieve substantial portions of this gain~\cite{Kulik2016Effectiveness}.

Learner modeling has evolved from simple skill tracking to rich cognitive models. Transformer-based approaches~\cite{li2024enhancing} can predict student knowledge states by analyzing interaction patterns. These models enable proactive support, intervening before learners become frustrated or disengaged.

The EduQate system~\cite{tio2024eduqate} uses reinforcement learning to optimize curriculum sequencing. By modeling learning as a sequential decision process, it adapts topic order based on individual progress. Such approaches could extend our systems: AIxSpeed might adjust not just playback speed but content sequence, while Profy could adaptively select practice materials.

\subsection{Balancing Automation and Learner Agency}
\label{subsec:relwork_agency}

The tension between AI automation and human agency represents a fundamental challenge in educational technology. Over-automation can lead to passive consumption, undermining the active construction of knowledge that deep learning requires. Recent frameworks for ``AI-augmented learning"~\cite{holstein2022designing} emphasize maintaining learner control while providing intelligent support.

Research on explanation interfaces shows how transparency can help users develop appropriate trust in AI systems~\cite{Khosravi2022Explainable}. The current prototypes expose outputs such as speed changes, summaries, and highlighted waveform regions, but they do not provide full explanations of model reasoning or uncertainty; this remains a design goal.

The concept of ``seamful design" deliberately exposes system limitations~\cite{Chalmers2003Seamful}. Rather than hiding imperfections, seamful interfaces help users understand when and why to trust AI recommendations. This approach proves particularly valuable in educational contexts, where understanding system limitations prevents over-reliance.

Studies of human-AI collaboration in creative tasks offer insights for learning applications~\cite{inie2023designing,hao2024collaborative}. Successful collaboration requires clear role division, with AI handling routine aspects while humans focus on creative decisions. This division inspired our approach: AI manages mechanical aspects (speed adjustment, content extraction, error detection) while learners control meaningful choices (what to study, how to practice, when to seek help).

\section{Design Principles from Commercial Systems}
\label{sec:relwork_commercial}

\subsection{Case Studies: Khan Academy, Coursera, Duolingo}
\label{subsec:relwork_cases}

Commercial educational platforms offer valuable lessons in scaling personalized learning. Khan Academy's mastery learning system requires proficiency before advancement, implementing Bloom's vision at scale~\cite{thompson2011khan}. Their exercise generation system demonstrates how procedural content can be created algorithmically while maintaining pedagogical quality.

Coursera's peer assessment system addresses the challenge of providing feedback at scale~\cite{Piech2013Peer}. By training learners to evaluate each other's work, they create a sustainable feedback ecosystem. This approach influenced our thinking about Profy: future versions might incorporate peer pronunciation assessment, with AI ensuring quality and consistency.

Duolingo's success demonstrates the power of mobile-first design and gamification~\cite{shortt2023gamification}. Their bite-sized lessons fit into busy schedules, while streak mechanics encourage daily practice. However, critics note that gamification can overshadow learning objectives. This tension informed our more subtle approach to engagement.

Analysis of edX viewing logs found that median engagement time was at most six minutes and that learners typically watched less than half of videos longer than nine minutes~\cite{Guo2014}. These observations motivate shorter, navigable presentations such as those produced by FastPerson.

\subsection{Gamification and Engagement Strategies}
\label{subsec:relwork_gamification}

Gamification in education extends beyond points and badges. Effective educational games create intrinsic motivation through well-designed challenges, clear feedback, and sense of progress~\cite{Plass2015Gaming}. The key lies in aligning game mechanics with learning objectives rather than applying superficial rewards.

Duolingo's approach demonstrates sophisticated gamification. Its ``skill tree" visualizes learning progress and prerequisite relationships. The ``health" system penalizes repeated errors, while timed challenges add time pressure. The educational effects of individual mechanics depend on implementation and learner context.

Research on flow states in educational games reveals design principles~\cite{Kiili2005Flow}. Optimal challenge requires matching difficulty to skill level—too easy breeds boredom, too hard causes frustration. Dynamic difficulty adjustment can help maintain this balance. The current AIxSpeed model adapts to acoustic features rather than learner skill, and Profy does not yet adjust feedback detail by proficiency; both are candidates for future learner-adaptive extensions.

Critics of gamification warn against ``chocolate-covered broccoli"—superficial game elements masking boring content~\cite{Habgood2011Intrinsic}. Effective educational games integrate learning deeply into core mechanics. Our systems avoid superficial gamification, instead focusing on intrinsic rewards: the satisfaction of understanding (AIxSpeed), efficiency gains (FastPerson), and skill improvement (Profy).

\subsection{Real-Time Processing and Edge Computing}
\label{subsec:relwork_realtime_edge}

Advances in model compression and optimization support faster-than-real-time speech processing on edge devices. Xu et al.~\cite{xu-etal-2024-conformer} reported a real-time factor of 0.19 for Conformer-based ASR on a small wearable device. A real-time factor does not by itself establish sub-200-ms end-to-end interaction latency, which also depends on buffering, feature extraction, and application design.

Browser-based implementation using Web Audio API and WebRTC provides accessible deployment without dedicated hardware requirements~\cite{w3c2021webaudio,w3c2021webrtc}. The Web Audio API enables sophisticated audio processing directly in the browser, including real-time pitch shifting, time stretching, and spectral analysis. WebRTC facilitates peer-to-peer audio streaming with minimal latency, enabling collaborative pronunciation practice sessions. These technologies underpin our systems' accessibility, requiring only a modern web browser for full functionality.

Edge computing strategies address both latency and privacy concerns. By processing sensitive voice data locally and only transmitting abstracted features or model updates, systems can maintain user privacy while benefiting from collective learning. This architecture is particularly important for Profy, where learners may be self-conscious about sharing pronunciation attempts.

\subsection{Scalability and Accessibility Considerations}
\label{subsec:relwork_scalability}

Scaling personalized learning requires careful system design. Technical challenges include handling concurrent users, storing interaction data, and computing real-time adaptations. Cloud infrastructure enables elastic scaling, though latency remains problematic for real-time interaction. Edge computing offers solutions, processing sensitive data locally while synchronizing insights globally.

Accessibility extends beyond technical compliance to inclusive design. The Web Content Accessibility Guidelines (WCAG) provide standards, but educational interfaces pose unique challenges. How do we make visual pronunciation feedback accessible to blind learners? Can speed-adjusted audio serve hearing-impaired students? Our systems address some concerns (captions in FastPerson, visual feedback in Profy) while acknowledging remaining gaps.

Multilingual support presents both technical and pedagogical challenges. Machine translation quality varies dramatically across language pairs. Cultural differences affect learning preferences and feedback interpretation. Pronunciation training must account for diverse L1 backgrounds. While our current implementation focuses on English, the framework design considers future multilingual extension.

The digital divide remains a crucial consideration. Not all learners have high-speed internet or powerful devices. Progressive enhancement strategies ensure basic functionality on limited hardware while enabling rich experiences where possible. AIxSpeed's client-side processing and Profy's lightweight feedback computation reflect this principle.

\section{Synthesis and Future Directions}
\label{sec:relwork_synthesis}

\subsection{The AI-Guided Learning Vision}
\label{subsec:relwork_vision}

Our review reveals converging trends pointing toward integrated learning support systems. Advances in speech processing enable nuanced pronunciation feedback. Multimodal AI can synthesize complex educational content. Adaptive algorithms personalize learning paths. The missing piece has been integration—connecting these capabilities into coherent learning experiences.

AI-Guided Learning represents our synthesis of these capabilities, grounded in robust theoretical foundations and cutting-edge technical innovations. By supporting the complete Consume-Understand-Imitate cycle, we address learning holistically rather than piecemeal. Self-supervised learning models like Wav2Vec2 enable personalized audio processing with minimal data requirements. Large-scale MOOC analytics inform content difficulty estimation. Psychoacoustic research guides perceptually-motivated speed adjustment. Second language acquisition theory shapes age-appropriate feedback strategies. Multi-dimensional assessment frameworks ensure comprehensive skill evaluation. Real-time edge computing makes interactive feedback feasible. Cognitive science insights optimize attention management. Transformer architectures enable sophisticated multimodal understanding. This integration creates emergent benefits: insights from viewing inform understanding, understanding guides imitation, and imitation reveals gaps requiring renewed viewing. The convergence of these technologies and theories enables a learning experience that adapts to individual needs while maintaining pedagogical rigor.

The framework respects both technological capabilities and human learning principles. We leverage AI's strengths (pattern recognition, consistent feedback, infinite patience) while preserving human agency (goal setting, meaning making, creative application). This balanced approach avoids both technophobic rejection and uncritical acceptance of AI in education.

\subsection{Open Challenges and Research Opportunities}
\label{subsec:relwork_challenges}

Our comprehensive analysis reveals critical gaps in current audio-visual learning support systems that define concrete research opportunities:

\textbf{1. Lack of Content-Aware Processing at Fine Granularity}
Many variable-speed playback systems operate at coarse temporal resolutions (utterance or segment level), without adapting to short-term variations in acoustic recognizability. The reviewed systems did not dynamically adjust playback speed from ASR confidence without semantic analysis. This gap motivates AIxSpeed's phoneme-level adaptation using ASR confidence as a proxy for acoustic recognizability.

\textbf{2. Loss of Multimodal Richness in Video Summarization}
Many video summarization approaches generate text outputs that do not retain the full multimodal context of educational videos. Preserving speaker voice, visual demonstrations, and temporal relationships remains challenging. This limitation motivates FastPerson's video-to-video summarization approach, which retains speaker and visual cues while reducing viewing time.

\textbf{3. Absence of Scalable, Fine-Grained Feedback for Skill Acquisition}
Many CAPT systems require annotated data or provide holistic scores. The challenge of deriving localized feedback from limited unannotated learner data motivates Profy's self-supervised visualization approach. Profy highlights classifier-emphasized waveform regions and displays model-derived distances; these outputs were not independently validated as phonetic error diagnoses or actionable corrections.

\textbf{4. Fragmented Support Across Learning Stages}
While individual technologies exist for content navigation, summarization, and skill practice, they are often studied as separate forms of support. This fragmentation motivates the Consume-Understand-Imitate framework used in this dissertation to organize the three prototypes.

\textbf{5. Limited Adaptation to Individual Learner Characteristics}
The prototypes evaluated in this dissertation do not model learner background, proficiency level, or learning goals. Personalizing support while maintaining scalability therefore remains an open direction for future extensions of the framework.

Together, these challenges motivate approaches that combine advances in deep learning with educational design. This dissertation examines that direction through three prototype systems and reports their measured outcomes and limitations.

\section{Conclusion}
\label{sec:relwork_conclusion}

This chapter has traced the evolution of technology-enhanced learning from early interactive systems to contemporary AI-powered platforms, revealing both remarkable progress and persistent challenges. Our analysis identifies three critical insights that motivate and position this dissertation:

\textbf{First, a fundamental mismatch exists between content delivery and human cognitive constraints.} While digital platforms have democratized access to audio-visual learning materials, the linear, time-bound nature of these media creates inefficiencies that current technologies fail to address. Uniform playback speeds ignore content complexity variations; text-based summaries lose multimodal richness; and skill practice lacks the immediate, detailed feedback essential for improvement. These limitations create an urgent need for AI systems that adapt to both content characteristics and learner needs.

\textbf{Second, recent advances in deep learning create unprecedented opportunities for educational innovation.} Self-supervised speech models enable fine-grained acoustic analysis without extensive labeled data. Multimodal transformers can process and synthesize information across visual and auditory channels. Voice cloning preserves speaker identity while enabling content manipulation. However, these technologies remain largely unexploited for educational applications, presenting a clear research opportunity.

\textbf{Third, effective learning support requires integration across the complete learning cycle.} Our review reveals that existing solutions address isolated challenges—video navigation OR content summarization OR pronunciation feedback—but fail to support the iterative nature of real learning. The Consume-Understand-Imitate framework emerges not as an arbitrary decomposition but as a natural structure reflecting how learners actually engage with audio-visual content.

This dissertation addresses these insights through three integrated systems:
- \textbf{AIxSpeed} bridges the gap in content-aware audio processing by implementing phoneme-level speed optimization based on ASR confidence
- \textbf{FastPerson} preserves multimodal richness while enabling efficient video consumption through chapter-based summarization with voice synthesis
- \textbf{Profy} provides the missing scalable feedback for skill acquisition through self-supervised error detection and visualization

Together, these systems instantiate the AI-Guided Learning paradigm: a comprehensive approach where AI augments rather than replaces human learning capabilities. By addressing each stage of the learning cycle with targeted technological innovation, we move beyond piecemeal solutions toward truly integrated learning support.

The next chapters detail how each system implements these principles and report the scope and limitations of its technical and user evaluation.

\chapter{AIxSpeed: Playback Speed Optimization Using Listening Comprehension of Speech Recognition Models}

\section{Introduction}
\label{sec:aixspeed_introduction}

This chapter presents the AIxSpeed system, specifically designed to address a key challenge within the broader AI-Guided Learning framework: the time-inefficient consumption of \textit{audio} content during the Consume-Understand learning stages. As discussed in Chapter 1, modern learners often face the necessity of efficiently acquiring knowledge from extensive long-form audio materials such as podcasts, audiobooks, and lecture recordings. This research focuses on optimizing the \textit{auditory learning experience} by proposing an innovative method for adaptive playback speed control at the phoneme level, guided by a speech recognition model's listening comprehension. However, due to the linear nature and variable information density of these contents, uniform playback speed cannot achieve optimal learning efficiency, potentially leading to decreased attention, increased cognitive load from processing overly dense or unnecessarily slow information, and ultimately, suboptimal learning outcomes. AIxSpeed presents an innovative approach of adaptive playback speed control at the phoneme level to address this challenge.

The core innovation of AIxSpeed lies in its ability to determine optimal playback speed for each segment based purely on acoustic features without requiring semantic understanding of the educational content. This is achieved by leveraging speech recognition models' confidence scores as a proxy for human listening ease. This content-agnostic approach enables the construction of a versatile system applicable to a wide range of audio content regardless of language or topic.

With the proliferation of video distribution services, opportunities for watching videos for various purposes, including information gathering, learning, and entertainment, have significantly increased. Humans possess the ability to comprehend naturally observed phenomena at rates faster than their original occurrence. Consequently, it has become common practice to increase the playback speed of videos and audio to understand more content in a shorter period. Research has demonstrated that, under specific conditions, differences in video playback speed do not affect learning effectiveness in educational video viewing and may even enhance performance~\cite{chen2024effect, Lang2020}. Furthermore, it has been reported that the proportion of viewers who prefer high-speed viewing of dramas ranges from 30\% to 80\%, varying by country~\cite{Duan2019}.

\begin{figure}[t]
    \centering
    \includegraphics[width=\textwidth]{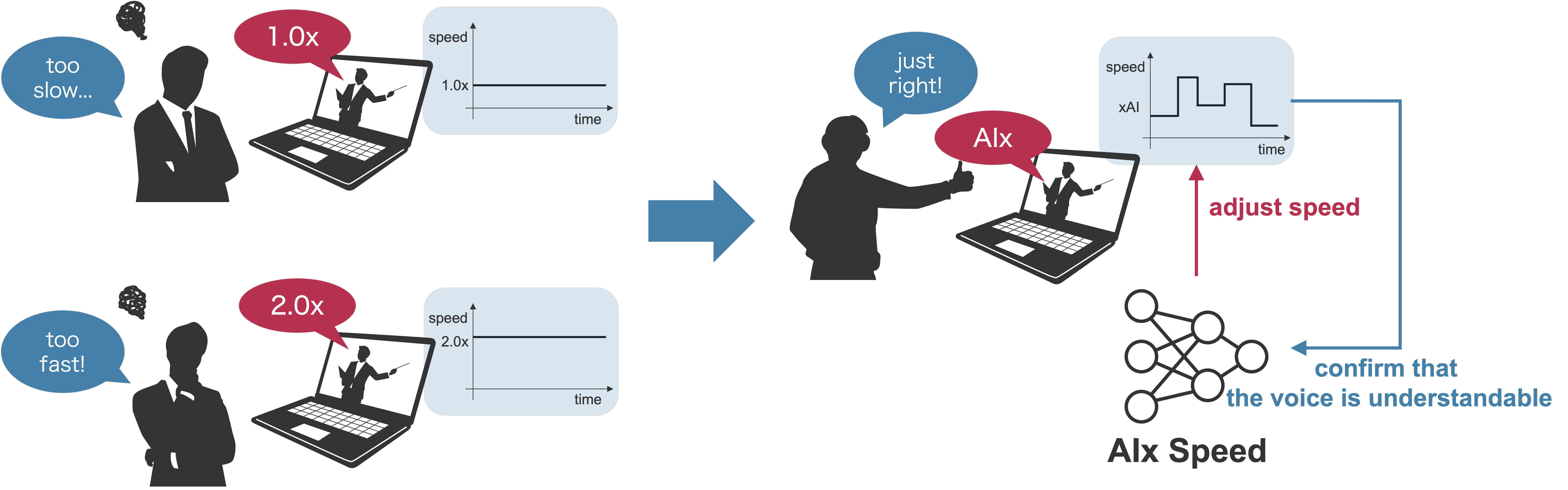}
    \caption[AIxSpeed optimizes video playback speed]{AIxSpeed optimizes the playback speed of a video in units as small as phonemes, maximizing the speed of audio to the extent that humans can hear it while allowing users to watch videos at a comfortable speed without having to adjust the playback speed.}
    \label{fig:aixspeed/overview}
\end{figure}

However, current high-speed playback technologies have fundamental limitations. As detailed in Chapter 1, the information density of audio content is not constant. While introductions, repetitions, filler words, and explanations of familiar concepts can be understood at high speed without hindering comprehension, sections with technical terminology, complex concepts, or unclear speaker pronunciation require listening at standard speed or slower. Existing uniform playback speed adjustments cannot accommodate this variable information density and listening difficulty, resulting in wasting time on easily understandable parts and missing content in difficult-to-understand parts.

Given the numerous advantages of high-speed listening to video and audio content, a multitude of methods have been proposed to automatically adjust playback speed according to content structure and user state, aiming to further enhance human high-speed listening capabilities. These methods are extensively studied as video summarization~\cite{Apostolidis2021} and audio summarization~\cite{vartakavi2021audio}. Existing approaches can be broadly classified into two strategies. The first strategy involves scanning user intentions and behavior to retain only the necessary parts for viewing or adjusting playback speed proportional to the user's concentration level~\cite{Kurihara2011, Kawamura2014}. The second strategy accelerates unnecessary parts of the content (i.e., parts without speech) or decelerates parts containing speech~\cite{Higuchi2017, Zhang2020}.

However, these existing methods face two significant challenges. Firstly, they do not explicitly model whether the resulting speech at different playback speeds remains intelligible to humans, making it difficult to estimate user comprehension across a wide range of speech types. Secondly, these systems vary playback speed for large chunks of speech, leaving room for adjustment at smaller time units. Fine-grained adjustment at the phoneme level enables more localized speed control.

To address these challenges, this research proposes the AIxSpeed system. This system adjusts audiovisual output speed while maintaining intelligibility by measuring speech intelligibility after increasing the playback speed. As illustrated in Figure~\ref{fig:aixspeed/overview}, AIxSpeed flexibly optimizes video playback speed at the phoneme level. By leveraging the listening ability of neural network-based speech recognition models, which are reported to rival human performance~\cite{Xiong2016}, the system simultaneously maximizes video playback speed and speech recognition rate after speed alteration.

This chapter first examines the validity of using speech recognizers as a proxy for evaluating human listening performance through the correlation of changes in human and speech recognition performance when speed is increased. Subsequently, it compares whether speech with playback speed controlled at the phoneme level, as generated by the proposed method, or speech played at a constant high speed is easier for humans to comprehend. Furthermore, the chapter experimentally confirms that AIxSpeed can transform non-native speakers' utterances into speech that is more comprehensible to native speakers by optimizing speech rate at the phoneme level.

The primary contributions of this research are threefold:

\begin{enumerate}
    \item Showing that human transcription accuracy and ASR accuracy changed similarly across the tested playback speeds above 1.0x ($r=0.9977$), supporting ASR as a proxy under those tested conditions.
    \item Proposing a method to increase playback speed while maintaining speech intelligibility at the phoneme level. This is achieved through a novel learning framework that leverages a pre-trained speech representation model (Wav2Vec2) but uniquely fine-tunes it with a \textbf{dual-objective loss function specifically designed for AIxSpeed}. This loss function aims to simultaneously maximize playback speed (by minimizing ${\rm Loss_{speed}}$) and preserve the intelligibility of the content (by minimizing ${\rm Loss_{ctc}}$ via the ASR proxy), a distinct approach from standard ASR fine-tuning or simple rule-based speed control. The fine-grained, phoneme-level adjustment guided by this framework is a core innovation.
    \item Improving the intelligibility of non-native speakers' speech through phoneme-level optimization of speech rate, potentially opening new applications in language learning and cross-cultural communication.
\end{enumerate}

These contributions enable AIxSpeed not only to support the enhancement of human high-speed listening ability but also to improve speech comprehension by generating variable-speed speech that considers the balance between playback speed and speech intelligibility. In particular, by performing optimization based purely on acoustic features without depending on content, the system achieves high versatility applicable to a wide range of audio content. The subsequent sections of this chapter present a detailed exposition of the AIxSpeed system's design, implementation, evaluation experiments, and analysis of results.

\begin{figure*}[t]
    \begin{minipage}[b]{0.5\columnwidth}
      \centering
      \includegraphics[width=\columnwidth]{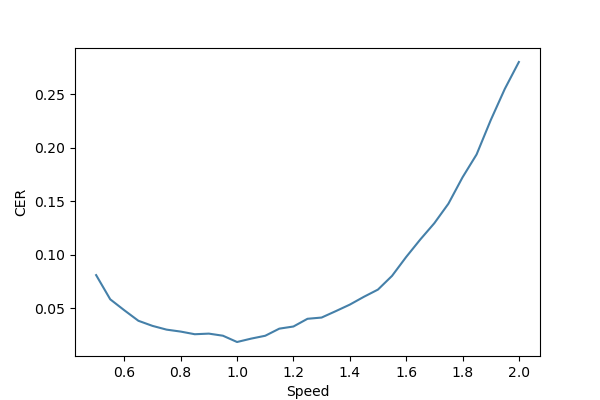}
      \subcaption{CER of ML models}
    \end{minipage}
    \hspace{0.04\columnwidth} 
    \begin{minipage}[b]{0.5\columnwidth}
      \centering
      \includegraphics[width=\columnwidth]{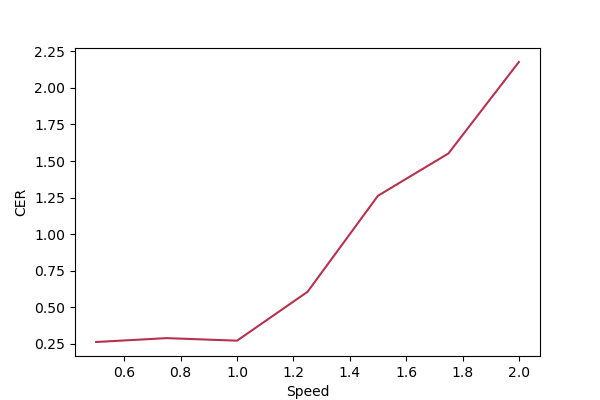}
      \subcaption{CER of humans}
    \end{minipage}\\
      \begin{minipage}[b]{0.5\columnwidth}
      \centering
      \includegraphics[width=\columnwidth]{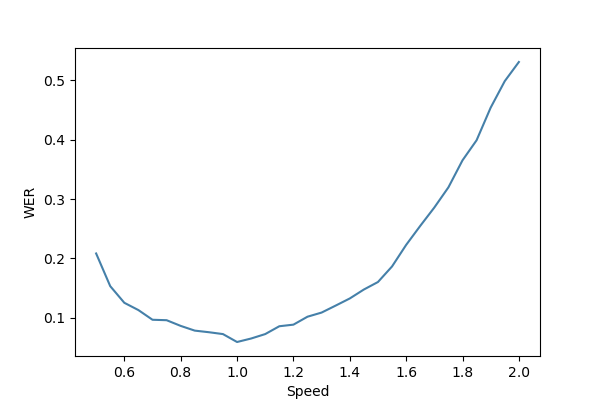}
      \subcaption{WER of ML models}
    \end{minipage}
    \hspace{0.04\columnwidth} 
    \begin{minipage}[b]{0.5\columnwidth}
      \centering
      \includegraphics[width=\columnwidth]{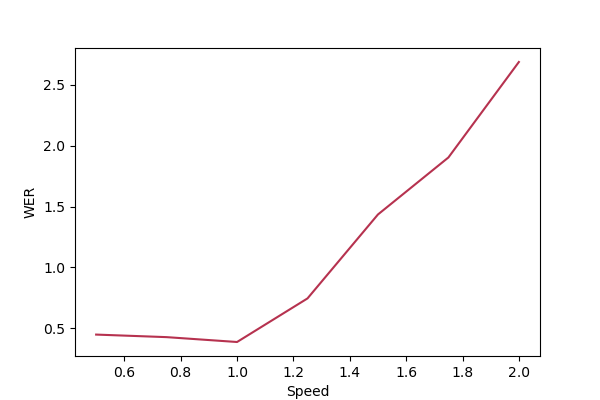}
      \subcaption{WER of humans}
    \end{minipage}
    \caption[Comparison of playback speed and listening comprehension]{Comparison of playback speed and listening comprehension in machine learning (ML)-based speech recognizers and humans, showing transitions in human and speech recognition model transcription performance at speeds greater than 1x correlated by a coefficient of 0.99.}
    \label{fig:aixspeed/pilot_study}
\end{figure*}
  
\section{Pilot Survey: Validating ASR as a Proxy for Human Listening}
\label{sec:aixspeed_pilot_survey}

The purpose of this study is to automate the maximization of speed to the extent that speech is understandable. To achieve this goal, we must validate the hypothesis that as playback speed increases, the recognition performance of both humans and speech recognizers will decrease in a similar manner. If this hypothesis is correct, we can evaluate how well a human can hear when playback speed is increased using a speech recognizer instead of a human.

Several attempts have been made to evaluate human hearing with speech recognizers in this way. For example, it has been shown that the results of human mean opinion score (MOS) listening tests correlate with the results of the speech recognition-based MOS estimation method introduced in~\cite{Jiang2002}, and in~\cite{Lionel2017}, the understanding and comprehension scores of a listener with simulated age-related hearing loss were highly correlated with a speech recognition-based system. A similar hypothesis has been used in speech learning support research to evaluate a learner's speech ability based on speech recognition performance~\cite{Cristian2021}. In other words, if a speech recognizer can recognize speech, it judges that the person speaks well.

However, the relationship between the machine learning model and the ability to understand human speech when the playback speed is varied, which is the focus of this study, has not been evaluated. Therefore, we first investigated whether this hypothesis is true. This study compared human listening performance and speech recognition performance for speech at 0.25, 0.5, 0.75, 1.0, 1.25, 1.5, 1.75, and 2.0x playback speeds.

To measure human listening performance, speech data of English sentences were prepared at each playback speed, and the subjects were asked to transcribe the data. The target English sentences were selected from LibriSpeech~\cite{Vassil2015}, a large English speech corpus created for the development and evaluation of automatic speech recognition systems. This corpus contains over 1,000 hours of audiobook readings and transcriptions. The participants were 140 English speakers who lived in the United States and had graduated from a US high school. All participants were recruited from the Amazon Mechanical Turk and were compensated for their time. Each subject was given 15 English sentences that had been randomly sped up by 0.5, 0.75, 1.0, 1.25, 1.5, 1.75, or 2.0x and asked to transcribe them.

Speech-recognizer transcriptions were collected by inputting 15 English sentences at each playback speed into a Wav2Vec2-based speech recognition model, similar to the human performance evaluation~\cite{baevski2020wav2vec}. Wav2Vec2 is a transformer-based model pre-trained through large-scale self-supervised learning, demonstrating state-of-the-art performance in speech recognition tasks.

Figure~\ref{fig:aixspeed/pilot_study} shows a graph of the change in listening performance of the human and machine learning models when the playback speed was changed. In these figures, the horizontal axis is the playback speed, and the vertical axis is the recognition performance. Recognition performance was evaluated using character error rate (CER) and word error rate (WER), which are commonly used in speech recognition (the lower these values are, the better).

The WER was calculated as follows:
{\footnotesize
\begin{align*}
\cfrac{\rm (\#~of~inserted~words + \#~of~replaced~words + \#~of~deleted~words)}{\rm (\#~of~correct~words)},
\end{align*}
}
and CER was calculated as:
{\footnotesize
\begin{align*}
\cfrac{\rm (\#~of~inserted~characters + \#~of~replaced~characters + \#~of~deleted~characters)}{\rm (\#~of~correct~characters)}.
\end{align*}
}

For both the human and machine learning models, listening performance decreased as playback speed increased from 1.0x. For playback speeds greater than 1.0x, the correlation coefficient between the change in listening performance for the human and machine learning models was 0.9977. This extremely high correlation strongly suggests that speech recognition models can function as a proxy for human listening ability.

On the other hand, when the playback speed was slowed down from 1.0x, the machine learning model showed a decrease in recognition performance, but the decline was barely observable for humans. In particular, while the performance of WER decreased slowly, the performance of CER showed almost no decrease. This indicates that human recognition performance on a character-by-character basis does not drop nearly as much when listening to slowed speech.

Thus, when the playback speed increased, the recognition performance of the human and machine learning models decreased similarly, but they exhibited different behavior when the playback speed was decreased. However, since this study focuses on increasing the playback speed of speech, the difference in behavior when the playback speed is slower than 1.0x has no effect. Therefore, by taking advantage of the fact that listening performance decreases as playback speed is increased for humans and speech recognition models alike, we replaced human listening performance with speech recognition performance to develop the desired system.

\section{System Design}
\label{sec:aixspeed_system_design}

\begin{figure*}[t]
  \includegraphics[width=0.95\textwidth]{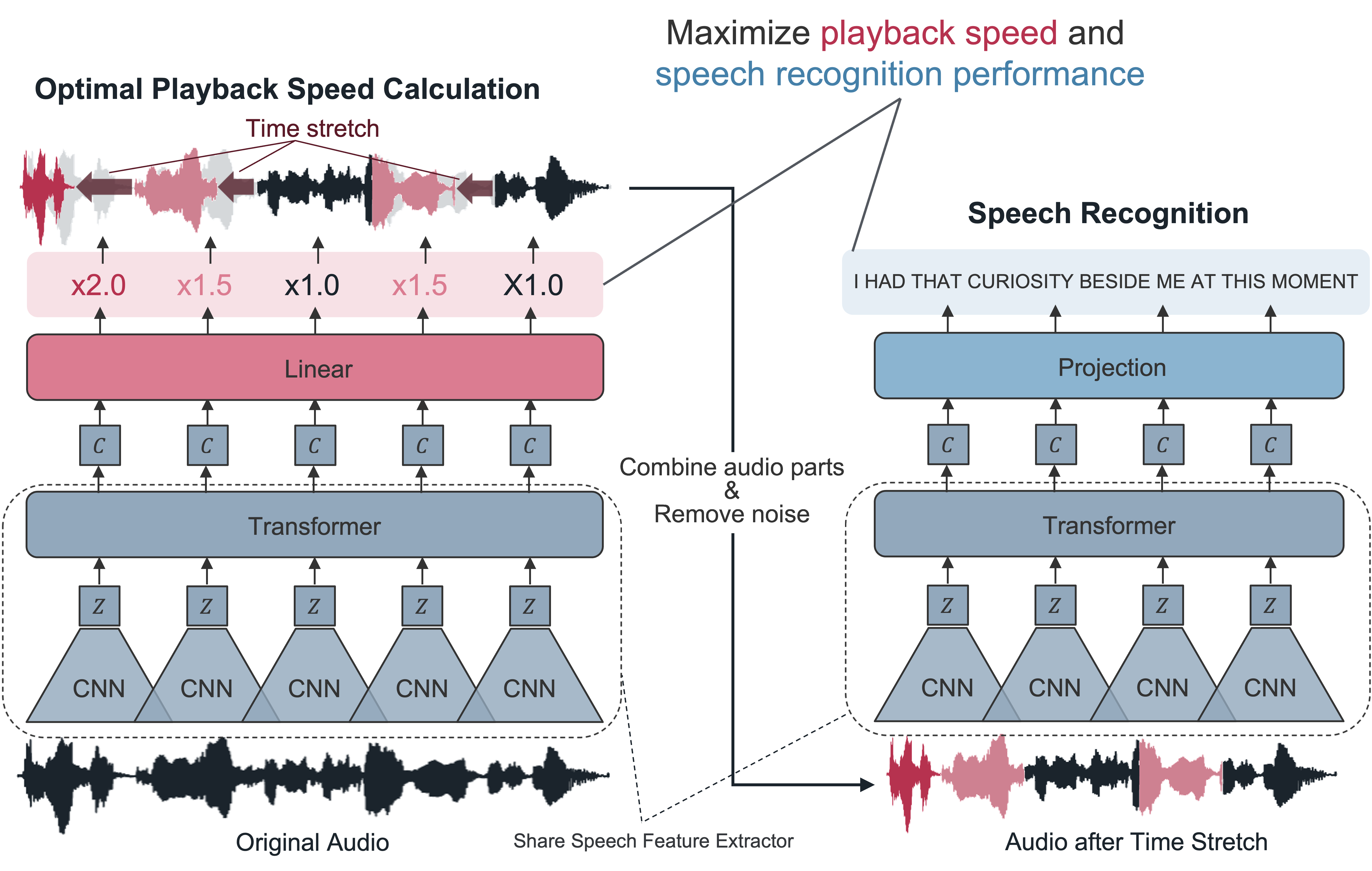}
  \caption[AIxSpeed architecture]{AIxSpeed architecture: Our system simultaneously optimizes the playback speed regulator (left) and the speech recognizer (right) to maximize the playback speed to the extent that the model can recognize.}
  \label{fig:aix_speed/architecture}
\end{figure*}

The AIxSpeed system (Figure~\ref{fig:aix_speed/architecture}) ingeniously combines the power of pre-trained speech models with a task-specific optimization strategy. 
While building upon the robust speech representations learned by Wav2Vec2~\cite{baevski2020wav2vec}, our primary innovation lies in the architectural design involving a playback speed adjuster and a speech recognizer that share features, and crucially, in the formulation of a \textbf{composite loss function (${\rm Loss} = {\rm Loss_{speed}} + \lambda {\rm Loss_{ctc}}$)}. This function guides the model to learn optimal phoneme-level speed modulations by explicitly balancing the drive for increased speed with the constraint of maintaining intelligibility as judged by the ASR proxy. 
This allows AIxSpeed to dynamically adjust speed based on localized acoustic intelligibility, moving beyond the capabilities of conventional playback controls or ASR systems not designed for this specific audio transformation task.
AIxSpeed increases the speed as much as possible, as long as the user can understand it. This system allows users to watch videos in a time-efficient manner without having to adjust the playback speed for each video. In addition, the system can automatically improve intelligibility by adjusting the speech speed to accommodate non-native speakers who are not proficient in the target language.

\begin{figure}[t]
  \includegraphics[width=\textwidth]{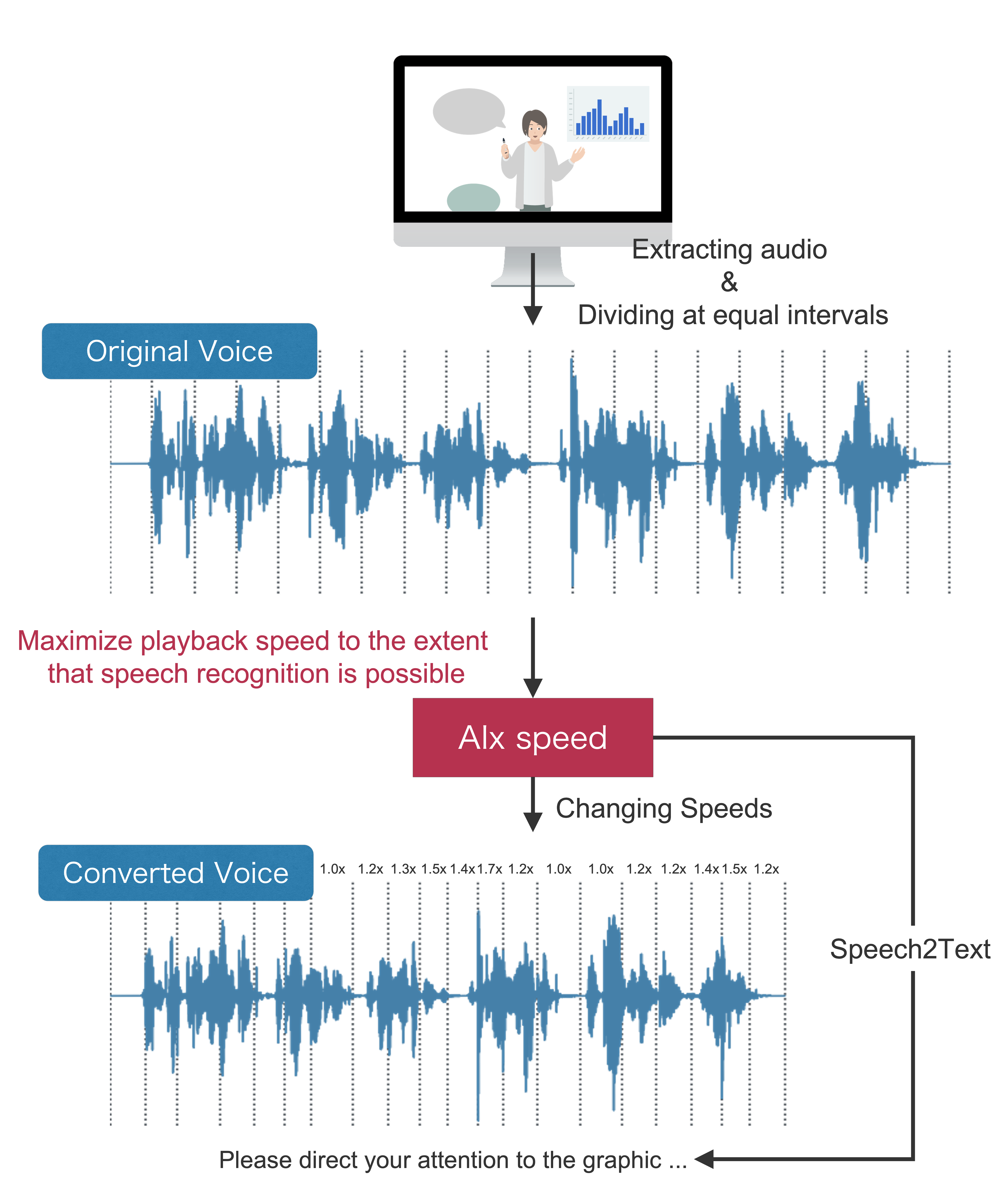}
  \caption[The work process of AIxSpeed]{The work process of AIxSpeed: extracting human voices from the target video, dividing them into specified equal intervals, calculating the optimal playback speed for each divided voice, changing each voice to the fixed playback speed and synthesizing it into one voice, and performing speech recognition on the synthesized voice to confirm understandability.}
  \label{fig:aix_speed/aix_speed}
\end{figure}

The working process of AIxSpeed is illustrated in Figure~\ref{fig:aix_speed/aix_speed}. The system first extracts the human voice from the target video. Next, it splits this voice into specified equal intervals. Next, using each segmented voice as input, the system calculates the optimal playback speed for each segment voice, taking into account the characteristics of the voice as a whole. Finally, the system changes each voice to the specified playback speed and combines them into a single voice. At the same time, the combined voice is recognized by speech recognition to confirm that the resulting single voice is understandable.

This system consists of two mechanisms, as shown in Figure~\ref{fig:aix_speed/architecture}. One is a playback speed adjuster (left), and the other is a speech recognizer (right). The former is used to maximize the playback speed of the input speech, while the latter is used to evaluate how understandable the input speech is. By training these two models simultaneously, it is possible to generate speech that plays back as fast as possible within the comprehension range.

\subsection{Playback Speed Adjuster}
\label{subsec:aixspeed_playback_speed_adjuster}

In this study, we use Wav2Vec2, a self-supervised neural network designed for speech signal processing systems, to optimize the playback speed. Wav2Vec2 is a model developed by Facebook's research team that can learn speech representations from large-scale unlabeled speech data. This model can be thought of as a speech version of masked language modeling like BERT, learning by masking parts of the input speech and predicting the original speech from the remaining parts.

The playback speed controller is divided into a feature extractor layer and a linear layer, as shown in Figure~\ref{fig:aix_speed/architecture}~(left). They are trained by pre-training through self-supervised representation learning on unlabeled speech data and regression learning, which outputs the playback speed based on the features after the representation learning.

The pre-training method for the feature extractor layer is similar to masked language modeling, as exemplified by bidirectional encoder representations from transformers (BERT)~\cite{devlin-etal-2019-bert} in natural language processing, where a portion of the input is masked and the corresponding utterance features are estimated from the remaining input. In this way, the model can learn good-quality features of the target language to which the rate of utterance should be adapted.

Specifically, Wav2Vec2 has the following architecture:
\begin{itemize}
  \item \textbf{Convolutional Neural Network (CNN) encoder}: Converts raw speech waveforms into 512-dimensional feature vectors.
    \begin{itemize}
      \item Number of layers: 7
      \item Channels per layer: 512
      \item Kernel sizes: $\{10, 3, 3, 3, 3, 2, 2\}$
      \item Strides: $\{5, 2, 2, 2, 2, 2, 2\}$
    \end{itemize}
  \item \textbf{Transformer encoder}: Processes CNN features through stacked self-attention and feedforward layers.
    \begin{itemize}
      \item Transformer blocks: 12
      \item Hidden dimension: 768
      \item Feed-forward dimension: 3072
      \item Attention heads: 8
    \end{itemize}
  \item \textbf{Quantization module}: Converts continuous features into discrete tokens, enabling mask prediction tasks.
\end{itemize}

\begin{figure*}[t]
  \includegraphics[width=1.01\textwidth]{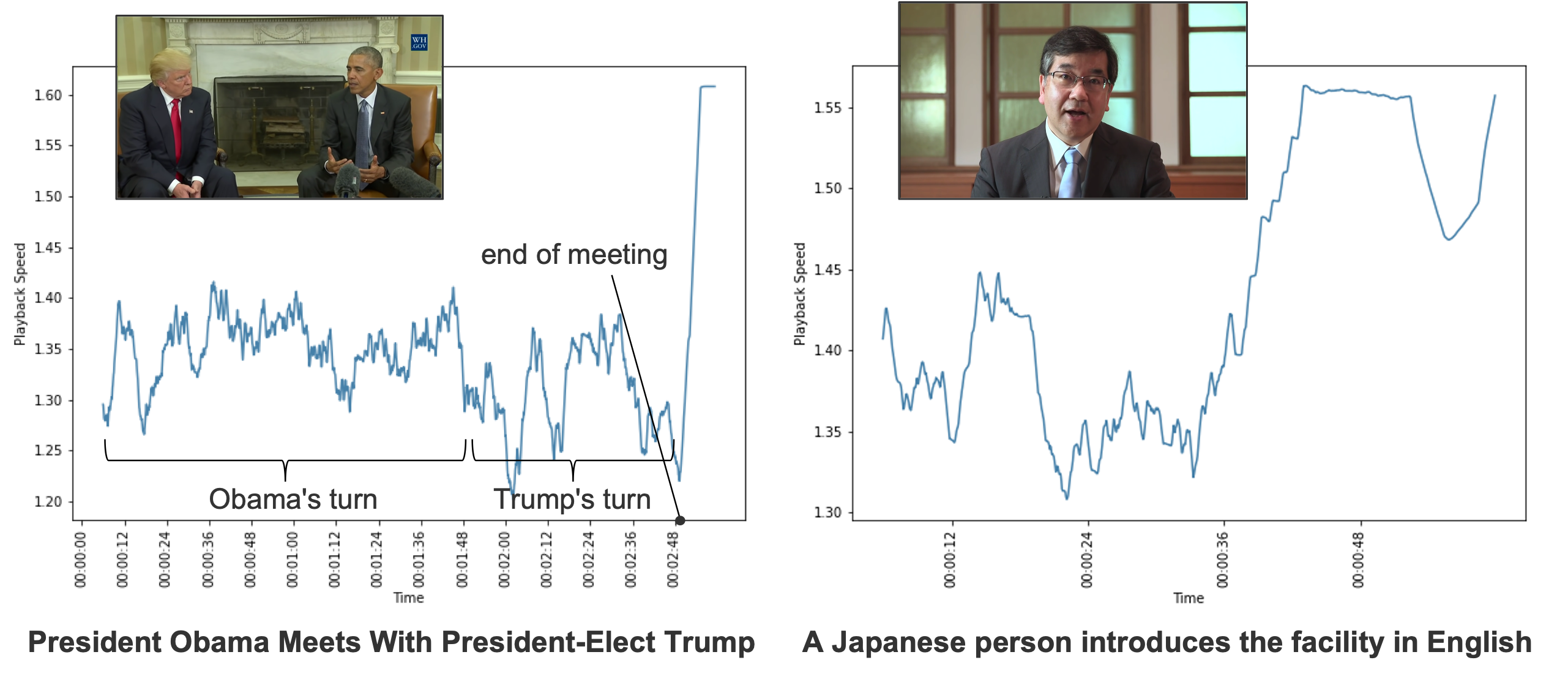}
  \caption[Application example of AIxSpeed]{Application example showing how the playback speed changed when AIxSpeed was applied to two YouTube videos, demonstrating flexible changes according to the speaker's speech.}
  \label{fig:aix_speed/application}
\end{figure*}

Typically, these self-supervised learners are used to tackle tasks such as speech recognition and speaker identification by pre-training and then fine-tuning with a small amount of label data. For example, in speech recognition, we have added a projection layer and a connectionist temporal classification (CTC) layer~\cite{Graves:06icml} to the output of self-supervised learners, such as Wav2Vec2 and HuBERT~\cite{Hsu2021a}, to enable transcription from speech waveforms.

Similar to these methods, we pre-train on unlabeled speech data and then connect and train a linear layer that outputs rates. In general speech processing tasks, such as speech recognition~\cite{Mishaim21} and speaker identification~\cite{Zx21, Kabir2021ASO}, the difference from the class label is minimized as an error function. On the other hand, the goal of the playback speed adjuster is to maximize the speed. Therefore, we designed the following function whose value decreases as the speed increases.

\begin {align*}
    {\rm Loss_{speed}} = \left(\frac{1}{10}\right)^{\frac{1}{S}\sum_{s=1}^{S}r_s},
\end {align*}

where $r_s$ is the playback speed for each segment obtained using Wav2Vec2 and a linear layer with audio $X$ as input as follows:

\begin {align*}
  C = c_{\left[1:T\right]}= {\rm Wav2Vec2}\left(X\right), \\
  R = r_{\left[1:S\right]} = {\rm Linear}\left(C\right).
\end {align*}

This loss function design trains the model to output as high a playback speed as possible. By using an exponential function, the loss decreases rapidly as speed increases, reinforcing the incentive for acceleration.

\subsection{Speech Recognizer}
\label{subsec:aixspeed_speech_recognizer}

The speech recognizer transcribes the speech converted to the playback speed obtained by the playback speed adjuster (Figure~\ref{fig:aix_speed/architecture}~(right)). When speech parts with different playback speeds are combined, noise is generated in the speech data and the sound quality is degraded. We use voice separation technology~\cite{mcfee2015librosa} to extract only the speaker's voice and reduce the effect of noise.

The resulting speech is then fed into a speech recognizer for speech recognition. The speech recognition process consists of the extraction of acoustic features from the speech waveform, estimation of the classes of acoustic features for each frame, and generation of hypotheses from the sequence of class probabilities. Since the speech recognizer can partially share the neural network with the playback speed adjuster, the overall network size can be reduced.

As shown in Figure~\ref{fig:aix_speed/architecture}, the dotted speech feature extractor is shared. The speech features obtained by this mechanism are used as input to generate text in the projection layer. In this process, the speech recognizer is trained to minimize CTC loss, as in normal speech recognition.

The CTC loss is defined as follows:
\begin{align*}
{\rm Loss_{ctc}} = -\log P(Y|X)
\end{align*}
where $Y$ is the correct character string and $X$ is the input speech. CTC can estimate the optimal character string considering all possible alignments even when the alignment between input and output is unknown.

In summary, the entire model is trained to minimize the following error function, which is a combination of this error function and the error function of the playback speed adjuster:

\begin {align*}
    {\rm Loss} = {\rm Loss_{speed}} + \lambda {\rm Loss_{ctc}}.
\end {align*}

Here, $\lambda$ is a hyperparameter that adjusts the importance of the playback speed calculation and speech recognition. By properly setting this hyperparameter, we can balance maximizing speed and maintaining recognition accuracy.

\section{Prototype Implementation}
\label{sec:aixspeed_prototype}

As a prototype of AIxSpeed, an application that optimizes audio playback speed using English as the target language has been implemented. This section describes the implementation and usage of the application.

\subsection{Implementation Details}
\label{subsec:aixspeed_implementation}

Wav2Vec2 was used for the prototype's shared utterance learning model (the utterance learning part shared by the speech recognizer and the playback speed adjuster). For pre-training, LibriSpeech~\cite{Vassil2015} was used as the dataset, train-clean-360 as the training data, and dev-clean as the validation data.

The dataset consists of clean speech data in LibriSpeech and was partitioned using the data partitioning method proposed in LibriSpeech for training/validation. The pre-training did not require the corresponding transcribed text, only the speech data. This is a major advantage of self-supervised learning, meaning that large amounts of unlabeled speech data can be utilized.

The hyperparameters for pre-training are as follows:
\begin{itemize}
  \item Batch size: 32
  \item Learning rate: 5e-5 (warmup steps: 32,000)
  \item Number of epochs: 400,000 steps
  \item Optimizer: Adam ($\beta_1=0.9$, $\beta_2=0.98$, $\epsilon=1\times10^{-6}$)
  \item Mask probability: 0.065
  \item Mask length: 10
  \item Temperature parameter (Gumbel softmax): Linear decay from 2.0 to 0.5
\end{itemize}

We then trained a speech recognizer and a playback speed adjuster using two sets of speech data, including the transcribed text. One was LibriSpeech's train-clean-100, and the other was the English speech database read by Japanese students (UME-ERJ)~\cite{Nobuaki2002}. The latter is a dataset of English spoken by non-native Japanese speakers, with 202 speakers (100 males and 102 females) reading simple English sentences.

The playback speed adjuster and the speech recognizer were trained separately. First, the speech recognizer was trained using LibriSpeech and UME-ERJ, and then the playback speed adjuster was trained using the same data with fixed weights for the speech recognizer. This staged training allows the playback speed adjuster to optimize speed while maintaining performance after the speech recognizer achieves stable performance.

All speech files used for training were normalized to a sampling frequency of 16 kHz. The Wav2Vec2 used the initial parameters implemented in PyTorch~\cite{Paszke2019}, and the final layers of both the playback rate adjuster and the speech recognizer were $768$-dimensional linear layers. The hyperparameter $\lambda$ for the loss function was set to $10^{-7}$, and the model was trained for 5 epochs with a batch size of 32 using the AdamW~\cite{loshchilov2018decoupled} optimization algorithm.

The detailed hyperparameters for fine-tuning are as follows:
\begin{itemize}
  \item Learning rate: 1e-4 (speech recognizer), 1e-5 (playback speed adjuster)
  \item Weight decay: 0.01
  \item Dropout rate: 0.1
  \item Gradient clipping: 1.0
  \item Warmup steps: 10\% of total steps
\end{itemize}

\begin{table*}[t]
  \caption{Performance comparison of models}
  \begin{minipage}{\linewidth}
  \centering
  \subcaption{LibriSpeech}
  \begin{tabular}{cccc}
    Model & Avg. Speed & CER & WER \\ \hline
    Wav2Vec2 (1.00x) & 1.00 & \textbf{4.38} & \textbf{12.57} \\
    Wav2Vec2 (1.30x) & 1.30 & 5.61 & 14.58  \\
    Wav2Vec2 (1.50x) & 1.50 & 6.83 & 17.19  \\
    \textbf{AIxSpeed} & 1.30 & \underline{5.21} & \underline{12.96} \\
  \end{tabular}
  \end{minipage}\\
  \begin{minipage}{\linewidth}
  \centering
  \subcaption{UME-ERJ}
  \begin{tabular}{cccc}
    Model & Avg. Speed & CER & WER \\ \hline
    Wav2Vec2 (1.00x) & 1.00 & \underline{26.74} & \underline{55.02}  \\
    Wav2Vec2 (1.29x) & 1.29 & 33.71 & 63.90  \\
    Wav2Vec2 (1.50x) & 1.50 & 36.93 & 66.51 \\
    \textbf{AIxSpeed} & 1.29 & \textbf{26.45} & \textbf{53.13} \\
  \end{tabular}
  \end{minipage}
  \label{table:aix_speed/technical_evaluation}
\end{table*}

\subsection{Usage of the Application}
\label{subsec:aixspeed_usage}

Examples of using AIxSpeed are shown in Figure~\ref{fig:aix_speed/application}. This is an example of the prototype applied to a video uploaded to YouTube. The horizontal axis is the playback time, and the vertical axis represents the playback speed output by the model for each playback time. The first is an example of speeding up a dialogue, movie, or lecture.

Two speakers appear in the video, and the optimal playback speed can be set for each speaker. Of particular interest is that the two speakers in the video speak at different speeds, so the average playback speed for the two speakers is different. It can also be seen that the playback speed increases drastically from the moment when the dialog between the two speakers ends and there is no more speech from the person. Although we did not intend to design this feature, we can see that our system, like conventional playback speed controllers, can speed up the playback speed in the parts where there is no speech.

The second example applies speed adjustment to non-native speech. The model selected a higher average playback factor than for the native-speaker speech in the first video. This example shows different model-selected speeds, but does not by itself establish improved human comprehension.

An important feature of this system is its fine-grained playback-speed control at the phoneme level. The experiments evaluated ASR performance and perceived speech quality, but did not separately evaluate the preservation of prosody or intonation.

\section{Evaluation}
\label{sec:aixspeed_evaluation}

\subsection{Technical Evaluation}
\label{subsec:aixspeed_technical_evaluation}

To evaluate whether the proposed method can optimize playback speed while maintaining ASR performance, we compared the CER and WER values for AIxSpeed-modified speech with those for constant-speed playback. The comparison conditions were 1.0x, 1.5x, and constant playback matched to AIxSpeed's average playback factor.

A standard Wav2Vec2-based speech recognition model, which was the speech recognizer used in our method, was used to compute the CER and WER for comparison. The model performance is shown in Table~\ref{table:aix_speed/technical_evaluation}~(a) and (b) for LibriSpeech and UME-ERJ, respectively.

AIxSpeed produces speech 1.30 times faster on average for LibriSpeech and 1.29 times faster on average for UME-ERJ. Both results show that the playback speed optimized by AIxSpeed has lower values for both CER and WER than constant-speed playback matched to the same average factor. Thus, the proposed method retained better ASR performance than the matched constant-speed baseline.

In addition, for UME-ERJ, speech generated by AIxSpeed had a lower WER than speech at 1.0x playback. This indicates that it was more recognizable to the ASR model used for evaluation; it does not directly establish improved intelligibility or comprehension for human listeners, which requires a dedicated human study.

\subsection{User Evaluation}
\label{subsec:aixspeed_user_evaluation}

User experiments were conducted to evaluate the perceived quality of the generated speech. Speech generated by the proposed method was compared with speech played at a constant speed matched to the same average playback factor. Quality was evaluated using the mean opinion score (MOS), which is commonly used in speech synthesis research~\cite{Vandenoord16, Yuxuan2017}. This measure rates speech quality on a five-point scale from 1 (poor) to 5 (excellent).

Participants were 50 US residents who used English on a daily basis. 20 sentences were extracted from each of the LibriSpeech and UME-ERJ datasets, and half were converted to speech with the proposed speed-up, while the other half were converted to speech with a constant speed-up. Participants were given a total of 40 sentences of audio and asked to rate the quality of the audio.

\begin{itemize}
  \item The source of the audio (AIxSpeed or baseline) was not disclosed to participants.
\end{itemize}

\begin{figure}[t]
  \includegraphics[width=\textwidth]{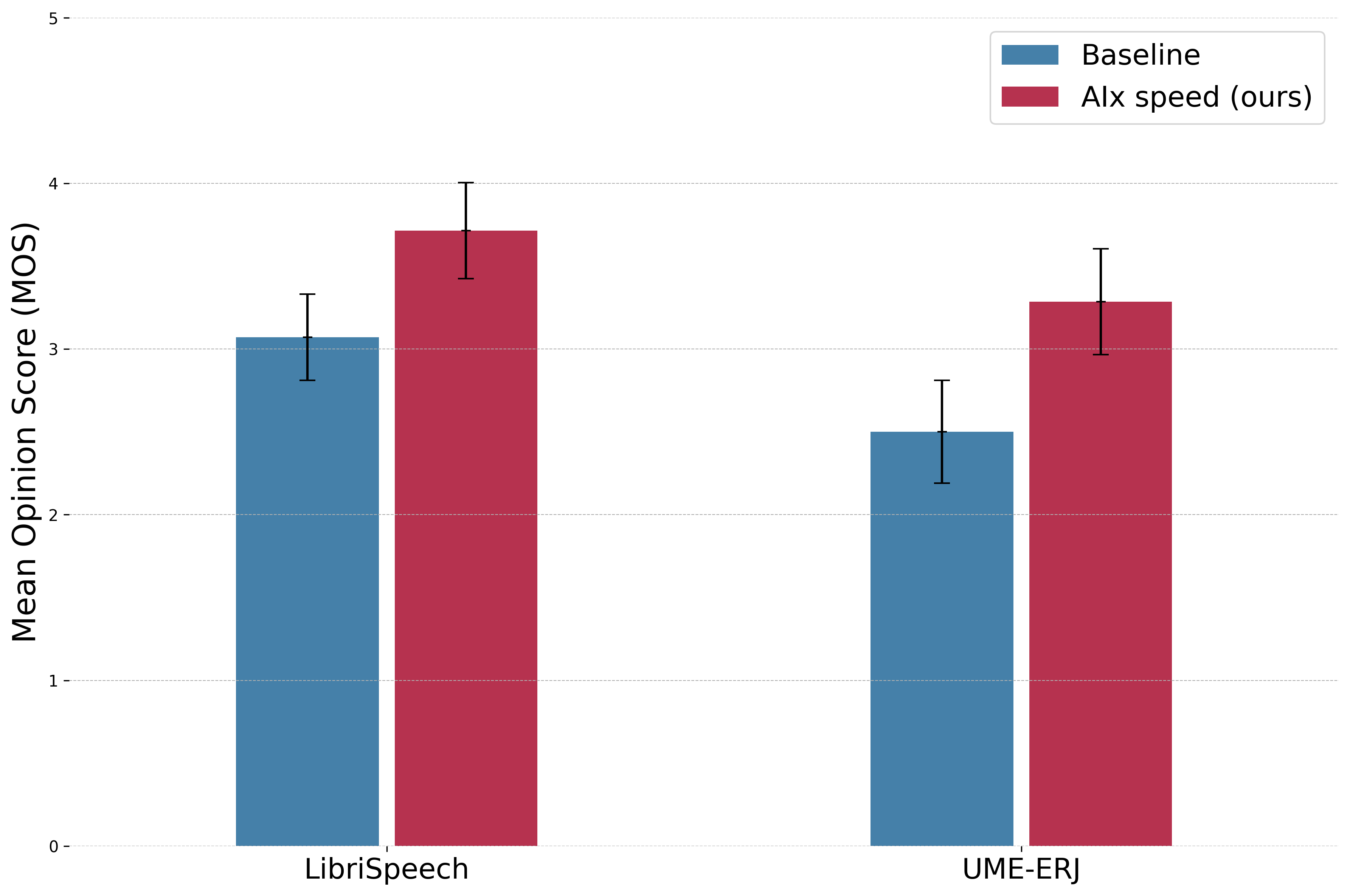}
  \caption[Voice quality comparison]{Voice quality comparison between baseline (constant playback speed increase throughout) and AIxSpeed (flexible playback speed increase).}
  \label{fig:aix_speed/user_evaluation}
\end{figure}

Figure~\ref{fig:aix_speed/user_evaluation} shows the perceived quality of speech at baseline and with AIxSpeed. The mean opinion scores for LibriSpeech and UME-ERJ were 0.5 and 0.8 points higher with the proposed method, respectively. The source paper did not report a significance test for these differences.

A larger MOS difference was observed for UME-ERJ than for LibriSpeech, but further evaluation is needed to determine whether AIxSpeed is especially effective for non-native-speaker speech.

\section{Discussion}
\label{sec:aixspeed_discussion}

\subsection{Evaluation Results and Implications}
\label{subsec:aixspeed_evaluation_results_implications}

The technical evaluation shows that the proposed method produces lower CER and WER than constant-speed speech at the same average playback factor. The user evaluation likewise shows higher mean opinion scores for AIxSpeed speech than for the matched constant-speed baseline. This average speed-up of approximately 1.3x (e.g., reducing a 60-minute lecture to about 46 minutes) represents a tangible time saving for learners.

This allows users to watch videos at a reasonable speed without having to adjust the playback speed for each video. However, the improvement in MOS values by using the proposed method is by no means sufficient. In the current model, the average conversion to a faster playback speed is about 1.3 times, but it is a future task to investigate whether it is possible to make this even faster.

In fact, many video playback services implement 1.5x and 2.0x playback speeds, and some people watch dramas and lectures at such speeds. Therefore, we expect that it will be possible to convert up to this speed and make the audio easy to understand. In addition, since each user has a different preferred playback speed, personalizing the model so that it plays at the optimal playback speed for users is also a future issue.

\subsection{Limitations and Future Work}
\label{subsec:aixspeed_limitations_future_work}

One key aspect of AIxSpeed is its content-agnostic approach, relying purely on acoustic features to determine playback speed. While this allows for broad applicability across languages and topics without needing semantic understanding, it also presents certain limitations, particularly in educational contexts. 
For instance, the system does not currently consider the \textit{semantic importance} of content segments or a speaker's pedagogical intent. 
A lecturer might intentionally slow down their speech to emphasize a critical point or explain a complex idea, even if their pronunciation is perfectly clear. 
In such cases, AIxSpeed, prioritizing acoustic intelligibility as determined by the ASR proxy, might inappropriately speed up these segments, potentially hindering deep comprehension or missing pedagogical cues. 
Future research could explore integrating natural language processing techniques to identify semantically important segments or to understand discourse structures (e.g., cues for emphasis or topic shifts). This would allow for a more nuanced speed adjustment that balances acoustic intelligibility with the preservation of pedagogically salient prosodic variations.
Furthermore, adapting the system to accommodate intentional pauses or variations in speech rhythm that contribute to the speaker's style or the listener's engagement, rather than solely optimizing for raw speed based on ASR confidence, remains an avenue for future enhancement.

The personalization mentioned in Section~\ref{subsec:aixspeed_evaluation_results_implications} could involve allowing users, or an overarching AI-Guided Learning system, to set a target intelligibility level (e.g., by adjusting the ASR confidence threshold that AIxSpeed uses as a proxy for human listening ease). 
For example, advanced learners or those familiar with the content might opt for a slightly lower intelligibility threshold to maximize speed further, while beginners or those encountering complex material might prefer a higher threshold to ensure thorough understanding. This would cater to the diverse needs, prior knowledge, and capabilities of learners, addressing the potential gap between a generic optimization and individual learner preferences.

While the current model achieves a commendable average speed-up of approximately 1.3x while improving or maintaining intelligibility compared to constant speed-up, future iterations could explore architectural or loss function modifications to safely push towards higher average speeds (e.g., 1.5x to 2.0x), which are common user-selectable options in many commercial players. 
Investigating the system's behavior and the perceived intelligibility limits at even more extreme speeds (e.g., up to 4x for highly redundant or exceptionally clear speech segments) could reveal further optimization potential or define an "intelligibility cliff" for adaptive speed methods. However, achieving high intelligibility at such significantly accelerated rates for general audio content would likely require substantial advancements in both the speed modification algorithm and the underlying intelligibility prediction model.

While AIxSpeed demonstrated effectiveness on English speech (both native and non-native), its performance across a wider range of languages, accents, and acoustic environments (e.g., noisy conditions, overlapping speech) requires further investigation to fully substantiate its claimed versatility. The current ASR proxy model is trained primarily on English, and its reliability as an intelligibility measure for other languages would need careful validation or language-specific adaptation.

\subsection{Listening Comprehension and Speech Recognition}
\label{subsec:aixspeed_listening_comprehension}

As discussed in the preliminary research chapter (Section~\ref{sec:aixspeed_pilot_survey}), it can be seen that there is a relationship between speech recognition performance and human transcription ability. Thus, we expect to build automated systems for various tasks and evaluations by replacing human speech comprehension ability with machine learning models, as in this system.

On the other hand, speech recognition performance based on playback speed does not perfectly match human speech comprehension. In other words, as the playback speed increases, the dictation performance decreases in both cases, but the performance values are not exactly the same. Thus, we anticipate that by training speech recognition models to match these relationships as closely as possible, it will be possible to use them more generally as alternatives to humans.

Distillation, a technique that learns to approximate an output that matches existing results, will be the technical key. Specifically, by using human listening data as a ``teacher" and training speech recognition models to show similar performance degradation patterns, it may be possible to build more accurate proxy models of human listening ability.

\subsection{Adjustment of Non-Native Speakers' Speech}
\label{subsec:aixspeed_non_native_adjustment}

The UME-ERJ ASR result suggests that moderate acceleration of non-native speech may be useful under some conditions. Because this study did not measure human intelligibility or comprehension for that comparison, this remains a hypothesis for future evaluation.

One motivation for this study was an informal observation that accelerated playback of some non-native speech could sound easier to follow. This observation is not treated as empirical evidence in the present study.

\subsection{Potential Applications Beyond Speed Optimization and Societal Implementation}
\label{subsec:aixspeed_beyond_speed_societal}

The technology of AIxSpeed has various application possibilities beyond simple high-speed playback. Particularly noteworthy is its application to hearing aids for the hearing impaired. By utilizing the technology for evaluating listening difficulty at the phoneme level, it is thought that it will be possible to develop adaptive hearing aids that automatically play difficult-to-hear parts slowly.

Such hearing aids could dynamically adjust playback speed according to environmental sounds and speaker characteristics. This is a possible research direction, not an evaluated application of the present study: studies with hearing-impaired users and noisy speech would be needed to determine whether it improves listening or comprehension.

Furthermore, applications to language learning can be expected. By automatically playing parts containing difficult phonemes or words slowly according to the learner's level, more effective listening practice becomes possible. It may also be used in presentation practice or lecture preparation, where speakers can objectively evaluate and improve the listening difficulty of their own utterances.

Furthermore, when compared to standard playback speed controls found in many video/audio platforms (e.g., YouTube, podcast players), AIxSpeed offers a significant advantage through its \textbf{adaptive and fine-grained (phoneme-level) speed adjustments}. 
Instead of a uniform speed-up that can make difficult passages unintelligible or slow passages tedious, AIxSpeed optimizes locally, aiming to maximize speed only where content remains clear, thus providing a more consistently comfortable and efficient listening experience. This contrasts with comprehensive language learning platforms like Duolingo, which focus on interactive exercises and curriculum, whereas AIxSpeed enhances the consumption of existing, often passive, audio learning materials.

Regarding broader societal implementation, AIxSpeed could be integrated as an advanced feature within existing Learning Management Systems (LMS), as a plug-in for popular video and audio streaming platforms, or even as a browser extension enhancing web-based audio content. 
Key technical challenges for widespread adoption include ensuring real-time or near real-time processing capabilities for on-the-fly adjustments, optimizing computational efficiency for deployment on diverse end-user devices (including mobile), and developing intuitive user interfaces that provide appropriate control and transparency over the automated speed adjustments. 
Furthermore, practical deployment would necessitate considerations regarding the processing of copyrighted material and ensuring user data privacy, especially if personalization features are implemented. 
Pilot studies in authentic educational settings would be crucial to gather user feedback on usability and perceived learning benefits, and to iteratively refine the system for robust, real-world application.

\section{Conclusion}
\label{sec:aixspeed_conclusion}

This chapter has presented AIxSpeed, a system that addresses rapid listening to audio content in the Consume stage of the AI-Guided Learning framework. AIxSpeed automatically adjusts playback speed at the phoneme level using ASR performance as a proxy derived from the pilot comparison with human transcription performance.

The key contributions of AIxSpeed are threefold. Firstly, the pilot study found a correlation of 0.9977 between changes in human transcription accuracy and ASR accuracy at tested playback speeds above 1.0x. Secondly, AIxSpeed introduces phoneme-level speed adjustment and achieved lower CER and WER than constant-speed playback matched to the same average factor. Thirdly, the UME-ERJ evaluation found lower WER than at 1.0x playback; this is an ASR result and does not establish improved human intelligibility for non-native speech.

The experimental results presented in this chapter provide evidence for the efficacy of AIxSpeed. At its average playback factors (1.30x for LibriSpeech and 1.29x for UME-ERJ), the system produced lower Character Error Rate (CER) and Word Error Rate (WER) than constant-speed playback at the same factors. In the user study, AIxSpeed speech also received higher mean opinion scores than the matched constant-speed baseline. These findings indicate the system's potential to enhance the efficiency of audio content consumption while maintaining intelligibility.

Despite these promising results, it is important to acknowledge the limitations of the current implementation of AIxSpeed. The system's performance may vary depending on factors such as the acoustic characteristics of the input speech, the diversity of speakers, and the complexity of the content. Further research is needed to assess and optimize the system's performance across a wider range of speech types and listening conditions. Additionally, the long-term effects of using AIxSpeed on listeners' cognitive load and information retention need to be thoroughly investigated.

Looking ahead, AIxSpeed holds considerable promise for a variety of applications. In the context of video distribution services, it could enhance user experience by providing personalized playback speeds that maximize comprehension and engagement. In language learning scenarios, AIxSpeed could be integrated into tools that help learners adapt to different speech rates and accents, potentially accelerating the acquisition of listening skills. Furthermore, the system's ability to improve the intelligibility of non-native speech could have significant implications for international communication and collaboration in educational and professional settings.

One possible future direction is adaptive playback for hearing-impaired users. A system could adjust playback speed according to the environment and speaker, but AIxSpeed was not evaluated with hearing-impaired users or as a hearing aid. Whether such adaptation improves accessibility requires dedicated clinical and usability evaluation.

Within the broader scope of this thesis, AIxSpeed complements the other two systems presented: FastPerson, which focuses on rapid comprehension of video content, and Profy, which addresses efficient feedback systems for skill acquisition. Together, these three systems form a comprehensive approach to enhancing time efficiency in learning through multimedia content. While AIxSpeed optimizes the auditory aspect of content consumption, FastPerson and Profy address the visual and interactive components respectively, collectively contributing to a more effective and personalized learning experience.

In conclusion, AIxSpeed represents a significant step forward in the field of time-efficient learning support systems. By enabling users to consume audiovisual content at optimal speeds without manual adjustment, it not only enhances the efficiency of information acquisition but also potentially improves comprehension, particularly for non-native speech. As research in this area continues to evolve, future work should focus on integrating AIxSpeed with other learning technologies, expanding its applicability across diverse content types and user groups, and conducting longitudinal studies to assess its impact on long-term learning outcomes. The insights gained from this research have the potential to reshape how we approach audio-based learning and information consumption in an increasingly fast-paced digital world.

\chapter{FastPerson: Enhancing Video-Based Learning through Video Summarization that Preserves Linguistic and Visual Contexts}

\section{Introduction}
\label{sec:fastperson_introduction}

This chapter presents FastPerson, a novel video summarization system that enables efficient comprehension of video content in the Consume-Understand stages toward realizing AI-Guided Learning, the overall goal of this thesis. As discussed in Chapter 1, lecture videos contain multiple information channels including speaker explanations (audio), slides and board writing (visual), and gestures, but conventional summarization methods cannot handle these integratively. FastPerson addresses this challenge through an innovative approach that integrates visual and audio information and generates summary videos in the original speaker's voice.

The core innovation of FastPerson lies in its novel ``video-to-video summarization" paradigm. Unlike conventional approaches that predominantly generate text-based transcripts or summaries from audio~\cite{Kim2016}, thereby sacrificing the rich visual context, or those that offer simple chaptering without true content distillation~\cite{Yang2023}, FastPerson processes both visual and audio streams to produce a condensed \textit{video} summary. This approach maintains the multimodal integrity of the original lecture. Furthermore, by employing voice cloning technology to retain the original speaker's vocal characteristics, FastPerson ensures a seamless and cognitively coherent learning experience, which is often disrupted when navigating between original content and disparate summary formats.

The proliferation of video-sharing platforms has significantly enhanced the use of videos across various fields, with education emerging as a particularly impactful domain. The adoption of video-based learning methods, such as flipped learning and Massive Open Online Courses (MOOCs), has revolutionized educational paradigms. Flipped learning inverts the conventional classroom model by encouraging students to engage with video lectures before class, allowing in-class sessions to focus on interactive discussions and hands-on learning experiences~\cite{Betihavas2016}. Similarly, MOOCs have democratized access to education, offering an extensive array of video lectures and educational resources to a global audience~\cite{Hew2014, Guo2014}.

Numerous studies have confirmed the effectiveness of video-based learning methods. Zhang et al. demonstrated that interactive videos improve learners' comprehension~\cite{Zhang2006}, while Kay underscored the positive impact of video-based learning on comprehension, adaptation, and achievement~\cite{Kay2012}. Despite these benefits, significant challenges persist in the realm of video-based learning.

Users often struggle with the sheer volume and length of educational videos, leading to time-consuming searches and difficulties in aligning learning pace with individual comprehension levels. As Chen et al.~\cite{Chen2013} found, users may spend up to 80\% of their time on video platforms browsing and viewing only parts of content, indicating a significant inefficiency in locating relevant information. Moreover, the linear nature of long-form video can impose substantial cognitive load and challenge sustained attention, particularly when learners must sift through extensive footage to find specific segments of interest or to grasp core concepts distributed across the timeline~\cite{sweller1988cognitive,Risko2012Wandering}.

To address these challenges, various video summarization methods have been proposed to help users grasp video content quickly. Existing approaches typically focus on either visual or audio information to determine important parts of the video and generate summarized versions~\cite{truong2007video}. However, these methods are limited in their effectiveness for lecture videos, where crucial information is often distributed across both visual and audio channels, and may not always overlap.

For example, while an instructor verbally explains complex concepts, slides may display complementary diagrams or equations. Audio-only summarization loses these visual elements, while visual-only summarization misses the instructor's detailed explanations and nuances. Furthermore, many existing summarization systems present results in text format, breaking continuity with the original video and fragmenting the learning experience.

In response to these limitations, this chapter introduces FastPerson, a novel video summarization technique that comprehensively considers both visual and audio information. Unlike conventional methods, FastPerson creates a summary text that integrates information from both modalities before generating a summary video with narration. This approach ensures a more holistic representation of the video content, reducing the risk of learners missing critical information from either the visual or audio components.

The key contributions of FastPerson are threefold:

\begin{enumerate}
    \item A comprehensive video summarization method that generates summary videos reflecting important information from both visual and audio elements of lecture videos. Particularly, it establishes a new paradigm of ``video-to-video summarization," providing a learning experience beyond conventional text-based summarization.
    
    \item A user-centered learning experience design that enables seamless switching between original and summary videos. By preserving the original speaker's voice quality through voice cloning technology, it minimizes cognitive dissonance and allows learners to freely choose between concise and detailed information acquisition for each video chapter.
    
    \item Empirical evidence that the proposed method significantly reduces viewing time, with no statistically significant difference in quiz-measured comprehension from conventional playback.
\end{enumerate}

These contributions position FastPerson as a crucial component within the \textit{Understand} phase of our proposed AI-Guided Learning framework (see Chapter 1). While AIxSpeed (Chapter 3) focuses on optimizing the \textit{Consume} phase for audio content through adaptive playback, FastPerson extends this pursuit of efficiency to the complex, multimodal nature of video lectures. By enabling learners to efficiently grasp core concepts from visual and auditory information, FastPerson not only addresses the specific challenges of video-based learning but also prepares learners for the subsequent \textit{Imitate} phase, supported by Profy (Chapter 5), which facilitates skill acquisition through targeted feedback. Together, these systems aim to create a synergistic learning environment that enhances time efficiency and learning outcomes across the entire spectrum of multimedia engagement.

The remainder of this chapter is organized as follows: Section 4.2 provides an overview of related work in video summarization and learning. Section 4.3 details the design and implementation of the FastPerson system. Section 4.4 presents the evaluation methodology and experimental setup. Section 4.5 discusses the results of the user study and their implications. Finally, Section 4.6 concludes the chapter with a summary of findings and directions for future research.

By addressing the challenges of efficient video content comprehension, FastPerson complements the audio-focused approach of AIxSpeed presented in the previous chapter. Together, these systems contribute to a more comprehensive solution for time-efficient learning in multimedia environments, paving the way for the skill acquisition support system, Profy, which will be introduced in the subsequent chapter.

\begin{figure}[t]
    \centering
    \includegraphics[width=\textwidth]{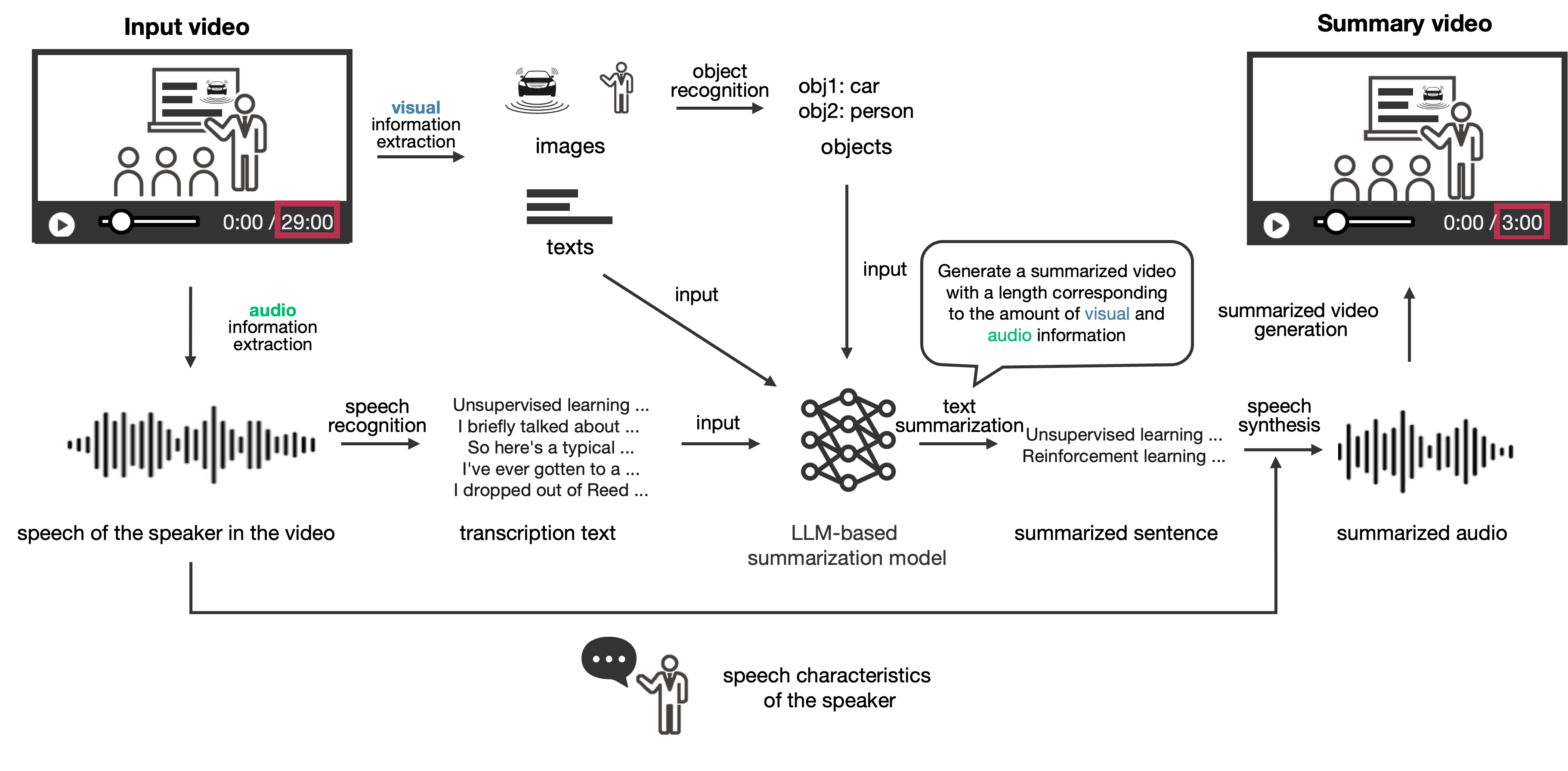}
    \caption[FastPerson video summarization method]{FastPerson: Video summarization method that generates a summary sentence using visual (objects and sentences on the frame) and audio information (speaker's voice) before synthesizing the sentence into speech using the speech characteristics of the original speaker.}
    \label{fig:fastperson/architecture}
\end{figure}

\section{System Design}
\label{sec:fastperson_system_design}

FastPerson can create a summarized video that considers both visual (text and objects on the screen) and audio information (speaker's voice). FastPerson automatically divides the original video into multiple coherent video segments, providing the user with the summarized video and original video for each segment.

The summary and original videos are aligned by combining the summary text with speech synthesized to resemble the original speaker and a corresponding clip from the original video, thereby supporting a seamless transition between the two versions. In addition, when generating the summary video, the summary ratio is adjusted by reflecting the amount of visual and audio information to avoid over- or under-information while preserving important content for the learner.

This method is expected to be applied to a wide range of lecture videos, and therefore, the use of LLM-based summarization is designed to eliminate the need for training with supervised data. We describe the summarization method by focusing on the FastPerson architecture, as shown in Figure~\ref{fig:fastperson/architecture}.

\subsection{Video Chapter Segmentation}
\label{subsec:fastperson_chapter_segmentation}

FastPerson first detects scene transitions and silence in a video and divides them into segments to summarize the video into coherent chapters and generate a video summary. For scene transition detection, we use a technique proposed by Nisreen et al.~\cite{Nisreen2012} to analyze the variation of the color distribution (histogram) between consecutive frames and determine the boundaries of the segments based on the magnitude of the variation.

According to Truong et al.~\cite{truong2007video}, distinct variations between consecutive frames indicate a change in scene, and histogram variations indicate changes in the mood or background of the video. Specifically, we calculate histograms in RGB color space with 64 bins and measure similarity between consecutive frames using Bhattacharyya distance. Changes in the Bhattacharyya distance exceeding a threshold of 0.6 between consecutive frames are detected as scene transitions. This threshold was empirically determined based on a pilot study involving diverse lecture videos, where it demonstrated a favorable balance between identifying meaningful segment breaks and avoiding over-segmentation due to minor visual shifts.

We use an algorithm that detects periods when the sound waveform has an amplitude below a certain threshold as silence because visual and audio breaks do not always coincide. This method is based on the work of Scheirer et al., which enables detecting periods of low amplitude or the absence of certain frequencies while identifying segments of silence or stillness~\cite{Scheirer1997}. Specifically, we calculate the RMS (Root Mean Square) of audio in 500ms windows and detect periods where the amplitude is below 5\% of the maximum amplitude for more than 1 second as silence periods. This 5\% threshold was selected based on preliminary experiments to effectively identify pauses in speech without misclassifying quiet speech segments as silence.

These visual scene transitions and periods of silence in audio are considered to separate and segment the video at the cut point. This functionality enables automatic chaptering (segmentation) without the need for the video creator to tag each contiguous segment.

\subsection{Visual and Audio Information Extraction}
\label{subsec:fastperson_information_extraction}

Our method converts visual and auditory information into textual information for generating a summary video that considers both. We use optical character recognition (OCR) and object detection techniques to obtain visual information.

OCR is used for digitizing imaged document information, and it can detect characters in images and videos and convert them into documents~\cite{Smith07}. This technology can effectively capture document information in video images, such as the content of slides or handwritten notes on a whiteboard in a lecture video. This information can then be used to reflect the topic and content of the documents in the video in the summary.

Our system uses the Tesseract OCR engine~\cite{Smith07} for OCR. Tesseract 4.0 employs LSTM-based neural networks and supports over 100 languages. As preprocessing, we apply grayscale conversion, noise removal (Gaussian blur), and binarization (Otsu's method) to improve OCR accuracy. These preprocessing steps are standard techniques known to enhance Tesseract's performance, particularly for text in video frames which can vary in quality.

Object detection is a technique for detecting and identifying specific objects or entities in each frame of a video~\cite{Cortes15}. The objects in the video are often closely related to the story focus or theme of the segment. If a car is shown on a slide in a scene from an online lecture on machine learning, the speaker is expected to talk about automated driving or image recognition, and the object ``car" is considered a key point in that video segment.

Therefore, key scenes and important entities in the video can be identified using object detection, and effective summaries can be generated based on this information. Our system uses the faster R-CNN~\cite{Cortes15} with ResNet-50-FPN as the object model backbone. This model is pre-trained on the MS COCO dataset and can detect 80 classes of common objects. The detection threshold is set to 0.7 to extract only high-confidence objects. This value was chosen after empirical testing to balance precision and recall for object detection in lecture video contexts.

The accurate extraction of speech information is necessary for academic content in videos, especially lectures and seminars. Audio information is extracted from a video as follows:

\begin{enumerate}[(1)]
    \item Audio stream is extracted from the video file.
    
    \item This audio stream is fed as an input to a speech recognition model to obtain a transcription of the document.
\end{enumerate}

The speech signal needs to be converted into document data with high accuracy because the accuracy at this stage can affect the quality of the summary. Recent advances in speech recognition have been attributed to models based on RNN~\cite{Graves2013, hannun2014} and Transformer~\cite{Vaswani2017, wang2020}.

In our method, we employ Whisper~\cite{radford2023} as the speech recognition model. Chosen for its state-of-the-art accuracy across diverse audio conditions, a result of its large-scale training (approx. 680,000 hours of multilingual speech), Whisper's deep learning architecture enables robust transcription essential for downstream summarization. Specifically, we use the Whisper Large model (1550M parameters) with a beam search width of 5 and a temperature parameter of 0 for inference. These parameters were chosen based on the recommendations in the original Whisper paper for achieving high accuracy in transcription tasks.

Despite the high performance of this method, its recognition accuracy is not perfect. Therefore, we use OCR to recognize text in the video, complementing technical terms and proper nouns missed by speech recognition.

\subsection{Summary Audio Generation}
\label{subsec:fastperson_summary_generation}

Video summarization uses data labeled with important points in a video, along with supervised learning methods to detect important locations in the video. However, our target educational videos cover a wide variety of domains, and it is difficult to use training data to extract such important points.

Recently, a number of LLMs have been proposed in the natural language processing domain based on Transformer architectures, including BERT~\cite{devlin-etal-2019-bert}, T5~\cite{raffel2020}, and GPT3~\cite{radford2018, radford2019}. These models are trained on documents and dialogues from different domains and have a wide range of domain knowledge, hence enabling them to perform various natural language processing, including summary generation~\cite{basyal2023}.

Our system can summarize lecture videos without the need for teacher data using this LLM. Usually, when a summary is generated from speech, a transcript of the speech is used as input to generate the summary. However, this method generates a summary based on both the transcript of the speech and visual metadata obtained through OCR and object recognition, thereby reflecting the visual information of the video in the summary.

The following instructions are provided to an LLM~\cite{ray2023} to generate a document summary for each video segment using both visual and audio information. In this study, we use GPT-3.5-turbo with a temperature parameter of 0.7 and top-p of 0.9 for inference.

\hspace{3mm}
\begin{itembox}[l]{Instructions}
    Using the provided transcription of spoken content, OCR-derived textual data, and object detection information, synthesize a comprehensive summary. This summary should highlight the key themes and actions depicted in the video. Focus on distilling the essence of the video by combining insights from the transcription (which captures the spoken words), the OCR data (which provides text found within the video), and object detection (which identifies significant objects and actions). 
\end{itembox}
\hspace{3mm}

Along with these instructions, the transcribed text of each video segment, OCR, and resulting visual information from object recognition are entered into the LLM as documents. This method enables creating a summary that emphasizes nuances not included in the speech content, speaker's movements, visual objects, and documents within slides and visual presentations.

\subsection{Summary Length Calculation}
\label{subsec:fastperson_length_calculation}

The length of the document summary must appropriately reflect the amount of information in the original video when summarizing a video. In other words, if the video is long or rich in content, it is necessary to provide sufficient information in the summary, and if the original video is short, the summary should be shortened by the same amount to eliminate redundancy.

Therefore, we adjust the number of characters in the output summary based on the length of the transcribed document. In addition, the number of output characters after summarization is adjusted based on the amount of visual information because our method generates video summaries based on visual information as well as audio. For example, if the amount of text on a slide is large even if the speaker's speech is short, it must be summarized to accurately reflect the information content of the slide.

From this point of view, the proposed system calculates the number of output characters \(N \) of the summary using Equation~\ref{eq:summary_length}:

\begin{equation}
    N = \max(N_{\min}, w_s \cdot L_t + w_v \cdot (L_o + L_c))
    \label{eq:summary_length}
\end{equation}

where
\begin{itemize}
    \item \( N \) represents the final output word count of the summary.
    \item \( N_{\min} \) is the minimum summary length (e.g., 50 words).
    \item \( L_t \) denotes the length of the transcribed text.
    \item \( L_o \) represents the number of objects identified on the screen.
    \item \( L_c \) represents the word count of OCR-processed visual elements.
    \item \( w_s \) and \( w_v \) represent the weight coefficients to adjust the importance between audio and visual information, respectively.
\end{itemize}

The weight coefficients \( w_s \) and \( w_v \) are selected based on experimental data and user preferences. These coefficients can be varied to adjust the emphasis on visual or auditory information. Although the balance can be altered according to user interest and comprehension levels to provide personalized summary videos, in the present study, fixed values of 0.3 and 2.5 are set for \( w_s \) and \( w_v \), respectively. We set \(N_{\min}=50\) words to ensure summaries are not overly terse.

These values were determined through preliminary experiments, confirming that they generate appropriate summary lengths for both lecture videos rich in visual information (slide-centered) and those centered on audio information (speaker-centered).

\subsection{Summary Video Generation}
\label{subsec:fastperson_video_generation}

A conventional audio summarization method used for audiobooks evaluates the importance of each audio transcription on a sentence-by-sentence basis and directly extracts and combines important audio segments. However, this method poses a challenge to the continuity of audio and video when merging the extracted segments. This method cannot produce a video summary that reflects the visual representation of the video.

Therefore, this method uses a speech synthesis~\cite{Shen2018, Vandenoord16} method, which uses a document summary with visual and audio information as the input and speech synthesis technology to generate its reading voice. After the synthesized speech is generated, the optimal video segment is selected from the original video based on the length of the speech.

The video segment is cut from the beginning, middle, or end of the original video, depending on the length of the summarized video. Based on the results of a simple user experiment, the video segment in the middle of the original video is used as the default joining method. A continuous and concise video summary is achieved by combining this selected video segment with the synthesized audio.

This method avoids unnatural connections caused by combining different parts of the video. In addition, users are expected to frequently switch between the original and synthesized videos when using this method, and therefore, it is necessary to provide a seamless experience.

Thus, we adopt a method that minimizes the change in the audio between the original and summarized videos using the speaker adaptation technique when synthesizing the audio~\cite{Arik2017}. The features of the speaker's voice in the original video are extracted, and a synthesized voice is generated that reflects the speaker's voice quality. This method reduces the mismatch of voice features when switching between the original and synthesized videos, thereby enabling seamless switching.

Specifically, this research uses the Coqui TTS (Text-to-Speech) system and adopts the VITS (Variational Inference with adversarial learning for end-to-end Text-to-Speech) model. For speaker adaptation, we extract clean speech samples of at least 30 seconds from the original video and learn speaker embeddings. This duration was found to be sufficient for VITS to capture the primary characteristics of the speaker's voice in our preliminary tests. Using these embeddings, we generate synthesized speech close to the original speaker's voice quality from the summary text.

\begin{figure*}[t]
    \centering
    \includegraphics[width=\linewidth]{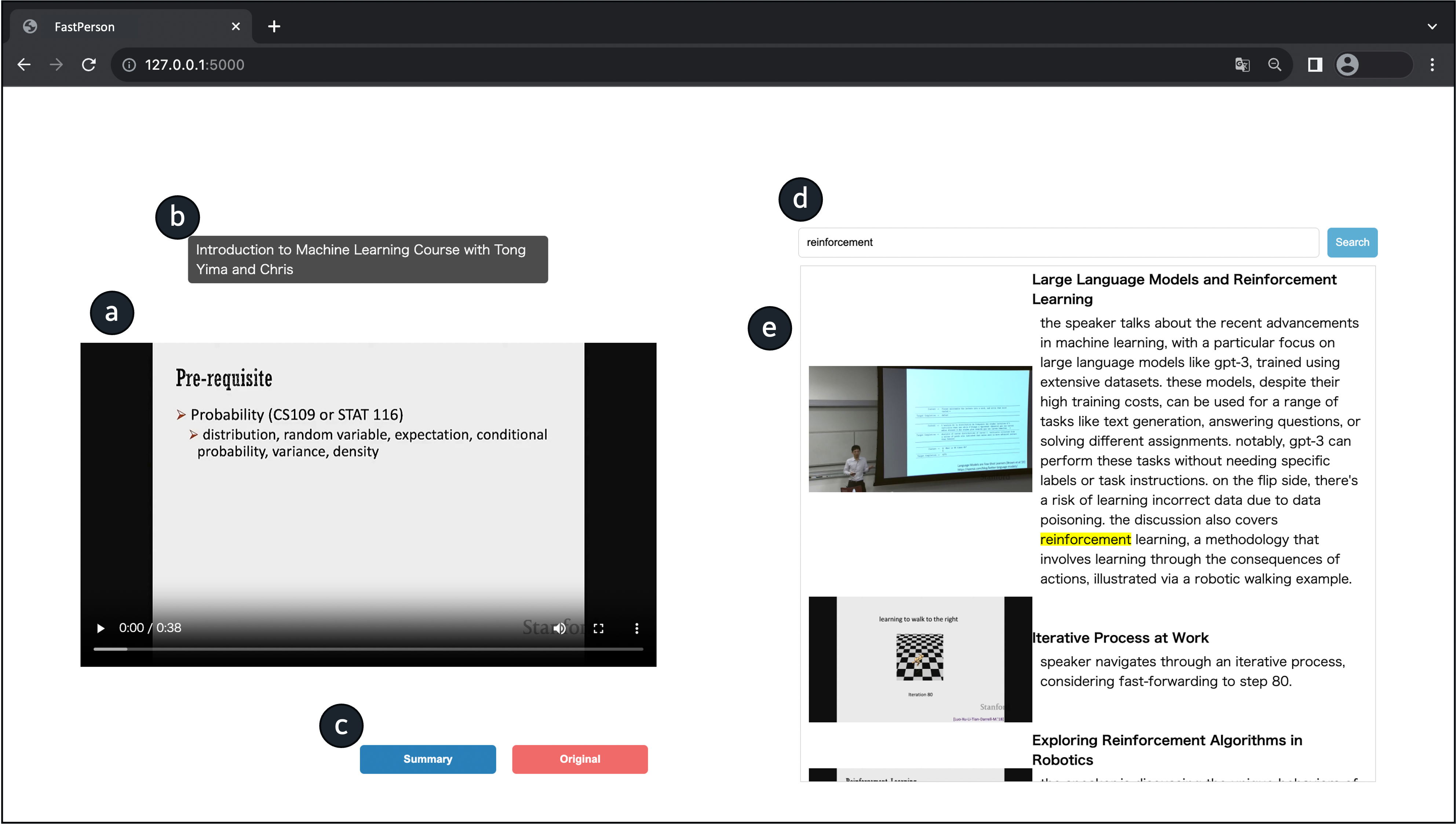}
    \caption[Overview of the FastPerson user interface]{Overview of the FastPerson user interface: A Video Window showcases the current video playback and offers user-controlled speed adjustments B Video Title provides segment-specific titles C Summary and Original buttons allow seamless toggling between video versions and facilitate segment switches D Search Container offers a text-based search mechanism E Segments Container displays the thumbnails of video segments, ensuring quick navigation.}
    \label{fig:fastperson/interface}
\end{figure*}

\section{Application and Interaction}
\label{sec:fastperson_application}

\subsection{Application Design}
\label{subsec:fastperson_application_design}

FastPerson enables users to switch between the original longer video and summarized video for each segment within each video, depending on their level of interest and understanding. In addition, a static component placed on the side of the video displays a thumbnail image and summary text for each video segment, enabling the user to quickly access the corresponding video chapter.

The entire user interface is shown in Figure~\ref{fig:fastperson/interface}; \textcircled{\scriptsize a} \textbf{Video Window} allows users to view videos in the window in either a summarized or original format. The playback speed can be adjusted based on preference, enabling flexible viewing of content.

\textcircled{\scriptsize b} \textbf{Video Title} is placed at the top of the video window and displays the title of the segment that is currently playing. \textcircled{\scriptsize c} \textbf{Summary and Original buttons} allow the user to quickly switch between the summary and original full version of the video. In this case, making the voice quality of the summary video similar to that of the original video allows the user to perceive continuity between the two versions.

Next to the video is a window displaying a pair of text summaries and thumbnail images representing the chapters of the video. These windows are effective in helping users quickly access the segment of interest in the video~\cite{chi2012, pavel2014}.

\textcircled{\scriptsize d} \textbf{Search Window} provides a document entry box and a search button to search for specific keywords in a video segment. This feature allows the user to see where their topics of interest are discussed in the video.

\textcircled{\scriptsize e} \textbf{Segment Window} displays thumbnails, titles, and summary text for each video segment. Users can quickly access the appropriate video segment by clicking on the appropriate thumbnail.

The application is built using Flask, a lightweight Python web framework. The FastPerson backend manages the files for each video segment and delivers them efficiently. The backend includes functions to load the associated thumbnail and text files based on the name of a specific segment in the video, enabling a quick delivery of the video segment requested by the user.

On the front end, the user interface is built using HTML, CSS, and JavaScript. The interface includes a video player, video title display, a button to toggle between the summarized and original videos, a search function, and thumbnails and summaries for each segment. JavaScript is used to control video playback, implement the search function, and respond to user interactions.

\begin{table*}[t]
    \centering
    \caption{Details about the videos used in the evaluation experiment.}
    \begin{tabularx}{\textwidth}{@{}l *{2}{>{\raggedright\arraybackslash}X}@{}}
    \toprule
    & \textbf{Video 1} & \textbf{Video 2} \\
    \midrule
    Title & Origin of Life & Market Revolution \\
    Thumbnail & \raisebox{-0.5\totalheight}{\includegraphics[width=5cm, height=2.5cm, keepaspectratio]{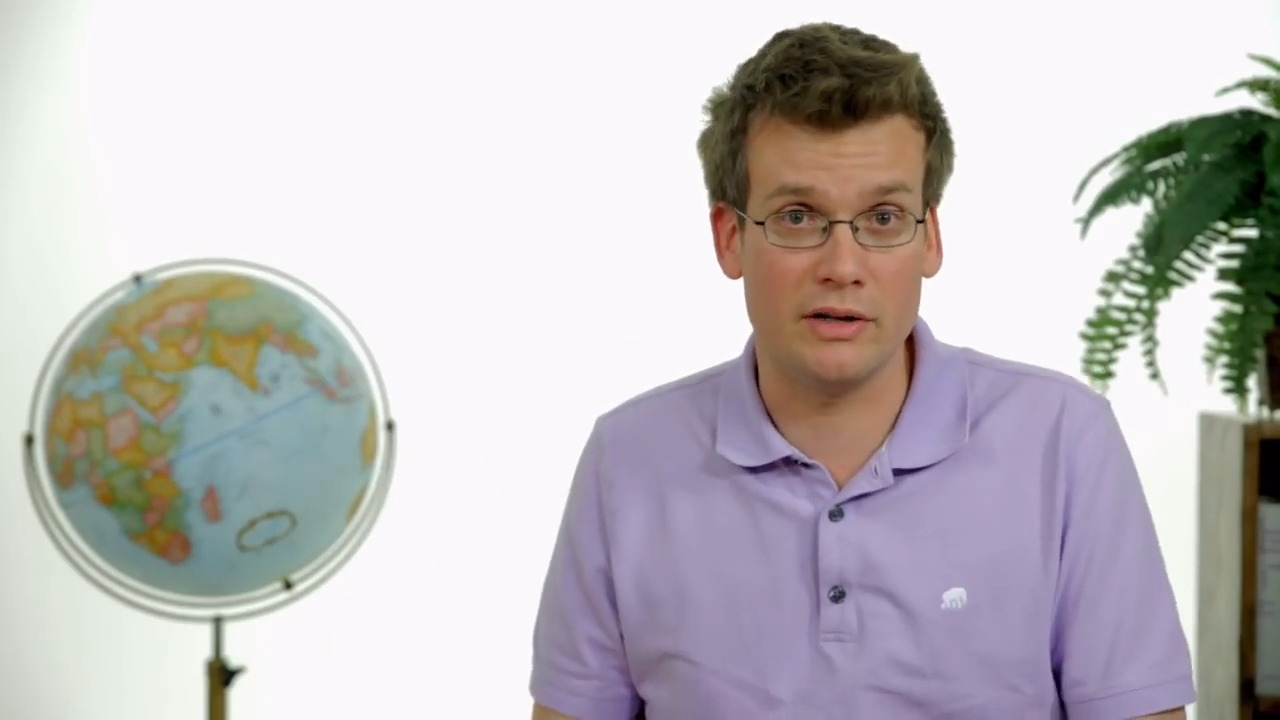}} & \raisebox{-0.5\height}{\includegraphics[width=5cm, height=2.5cm, keepaspectratio]{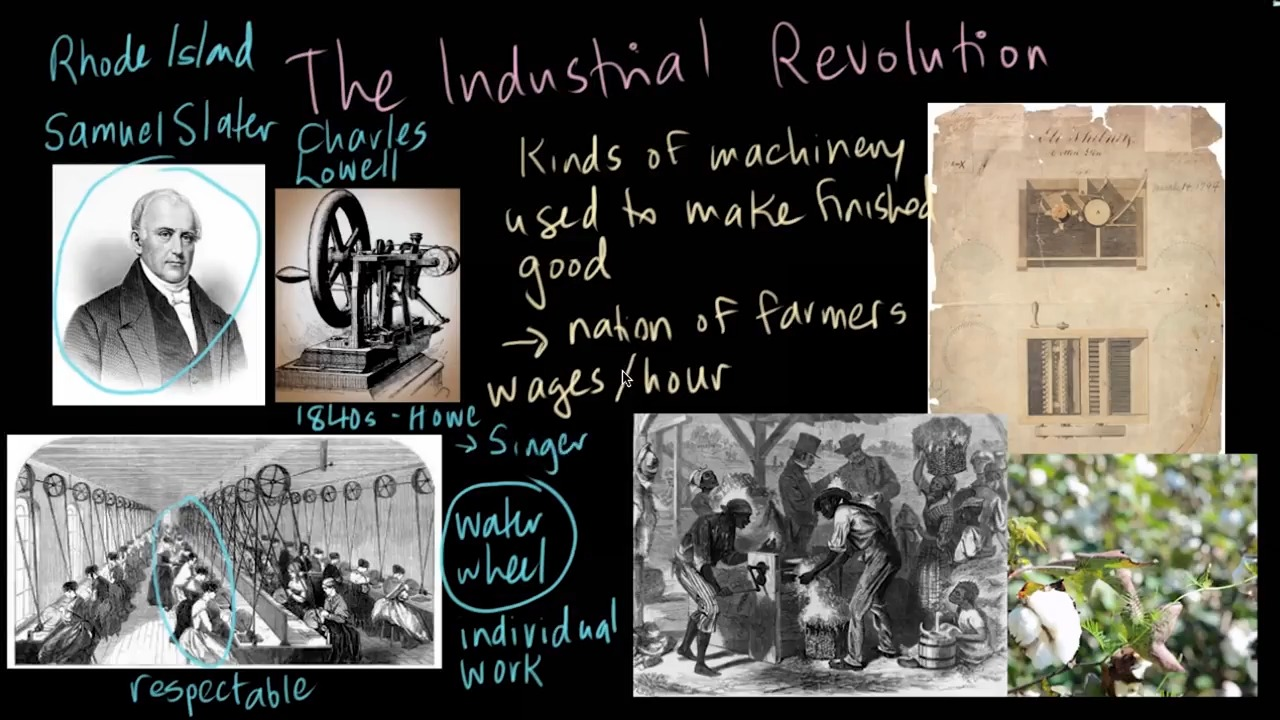}} \\
    Length & 12:58 & 21:55 \\
    Question 1 & What molecule, also called the biological blueprint for life, do all living organisms possess? & The activity shown in the image contributed most directly to which of the following? \\
    Question 2 & All living things on Earth share important processes. What are those four processes? & Which of the following developments most directly relates to the overall trend from 1800 to 1840 depicted on the graph? \\
    Question 3 & In 1828, a German chemist, Friedrich Wohler, proved that organic life is composed of the same components as inorganic matter. How did he do this? & According to the passage, which of the following best explains the most important effect that technological developments had on the American society? \\
    Question 4 & What ingredient(s) are considered necessary for the creation of life? & Which of the following had a significant long-term result of the major pattern depicted on the map? \\
    Question 5 & What quality of life allows for its slow diversification? & None \\
    \bottomrule
    \end{tabularx}
    \label{tab:fastperson/video_info}
\end{table*}
    
\subsection{User Interaction}
\label{subsec:fastperson_user_interaction}

FastPerson provides a personalized learning experience by allowing users to adjust the amount of time they spend watching a video and where they spend time. For example, users can quickly grasp the gist of an entire video by watching only the summary video. Alternatively, users can view the summary video for the parts they understand and the original detailed video for the parts they do not.

The design of this learning experience is inspired by the flexibility of the information acquisition experience when reading a book. An example of how the system can be used to view a video is presented below.

\begin{enumerate}
    \item \textbf{Skimming through the entire summary}: By clicking the \textbf{Summary button} (C) and playing the video, users can watch the summarized version of the video. This is similar to the action of quickly skimming through the entire text content.
    
    \item \textbf{Skimming over parts that are interesting}: Using the \textbf{Segments Container} (E), users can obtain a brief overview of each segment based on the associated thumbnail and text. Clicking on any segment will play its respective video after clicking the \textbf{Summary button} (C), enabling users to sample content before deciding to watch the detailed video.
    
    \item \textbf{Watching only the parts of interest repeatedly}: Users can replay any segment multiple times for a thorough understanding, which mirrors the experience of rereading one's favorite section of a book.
    
    \item \textbf{Searching for specific content}: The \textbf{Search Container} (D) allows users to search for specific terms or keywords across segments, providing direct access to desired content, which is similar to using an index in a book.
    
    \item \textbf{Switching between the summary and the detailed view}: At any given point, users can toggle between the summarized and original videos using the \textbf{Summary and Original buttons} (C), catering to their comprehension and interest levels.
\end{enumerate}

\section{Evaluation}
\label{sec:fastperson_evaluation}

We conducted a series of quantitative and qualitative evaluation experiments to assess the learning effectiveness and user experience of the FastPerson system. The specific objectives were as follows:

\begin{itemize}
    \item To compare the learning effects when using the FastPerson system versus a standard video playback player.
    
    \item To evaluate the effectiveness of summarization using the FastPerson system and its feature of switching between the summarized and original videos in aiding understanding.
    
    \item To assess the usability and user satisfaction of the FastPerson interface.
\end{itemize}

\begin{figure*}[t]
    \centering
    \includegraphics[width=\linewidth]{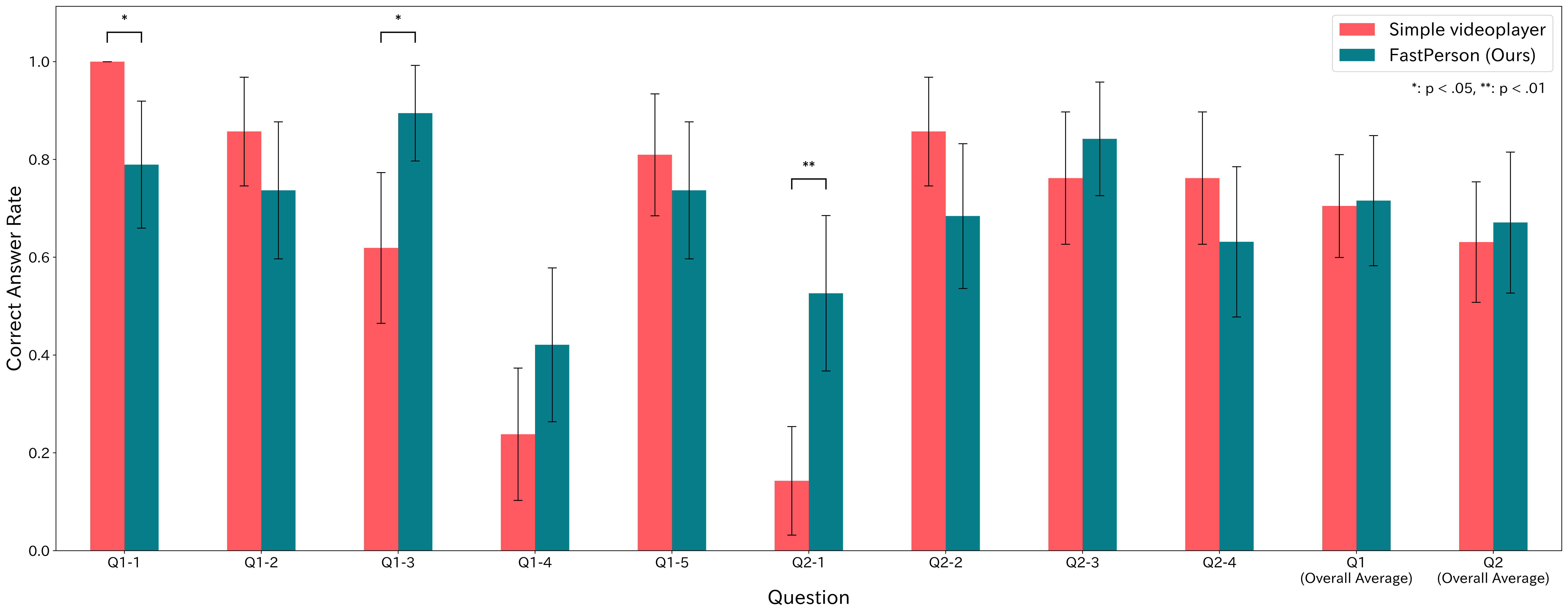}
    \caption[Correct answer rates for each video question]{Correct answer rates for each question of the video.}
    \label{fig:fastperson/exp1_1}
\end{figure*}

\subsection{Participants}
\label{subsec:fastperson_participants}

Participants from diverse backgrounds were recruited through the crowdsourcing platform Prolific. A total of 40 people participated in the study and were randomly assigned to an intervention group that used FastPerson (19 participants) or a control group that used a generic video player (21 participants).

The participants were selected based on their fluency in English and completion of at least secondary education, because the videos were in English and the participants needed to be able to understand the content of the videos. The participants' demographics (gender, age, educational background, etc.) were collected as additional information to assess the effect of the experiment.

The participants comprised 23 males and 17 females, with an average age of 26.05 years ($\sigma=7.05$). Among the participants, 35, 35, 15, and 15\% had a high school diploma or GED, university bachelor's degree (BA/BSc/other), technical school or community college, and graduate degree (MA/MSc/MPhil/etc.), respectively.

A post-experiment questionnaire measured the participants' prior knowledge of the videos they watched on a 5-point scale (1 being ``no prior knowledge" and 5 being ``extensive knowledge"), and the results showed that, for the first video, the intervention and control groups had means of 2.63 ($\sigma=1.16$) and 2.10 ($\sigma=0.94$), respectively.

For the second video, the intervention and control groups had means of 1.32 ($\sigma=0.58$) and 1.38 ($\sigma=0.59$), respectively. A t-test between the two groups revealed t-statistic values of 1.607 with a p-value of 0.116 and -0.351 with a p-value of 0.727 for the first and second videos, with no statistically significant difference between the two groups at the 0.05 level of significance.

Thus, there was no clear difference in prior knowledge between the two groups. Further, the participants were motivated to achieve high scores because it was clearly stated that those with the highest scores would receive a bonus reward. All participants understood and agreed that the process was voluntary and that they had the right to withdraw at any time.

\subsection{Procedure}
\label{subsec:fastperson_procedure}

In this experiment, participants gathered through crowdsourcing were randomly assigned to an intervention group using FastPerson and a control group that used a general video player. Two videos were selected from Khan Academy, an online learning service, for viewing, and the accompanying questions were used to measure the learning effect of the videos.

To prevent the participants' understanding of the videos from changing drastically based on their prerequisite knowledge, we selected lecture videos with a level of difficulty that could be understood from the video alone. The selected videos were from the fields of biology and history, as shown in Table~\ref{tab:fastperson/video_info}. The videos were accompanied by questions used to assess the participants' understanding of their content.

Several questions for the first video could not be answered correctly without looking at materials besides the video, and therefore, five questions excluding these ones were used to check the participants' understanding of the video. For the second video, the four questions included in the video were used to assess comprehension.

The first video is a lecture video that focuses on the speaker and visual elements, while the second focuses on the board. The participants viewed these videos under their assigned condition and subsequently answered the accompanying questions.

The intervention group also completed an additional questionnaire about FastPerson. The experiment was conducted through Prolific, with a median working time of 46 min and 26 s, and each participant was paid £7.

\begin{figure}[t]
    \centering
    \includegraphics[width=\linewidth]{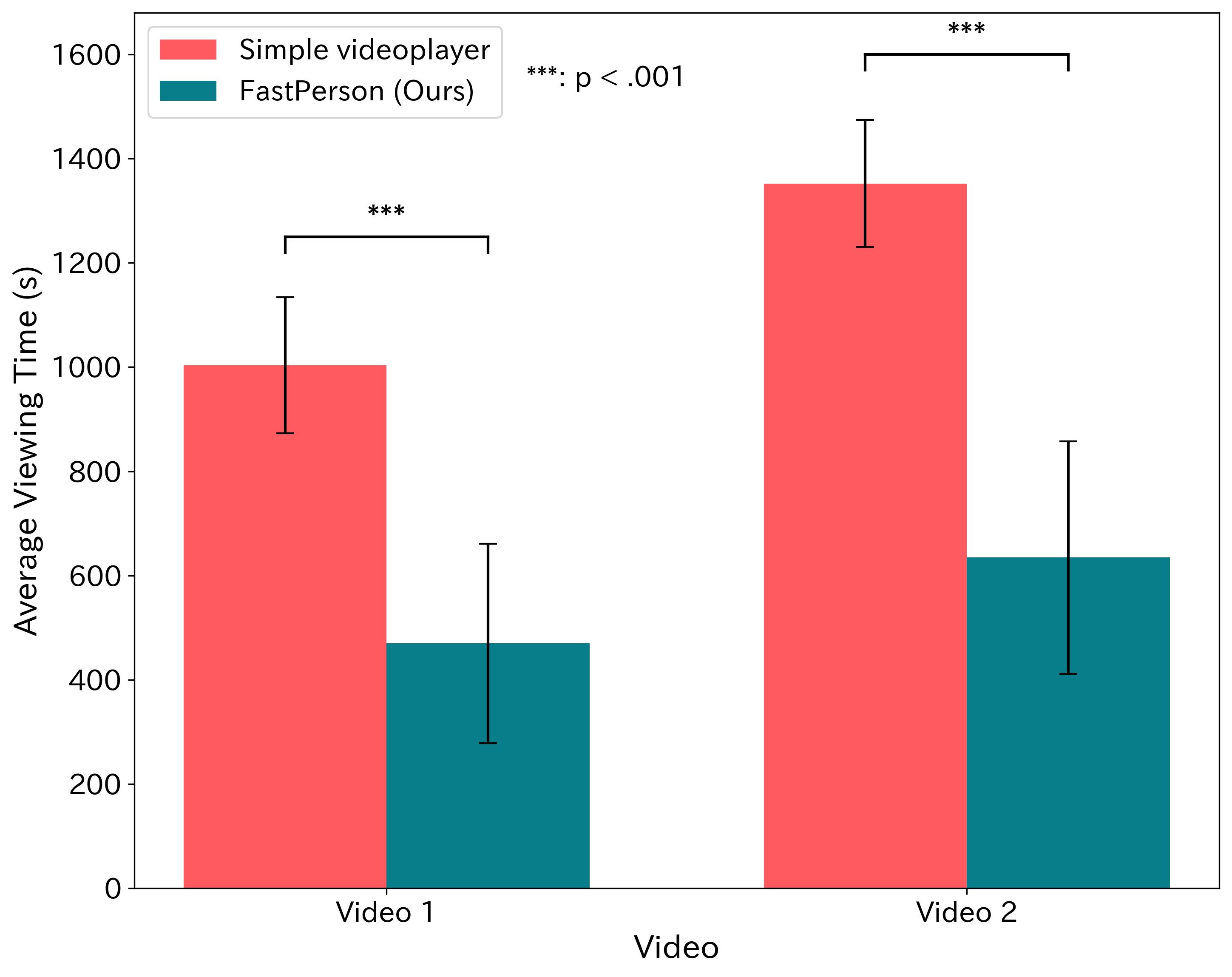}
    \caption[Average video viewing time by viewing method]{Average video viewing time by viewing method.}
    \label{fig:fastperson/exp1_2}
\end{figure}

\subsection{Evaluation of Learning Effectiveness}
\label{subsec:fastperson_learning_effectiveness}

We analyzed the percentage of correct answers to questions about the videos and the time taken to view the videos after watching the same two videos for the control and the intervention groups to evaluate the learning effectiveness. The right side of Figure~\ref{fig:fastperson/exp1_1} compares the average correct response rates for questions related to Video 1 (``Q1") and Video 2 (``Q2").

For Video 1, the correct answer rate for the control group (red bar) was 0.70 ($\sigma=$ 0.33) and that of intervention group (blue bar) was 0.71 ($\sigma=$ 0.42) while those for Video 2 are 0.63 ($\sigma=$ 0.39) and 0.67 ($\sigma=$ 0.46), respectively. Therefore, there is no clear difference in the percentage of correct answers between both groups.

In fact, a t-test between the two groups yields t-statistic values of 0.170 with a p-value of 0.864 and 0.528 with a p-value of 0.598 for the first and second videos, respectively, neither of which is statistically significant when the significance level is set at 0.05.

Further, the percentage of correct answers for each question (left side of Figure~\ref{fig:fastperson/exp1_1}) shows that the score for viewing on a general video player was higher for question 1 of video 1 (Q1-1), while the score for viewing on FastPerson was higher for question 3 of video 1 (Q1-3) and question 1 of video 2 (Q2-1).

Q1-1 is ``What molecule, also called the biological blueprint for life, is found in all living organisms?" and the correct answer is ``DNA," which is picked up as a keyword in the FastPerson summary, indicating that it is a keyword that ``all living organisms possess." However, the percentage of correct answers is likely to have decreased because of a lack of explanation of the term ``biological blueprint for life."

Meanwhile, although Q1-3 and Q2-1 are adequately covered in both the simple video playback and FastPerson, the topics of the other incorrect choices in the question are relatively fewer in FastPerson. Therefore, users who watched FastPerson are not expected to select these incorrect choices, resulting in a higher percentage of correct answers.

The time taken to watch the videos is summarized in Figure~\ref{fig:fastperson/exp1_2}, where the average time taken by users to watch Videos 1 and 2 was 1003 ($\sigma=312$) and 1352 ($\sigma=270$) s, respectively, using the regular video player. FastPerson, which provides summary videos, took users an average of 469 ($\sigma=424$) and 634 ($\sigma=496$) s to watch Videos 1 and 2, respectively.

Using FastPerson reduced the viewing times for Videos 1 and 2 by 53.24 and 53.11\%, respectively. A t-test between the two groups showed that the t-statistic values for the first and second videos were 4.409, with a p-value of $1.063\cdot10^{-3}$, and 5.281, with a p-value of $9.955\cdot10^{-6}$, respectively. Both differences were statistically significant at $p<0.01$.

Therefore, FastPerson reduced viewing time without a statistically significant difference in quiz-measured comprehension compared with the original-video condition. This 53\% time reduction represents a significant improvement in time efficiency, which has important implications especially for time-constrained learners.

\subsection{Evaluation of User Experience}
\label{subsec:fastperson_user_experience}

We conducted a questionnaire survey on the proposed method after 19 participants in the FastPerson intervention group finished learning from the videos to evaluate the usability, learning effectiveness, and user experience of the system.

The survey included the following aspects:

\begin{enumerate}
    \item Interface Usability
    \begin{itemize}
        \item[Q1:] Was the system interface intuitive to use? (1: Very Difficult - 5: Very Intuitive)
        \item[Q2:] Was it easy to find the required information? (1: Very Difficult - 5: Very Easy)
        \item[Q3:] How was your experience of using the Summary and Original buttons? (Open-ended Response)
        \item[Q4:] What improvements would you suggest for the user interface design? (Open-ended Response)
    \end{itemize}
    \item Learning Experience Satisfaction
    \begin{itemize}
        \item[Q5:] Was learning with our system enjoyable? (1: Not Enjoyable At All - 5: Extremely Enjoyable)
        \item[Q6:] Compared with regular video players, how satisfied are you with learning using our system? (1: Not Satisfied At All - 5: Extremely Satisfied)
        \item[Q7:] How would you rate your learning experience with our system compared to that with other methods? (Open-ended Response)
    \end{itemize}
    \item Usefulness of Summary and Original Videos
    \begin{itemize}
        \item[Q8:] Was the summarized video helpful in understanding the information? (1: Not Helpful At All - 5: Extremely Helpful)
        \item[Q9:] Was the feature to switch between the Original and Summary videos useful? (1: Not Useful At All - 5: Extremely Useful)
        \item[Q10:] How did you perceive the quality of the summarized videos? (Open-ended Response)
    \end{itemize}
    \item Evaluation of Specific Features
    \begin{itemize}
        \item[Q11:] Was it convenient to access specific video segments using thumbnails and summary texts? (1: Not Convenient At All - 5: Extremely Convenient)
        \item[Q12:] Was it easy to find information in the videos using the search window? (1: Very Difficult - 5: Very Easy)
    \end{itemize}
\end{enumerate}

\begin{figure*}[t]
    \centering
    \includegraphics[width=\linewidth]{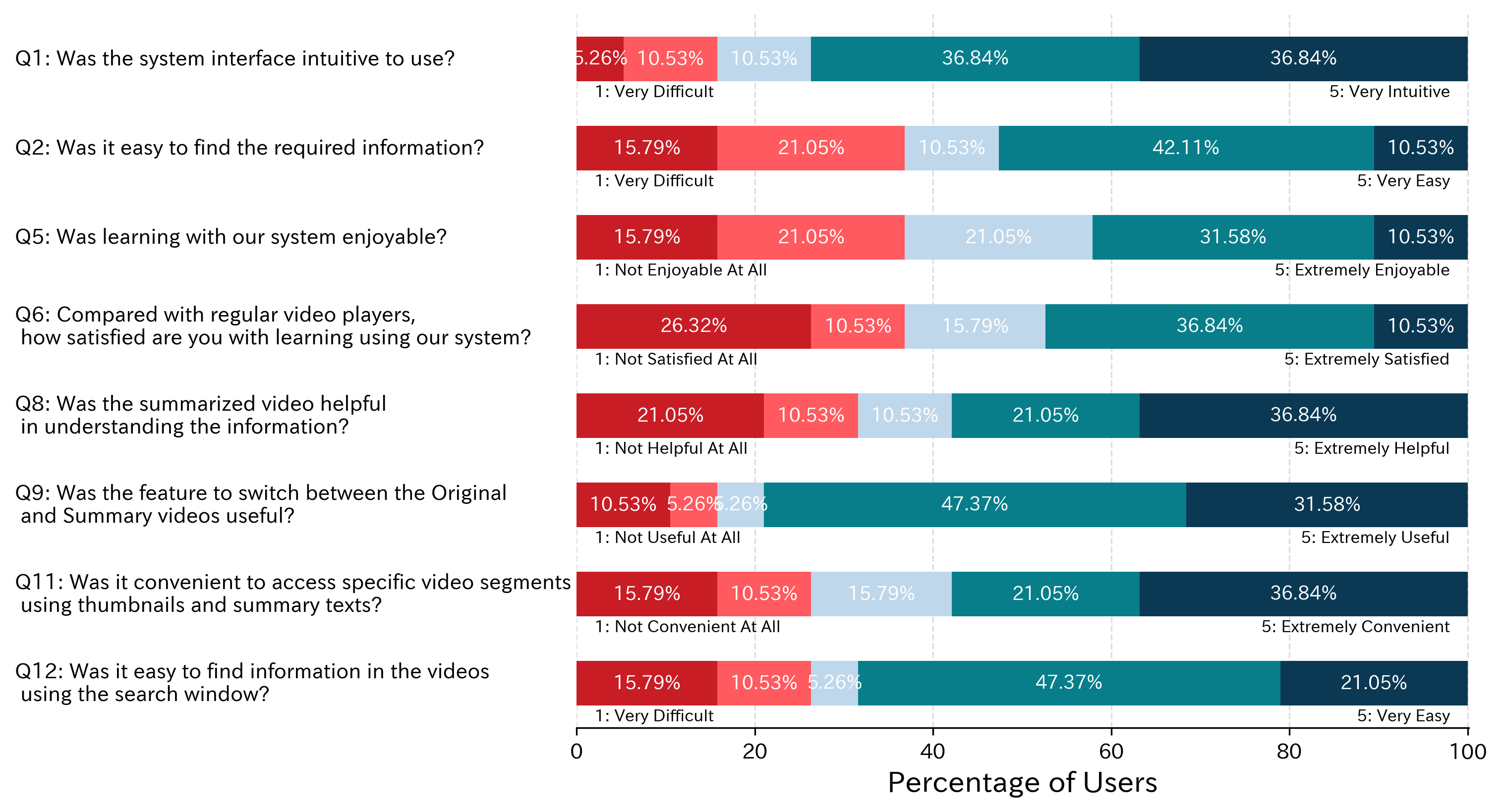}
    \caption[Survey results on user experience]{Survey results on user experience.}
    \label{fig:fastperson/exp2_1}
\end{figure*}

The questions included both numerical and open-ended questions, and the results for the numerical questions are shown in Figure~\ref{fig:fastperson/exp2_1}. For each question, the percentage of users who responded to each of the five levels is presented as a bar graph.

\subsubsection{Evaluation of Interface Usability}
\label{subsubsec:fastperson_interface_usability}

The average rating was 3.89 ($\sigma=1.20$) for the question about the system's ease of use, especially intuitiveness (Q1). This indicates that, on average, the users found the interface intuitive.

For the question about the ease of finding information in the system (Q2), the average rating was 3.11 ($\sigma=1.33$). This rating reflects that users found it moderately easy to find the required information. However, the relatively high standard deviation highlights variation in the user experience.

The open-ended questions provided additional insights. For example, in terms of the usability of the Summary and Original buttons, many participants stated that they were ``very easy to use and understand" and appreciated the usefulness and convenience of switching between the different video modes (Q3). However, one user commented, ``The interface is all over the place, and you have to spend a lot of time to understand it."

This suggests that there is room for improvement in the placement of buttons and components. Suggestions for interface design improvements received a wide range of responses (Q4). Although some participants felt that the humanization of the summary descriptions and voice reading needed further improvements, others called for improvements to the overall visual aspect of the site.

\subsubsection{Evaluation of Learning Experience Satisfaction}
\label{subsubsec:fastperson_learning_satisfaction}

A quantitative measure of the enjoyment of learning with FastPerson was obtained with a mean rating of 3.00 (Q5). Although this rating indicates a moderate level of enjoyment among users, the standard deviation of 1.29 suggests that there is a wide range of variation in the participants' experiences.

Next, we measured satisfaction with learning using FastPerson compared with a regular video player and found a mean rating of 2.95 ($\sigma=1.43$) (Q6). This could be attributed to some users preferring or being more familiar with traditional video players.

In the qualitative feedback received through the open-ended question Q7, users who rated their experience positively indicated that the features of FastPerson that allowed them to quickly obtain the information they needed and the interactive nature of the learning experience contributed to their enjoyment and satisfaction.

\subsubsection{Evaluation of the Usefulness of Summary Videos}
\label{subsubsec:fastperson_summary_usefulness}

The mean rating for the usefulness of the summary videos in understanding the information was 3.42 ($\sigma=1.61$) (Q8). This rating indicates a moderately positive acceptance of the effectiveness of the summary videos in conveying information. Similarly, the average rating for the ability to switch between summary and original video formats was 3.84 ($\sigma=1.26$) (Q9).

Users positively rated the ability to switch between in-depth and summary video formats. In particular, 78\% of FastPerson users rated the chapter-switching function as ``useful" or ``very useful." This reports perceived usefulness; the study did not separately test the feature's causal effect on learning outcomes.

In the qualitative feedback (Q10), many users praised the concept of the summary videos, noting that they could help them quickly understand key concepts, especially when reviewing a topic or when time is limited. However, they raised concerns about the quality of the content in the summaries.

The main issues were related to the depth of content and the narration of the summary video. The summaries were found to be useful; however, they sometimes lacked sufficient detail to fully understand complex topics. In addition, several users pointed out that the pronunciation of the words in the summary video was different, confirming that the quality of the summary video could be improved by using a more powerful speech synthesizer.

\subsubsection{Evaluation of Specific Features}
\label{subsubsec:fastperson_specific_features}

The ease of accessing video segments using thumbnails and summary text received an average rating of 3.53 ($\sigma=$1.50) (Q11). Similarly, the ease of retrieving information in the video using the search window received an average rating of 3.47 ($\sigma=$1.39) (Q12).

These features are not unique to FastPerson; however, they are compatible with the ability to switch between the summary and original videos. The majority of users rated both the thumbnail and summary text as well as the ability to switch between the original and summarized videos (average rating 3.84) as useful, which indicates that presenting the user with both the summary video and summary text, rather than just the summary text of the video, improves usability and convenience.

\section{Discussion}
\label{sec:fastperson_discussion}

\subsection{Advantages of FastPerson in Educational Content Summarization}
\label{subsec:fastperson_advantages}

FastPerson reduced video viewing time by 53\%, with no statistically significant difference in quiz-measured comprehension from the original-video condition. For some questions related to important parts of the video, FastPerson produced a higher percentage of correct answers than the original full-length video; for other questions, the original-video condition performed better.

The ability to switch between summary and original video formats was particularly appreciated by users, offering flexibility in learning and content consumption that is not available in traditional methods. This feature allows learners to quickly grasp essential concepts through summaries and delve into detailed explanations as needed, thereby catering to diverse learning preferences and needs.

In addition, FastPerson's use of visual elements synthesized with summarized audio enhances the educational value of the content, making complex information more accessible and engaging, especially for visual learners. The preservation of the original speaker's voice quality through voice cloning technology is also an important advantage, minimizing cognitive load and maintaining continuity in the learning experience.

\subsection{Identified Limitations and Areas for Improvement}
\label{subsec:fastperson_limitations}

FastPerson has some limitations. For some questions, the percentage of correct answers was higher when the original video was viewed, suggesting that information may be missed in the summarized video. We believe that the system can be improved by encouraging users to view the original video when information is missing.

Further, users have also indicated that transitions between video segments are not seamless and that the sound quality needs improvement. We anticipate that these issues could be addressed by improving the implementation and replacing it with a higher quality speech synthesis method. In particular, adopting the latest neural voice cloning technology could potentially generate more natural synthesized speech.

Other users also pointed out the need for a more intuitive user interface because they found it difficult to understand the functionality of the various features. In addition, we observed that some users prefer the original simple video playback player, and we believe that simplifying the overall structure of FastPerson can help improve user accessibility.

\subsection{Technological Enhancements and Future Directions}
\label{subsec:fastperson_future_directions}

FastPerson extracts visual elements to be synthesized with the summarized audio from the original video. Using a text-to-video model, other methods can be employed to generate a video from the summarized text; Runway's Gen-2 and OpenAI's Sora are possible candidates for these methods.

These methods focus on generating object-centered videos, and therefore, it currently requires more work to create lecture videos that are text-heavy and slide-centered. However, it may be possible to create helpful summary videos for generating lecture videos focused on some topics.

For example, if a lecture is centered around the intricacies of plant biology, a text-to-video model could be used to generate video segments illustrating key concepts such as photosynthesis, cellular structures of plants, or growth cycles. This approach would not only make the content more accessible but also more engaging for viewers, especially for visual learners.

Additionally, future developments should consider creating adaptive summarization systems that automatically adjust the level of detail in summaries based on users' learning history and comprehension levels. By analyzing user learning patterns using machine learning and generating summaries optimized for individual needs, more effective personalized learning could be achieved.

\subsection{Broader Implications and Applications of FastPerson}
\label{subsec:fastperson_broader_implications}

The broader implications of FastPerson go beyond its immediate application in educational video summarization. FastPerson has the potential to revolutionize how we interact with a wide range of video content. For example, in professional settings, FastPerson could be used to distill long training sessions or meetings into concise summaries, allowing for efficient review and better time management.

Further, FastPerson has potential applications related to media consumption, where users can quickly view news or entertainment highlights. In addition, its application could extend to content creators and marketers, giving them a video editing tool to create engaging, summarized versions of their content for better audience engagement.

FastPerson's adaptability could enable its integration into various platforms, making information more accessible and digestible for a broader audience. In particular, its application is expected in diverse fields such as news distribution, corporate training, online education, and even the entertainment industry. This new paradigm of ``video-to-video summarization" could become a new standard for efficient information consumption in the age of information overload.

\section{Conclusion}
\label{sec:fastperson_conclusion}

This chapter has presented FastPerson, an innovative video summarization system that addresses the challenge of enabling rapid comprehension of video content using visual information in the Consume-Understand stages, which is one of the three primary objectives of this thesis aimed at achieving time-efficient learning through audio and visual content under the AI-Guided Learning framework. FastPerson demonstrates a novel approach to video summarization by comprehensively considering both visual and auditory information in lecture videos, thereby enhancing the efficiency of the learning experience.

The key contributions of FastPerson are threefold. Firstly, it introduces a comprehensive video summarization method that generates summary videos reflecting important information from both visual and audio elements of lecture videos. Particularly, it establishes a new paradigm of ``video-to-video summarization," preserving context and speaker characteristics that are often lost in conventional text-based summarization. Secondly, FastPerson presents a user-centered learning experience design that enables seamless switching between original and summary videos. Reproducing the original speaker's voice through voice cloning supports continuity between the two versions. Thirdly, the system demonstrates empirically that it significantly reduces viewing time, with no statistically significant difference in quiz-measured comprehension from conventional playback.

The experimental results presented in this chapter provide evidence for the efficacy of FastPerson. The evaluation experiments indicated no significant difference in video comprehension when using FastPerson compared to conventional video playback methods. Crucially, FastPerson reduced viewing time by approximately 53\%. This substantial time saving, together with the absence of a statistically significant quiz-score difference, indicates the system's potential to enhance the efficiency of video-based learning. Furthermore, 78\% of FastPerson users rated the ability to switch between summary and original videos as useful or very useful.

Despite these promising results, it is important to acknowledge the limitations of the current implementation of FastPerson. The evaluation results indicate room for improvement in terms of the depth of the summary content and the quality of the audio in the generated summaries. These areas present opportunities for future research and development. Additionally, further investigation is needed to assess the long-term effects of using FastPerson on knowledge retention and learning outcomes across various subject areas and learner demographics.

Looking ahead, FastPerson holds considerable promise for a variety of applications in educational technology and beyond. In the context of online learning platforms and MOOCs, it could significantly enhance the learning experience by allowing students to quickly grasp the key points of lengthy lecture videos and efficiently review course materials. In corporate training scenarios, FastPerson could help employees quickly assimilate information from training videos, potentially improving the efficiency of professional development programs.

Furthermore, the system's approach to integrating visual and auditory information could have applications in fields such as multimedia content analysis and automated video captioning. In particular, the innovative approach of ``video-to-video summarization" could open new applications that were not possible with conventional text-based summarization.

Within the broader scope of this thesis, FastPerson complements the other two systems presented: AIxSpeed, which focuses on optimizing audio playback speed, and Profy, which addresses efficient feedback systems for skill acquisition. While AIxSpeed enhances the auditory aspect of content consumption and Profy facilitates interactive learning, FastPerson optimizes the visual component of educational videos. Together, these three systems form a comprehensive approach to enhancing time efficiency in learning through multimedia content, collectively contributing to a more effective and personalized learning experience.

In conclusion, FastPerson represents an advancement in video summarization technology and its application to educational contexts. By reducing viewing time without a statistically significant difference in quiz-measured comprehension in the present experiment, it addresses a critical challenge in the era of abundant online video content. As research in this area continues to evolve, future work should focus on refining the summary generation algorithms, improving the quality of synthesized audio, and exploring the integration of FastPerson with other learning technologies.

Furthermore, longitudinal studies to assess the impact of FastPerson on long-term learning outcomes and its effectiveness across different subject domains would provide valuable insights. The findings from this research have the potential to reshape how we approach video-based learning and content consumption in educational and professional settings, contributing to more efficient and effective knowledge acquisition in our increasingly digital world.

\chapter{Profy: Pronunciation Evaluation and Visualization of Differences from Native Speaker Groups Using Unannotated Audio Data}

\section{Introduction}
\label{sec:profy_introduction}

This chapter presents Profy, a skill acquisition feedback system in the Imitate stage toward realizing AI-Guided Learning, the overall goal of this thesis. As discussed in Chapter~1, it is difficult for learners to evaluate differences between their own performance and expert models in imitation-based learning. Profy addresses this challenge by learning a proficiency classifier from small amounts of unannotated audio data and visualizing waveform regions emphasized by that classifier.

The core innovation of Profy lies in its ability to learn a proficiency classifier end-to-end without specific predefined rules or detailed annotations. While many conventional CAPT (Computer-Assisted Pronunciation Training) systems use detailed phoneme-level annotations or linguistic expertise, Profy learns a classifier and visualizes model-emphasized regions from small amounts of unannotated data.
Pronunciation proficiency is a cornerstone of effective communication in a foreign language, yet it remains one of the most challenging aspects for learners in online environments due to the lack of scalable, individualized feedback~\cite{RogersonRevell2021}. Profy specifically targets this bottleneck within the broader scope of AI-Guided Learning. By enabling data-driven assessment and detailed feedback on pronunciation, Profy aims to make the skill acquisition phase of language learning more efficient and effective, thereby contributing to the overall goal of optimizing online learning experiences. The ``proficiency'' learned by Profy primarily refers to the acoustic characteristics that distinguish native-like speech from beginner speech in the context of target utterance repetition, rather than evaluating the semantic content or the pedagogical clarity of, for instance, a lecturer's intentional slow speech, which might be a consideration for systems like AIxSpeed.

Importantly, Profy's methodology may extend beyond pronunciation learning. The framework of learning proficiency from data and visualizing model-emphasized regions could be investigated for skills acquired through imitating expert performance, such as sports form correction, musical instrument instruction, and professional skill acquisition. This chapter evaluates pronunciation only; transfer to other domains remains future work.

Learning a non-native language presents numerous challenges, particularly in the domain of speaking proficiency. The acquisition of proper pronunciation is a critical aspect of language learning, as it significantly impacts communication effectiveness and overall language competence. Traditional methods of pronunciation instruction often rely heavily on direct feedback from native speakers or language experts. However, this approach is limited by the availability of such experts and the time-intensive nature of individual feedback sessions.

To address these limitations, there has been a growing interest in the development of Computer-Assisted Pronunciation Training (CAPT) systems. Existing CAPT systems have made significant strides in automating pronunciation feedback. These systems typically compare the user's speech with that of a specific native speaker as a model, analyzing units of rhythm, phonemes, or words to calculate differences~\cite{witt2000phone, Li2016, Qian2010}.

While effective, these approaches face several limitations. Firstly, they often require extensive speech data with detailed annotations, which can be costly and time-consuming to produce. Secondly, many systems are limited to comparisons with a single specific native speaker, potentially narrowing the range of acceptable pronunciations. Thirdly, the development of such systems for new languages or dialects can be challenging due to the need for specialized linguistic expertise and resources.

To address these challenges, this chapter introduces Profy, a language learning support system that calculates model-derived speech scores from a small amount of unannotated speech data without comparison to a specific person. Profy displays a proficiency score, classifier-emphasized waveform regions, and distances between learner speech and a group-level native-speech representation.

This approach lets learners compare how the model-derived score, highlighted regions, and distance change across repeated attempts. The attention mechanism highlights regions that contributed to the classifier output, providing more localized information than an overall score; it does not by itself establish that a region is a phonetic error or explain how to correct it.

The key contributions of Profy are threefold:

\begin{enumerate}
    \item A technical framework for assessing learners' speech proficiency using \textit{unannotated} speech data. It (a) learns to distinguish acoustic patterns in native and learner speech for specific utterances in an \textit{end-to-end, data-driven manner}, without explicit rules for each target phoneme or prosodic element, and (b) visualizes \textit{classifier-emphasized regions} through an attention mechanism, moving beyond a holistic score while not claiming phonetic error diagnosis.
    
    \item A language learning support system for non-native English speakers that lets learners inspect how model-derived waveform highlights and latent-space distances change across repeated attempts.
    
    \item Empirical evidence from a user study in which the Profy condition showed a larger observed intelligibility improvement than elicited imitation (EI), with non-overlapping pre- and post-practice confidence intervals within the Profy condition.
\end{enumerate}

These contributions address pronunciation-learning feedback within the broader objectives of this thesis. The present chapter does not measure practice-time efficiency or establish generalization to imitation tasks beyond pronunciation.

The remainder of this chapter is organized as follows: Section 5.2 provides an overview of related work in Computer-Assisted Language Learning (CALL) and pronunciation assessment. Section 5.3 details the design and implementation of the Profy system. Section 5.4 presents the evaluation methodology and experimental setup. Section 5.5 discusses the results of the user study and their implications. Finally, Section 5.6 concludes the chapter with a summary of findings and directions for future research.

By addressing feedback in the \textit{Imitate} stage, Profy complements AIxSpeed (focused on the \textit{Consume} stage for audio) and FastPerson (focused on the \textit{Consume} and \textit{Understand} stages for video) presented in the previous chapters. Together, these three systems instantiate the proposed AI-Guided Learning framework. Profy's contribution is a scalable, model-derived mechanism for skill practice. This chapter details Profy's methodology, implementation, pronunciation evaluation, and possible—but not yet evaluated—extensions to other imitation-based skills.

\begin{figure}[t]
    \centering
    \includegraphics[width=0.8\textwidth]{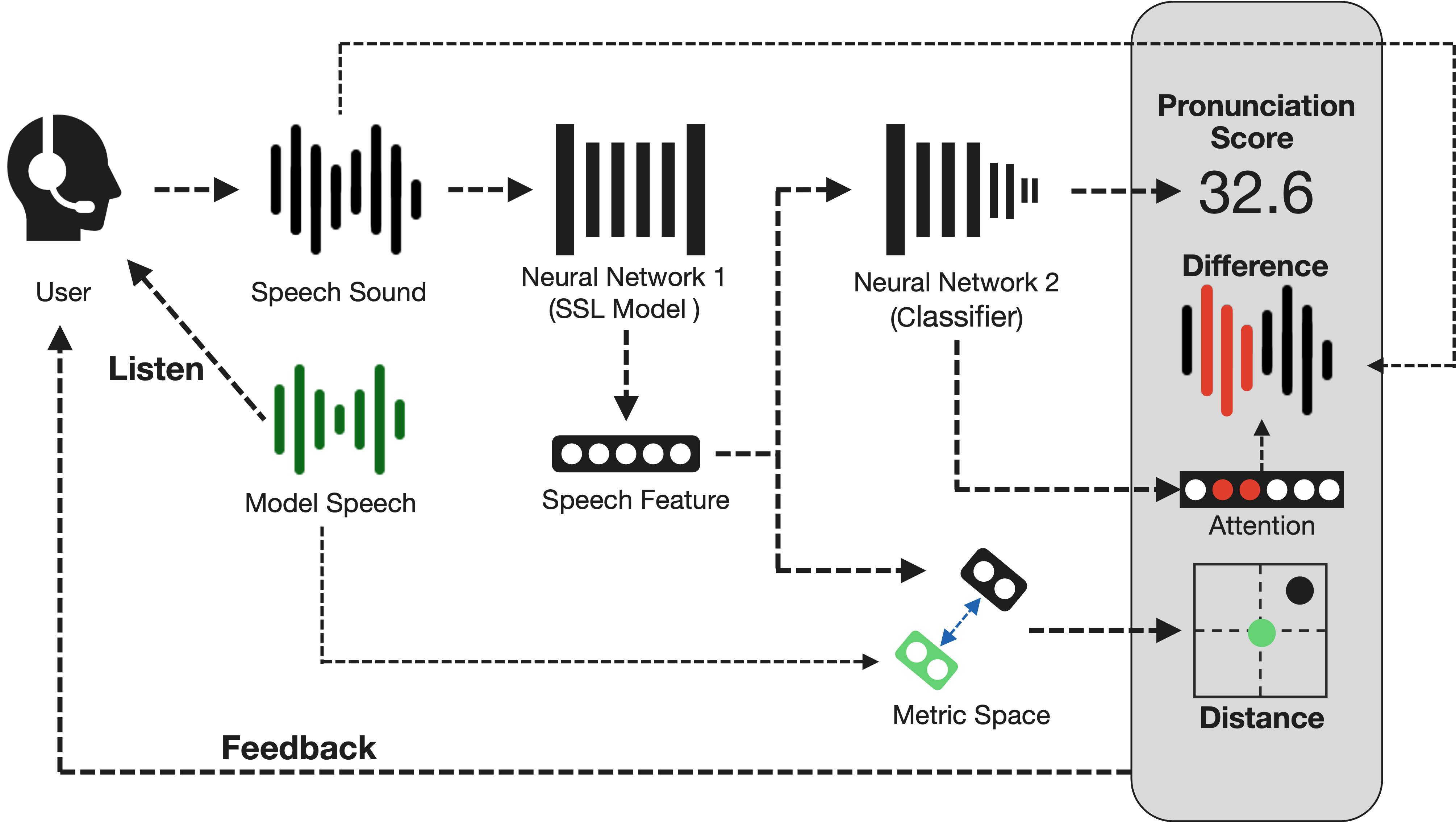}
    \caption[Profy system architecture]{Profy architecture: Overview of the system's components and data flow.}
    \label{fig:profy/overview}
\end{figure}

\section{System Design}
\label{sec:profy_system_design}

\subsection{Overview}
\label{subsec:profy_overview}

Profy has three main functions:
\begin{itemize}
  \item Pronunciation Score: a measure of how well the user speaks.
  \item Difference: different positions of the user's speech and a model's speech on the waveform.
  \item Distance: two points on the two-dimensional coordinates that indicate how far the user's speech is from the model's speech.
\end{itemize}

The proposed system uses mainly deep neural network techniques to achieve these functions, as shown in Figure~\ref{fig:profy/overview}. If the speech dataset contains values such as proficiency, difference, and distance as annotations, then a simple supervised classifier can be trained to build a model that finds these values in the user's speech.

However, such annotated data must usually be manually created, which is expensive. To address this problem, our system was designed to find the three desired values only from raw (unannotated) speech data from non-native and native speakers of the target language.

As shown in Figure~\ref{fig:profy/overview}, the pronunciation score is calculated using the classification of the unannotated data, the difference is calculated using attention, and the distance is calculated using metric learning. A further problem is the difficulty of collecting speech data from non-native speakers.

\begin{figure*}[t]
    \begin{center}
    \includegraphics[width=\textwidth]{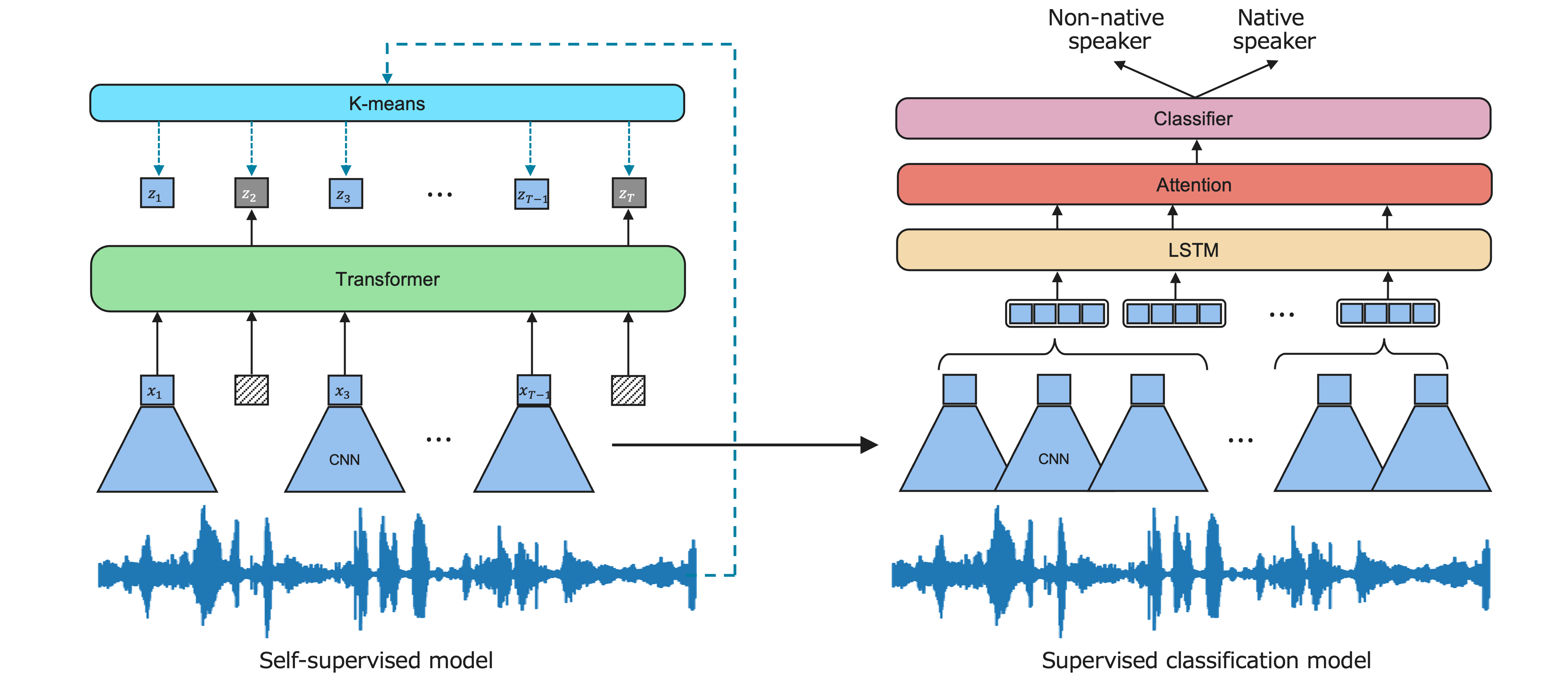} 
    \caption[Pronunciation score determination in Profy]{Profy pronunciation score determination: Process of classifying speech data and calculating proficiency value.}
    \label{fig:profy/proficiency}
    \end{center}
\end{figure*}

In contrast, speech data from native speakers can be easily collected from various sources, such as the radio, TV, and video streaming services~\cite{Ardila2019,Vassil2015,Zen2019}. For this reason, the system was designed to effectively use large amounts of data from native speakers using self-supervised learning so that it does not need to collect large amounts of speech data from non-native speakers.

The technical challenges in designing this system are:
\begin{description}
  \item[TP1] We need to train a deep learning model with minimal speech data from non-native speakers.
  \item[TP2] We need to obtain the desired values from speech data without human evaluation annotations.
\end{description}

\subsection{Pronunciation Score}
\label{subsec:profy_pronunciation_score}

As shown in Figure~\ref{fig:profy/proficiency}, the model for calculating the pronunciation score of a user's speech is trained to classify speech data from non-native and native speakers of the target language. Importantly, this model learns the abstract concept of proficiency from data without predefined rules.
In this context, proficiency refers to the learnable acoustic and temporal patterns that statistically differentiate native speaker utterances from those of non-native speakers for a given phrase, rather than relying on explicit phonetic correctness rules or subjective perceptual judgments of clarity alone. The model aims to capture a holistic set of features that characterize well-formed target language pronunciation as exemplified by a group of native speakers.

The model consists of an LSTM and a linear classifier for learning temporal features of speech.

The linear classifier uses the LSTM's hidden states $H = [h_{1}, h_{2},  \ldots, h_{T}]$ as input:

\begin{align}
s &= W^{(S)}H\alpha + b^{(S)}\\
\hat{p}(y \mid \mathcal{X}) &= \operatorname{softmax}\left(s\right), \\
\hat{y} &= \arg \max _{y} \hat{p}(y \mid \mathcal{X}),
\end{align}

where $W^{(S)}$ and $b^{(S)}$ are trained parameters, and $\alpha$ is the attention weight introduced in Section~\ref{sec:profy/visualizing_speech_differences}.

Subsequently, the calibrated value~\cite{Nixon2019} that brings the output value of the model closer to the probability of belonging to each class is used as the proficiency value.

Here, as shown in TP1, the problem of insufficient training data for the classes of non-native speakers arises. Recently, self-supervised learning, a method in which the model provides teacher labels even if the data do not have them, has been actively studied~\cite{Schneider2019,baevski2020wav2vec,Shuo2022}.

Training such a model on a large amount of unlabeled data and fine-tuning it with a small amount of labeled data is known to effectively perform downstream tasks~\cite{Shu-wen2021}. In our system, a self-supervised learning model is trained on a large amount of unlabeled speech data from native speakers, and then a binary classifier is trained on a small amount of labeled speech data from native and non-native speakers.

We use HuBERT~\cite{Hsu2021a} as a self-supervised model for training with a large amount of speech data from native speakers, obtaining latent representations $\mathcal{Z}$ from the input speech $\mathcal{X}$.

Then, the time series data $[z^{\prime}_{1:k}, z^{\prime}_{k+1:2k}, \ldots, z^{\prime}_{T-k+1:T}]$ obtained by combining the representations $\mathcal{Z}$ at each step $k$ are fed into the LSTM.

\subsection{Difference}\label{sec:profy/visualizing_speech_differences}
\label{subsec:profy_difference}

Suppose that the non-native speakers' speech data have human-rated labels indicating which parts of the speech differ from that of native speakers. In this case, we can train a supervised learning model that inputs the speech data and outputs the parts that differ from the native speakers' speech.

However, as stated in TP2, such labeled data were not available for this study. Therefore, our system uses an attention mechanism to localize regions that contribute to classification from unannotated speech data. These regions are model-derived and were not independently validated as phonetic error locations.

Attention has been successfully used in various tasks, such as question answering, machine translation, and speech recognition~\cite{Bahdanau2014, Chorowski2015, Hermann2015}, to improve model performance and the interpretability of deep learning models. Specifically, it visualizes where in the data the model focuses its attention when performing a particular task.
The core idea is that if the classifier (trained as described in Section~\ref{subsec:profy_pronunciation_score}) deems an utterance as ``non-native" or ``less proficient," the attention weights will highlight the specific time-frames in the input speech that contributed most to this classification. These highlighted segments are interpreted as the regions where the learner's pronunciation deviates most significantly from the learned patterns of native speech.

\begin{figure}
  \includegraphics[width=\textwidth]{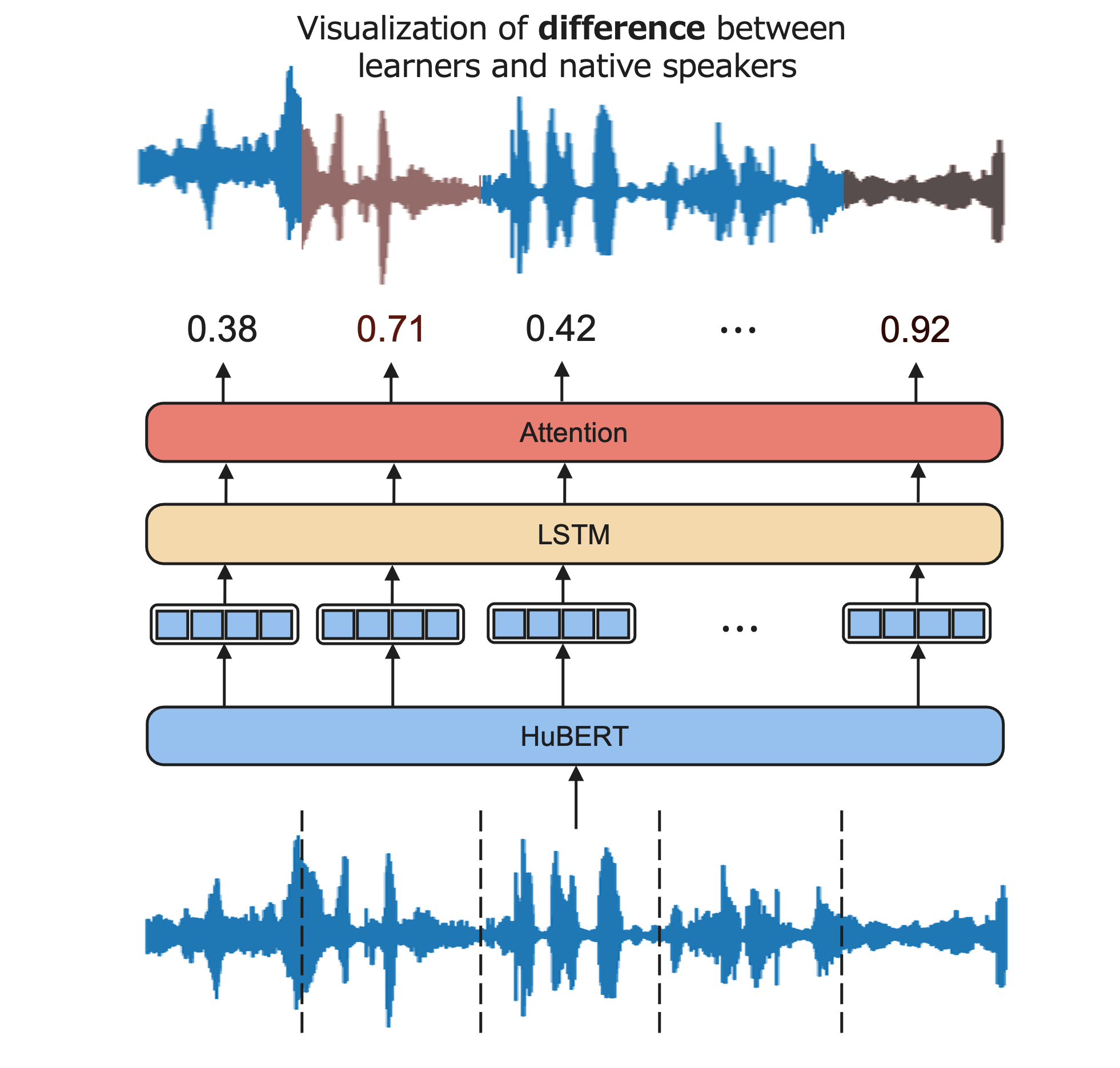}
  \caption[Difference visualization in Profy]{Profy difference visualization: Method for highlighting parts of speech that differ from native speakers.}
  \label{fig:profy/difference}
\end{figure}

As shown in Figure~\ref{fig:profy/difference}, we apply this mechanism to the classifier to determine the parts of the data on which the classifier focuses when it determines that the learner is not sufficiently proficient. The attention weight $\alpha$, which indicates the importance of the time-series feature, is calculated as follows:

\begin{align}
e &= \tanh(W^{A1}H)\\
\alpha &= \mathrm{softmax}(W^{A2}e)
\end{align}

where $W^{A1}$ and $W^{A2}$ are trained parameters. The weight $\alpha$ takes the value $[0,1]\in\mathbb{R}$. This value is standardized, and the waveform in the corresponding area is displayed in red when it exceeds a certain threshold.

The color is darker in segments that require greater attention to encourage the user to focus on them.

\subsection{Distance}
\label{subsec:profy_distance}

Before calculating the speech distance between the user and the native speakers, it is necessary to design an optimal distance function to measure it. One of the main approaches to learning the distance between data is triplet loss~\cite{Elad2015}.

\begin{figure}
  \includegraphics[width=\textwidth]{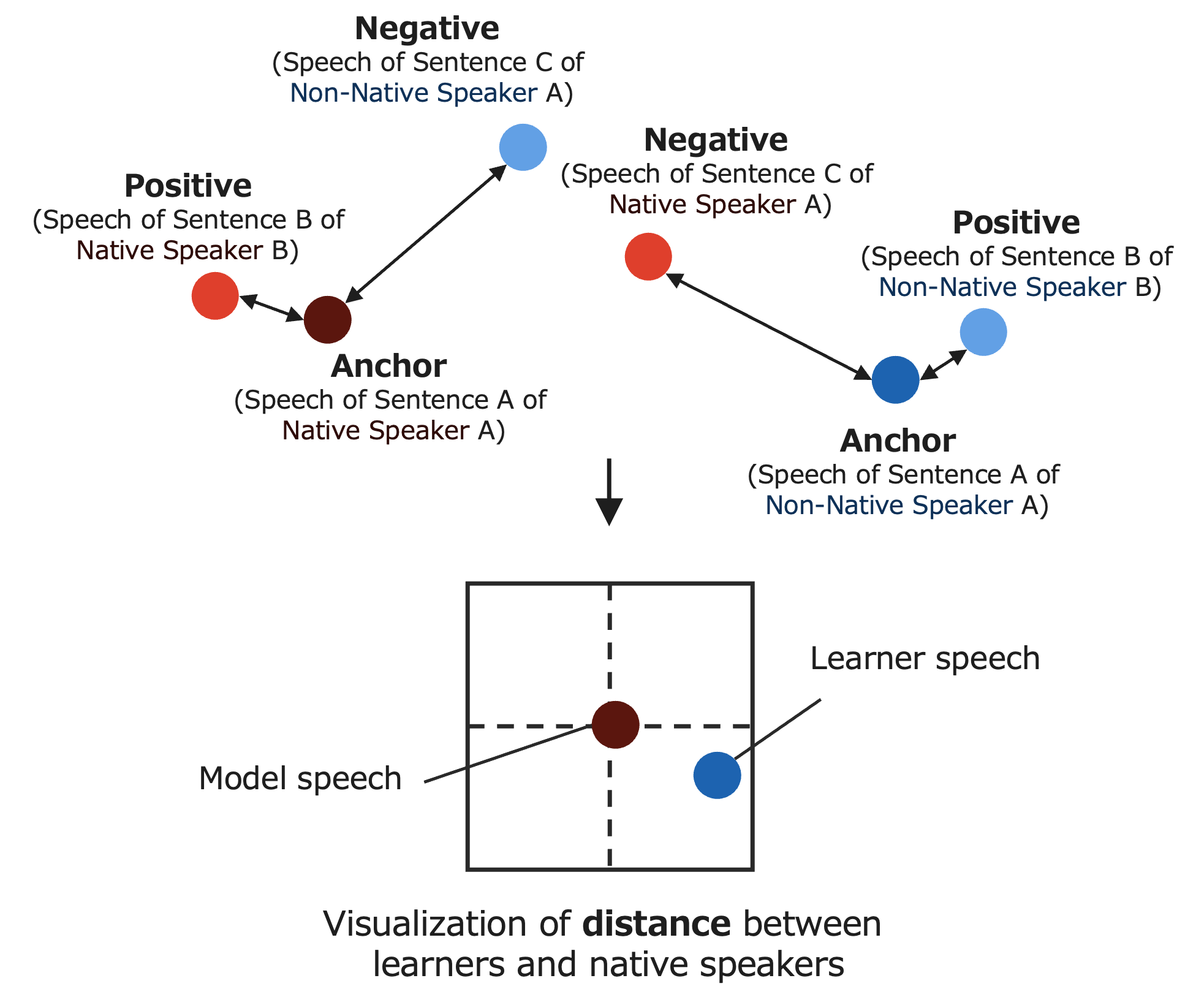}
  \caption[Distance visualization in Profy]{Profy distance visualization: Representation of speech data in two-dimensional space using triplet loss.}
  \label{fig:profy/distance}
\end{figure}

Triplet loss is learned using the following three types of speech data, as shown in Figure~\ref{fig:profy/distance}:
\begin{itemize}
    \item Anchor: speech data from a non-native speaker.
    \item Positive: speech data from a non-native speaker different from anchor.
    \item Negative: speech data from a native speaker.
\end{itemize}

In obtaining the distance function, the distance from the anchor data to the positive data ($d_{a p}$) is trained to be smaller, while the distance from the anchor data to the negative data ($d_{a n}$) is trained to be larger:

\begin{align}
  L_{\text {triplet }}=\left[d_{a p}-d_{a n}+m\right]_{+},
\end{align}

where $m$ is the margin---that is, the desired difference between the anchor-positive and anchor-negative distances.

The learned distance function is then used to calculate the two-dimensional vector difference between the user's speech data and the native speaker's speech data. In this case, the latter are the average of the native speaker's speech included in the data. Finally, the user's and native speaker's speech data are represented as points on the two-dimensional coordinates, with the native speaker's point at the center (Figure~\ref{fig:profy/distance}).
The axes of this learned two-dimensional space are not assigned specific phonetic or acoustic meanings. The visualization therefore indicates relative proximity to or divergence from the average native-speaker representation, rather than explaining which articulatory change a learner should make.

For metric learning, we use a linear layer with 512 hidden units and an LSTM with 128-dimensional hidden states.

In this way, the distance can be automatically calculated from the raw speech data, even if the data do not include a detailed human evaluation of how the native and non-native speech data differ, which is the challenge described in TP2.


\begin{figure*}[t]
  \centering
  \includegraphics[width=\textwidth]{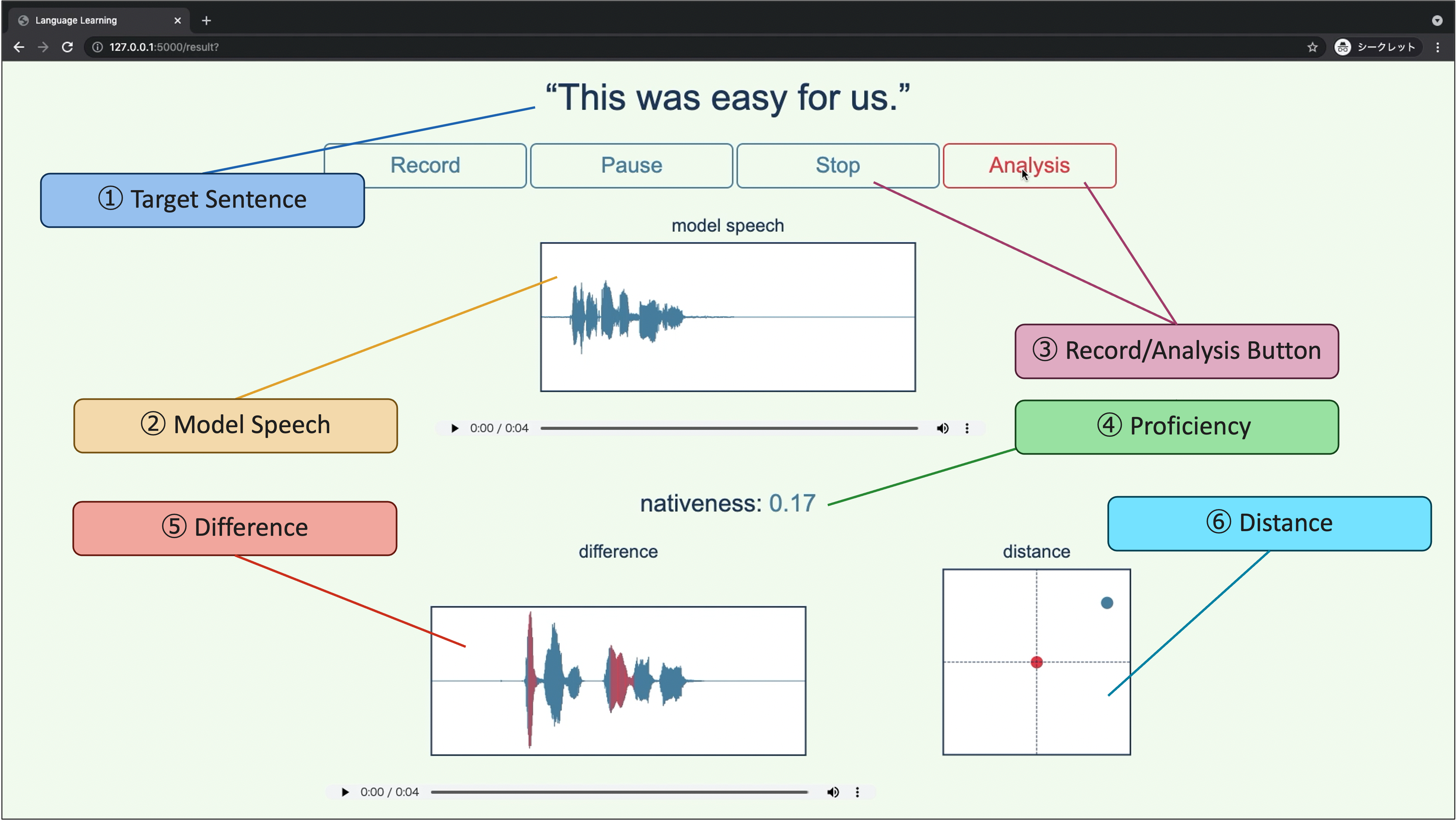}
  \caption[Profy user interface]{Profy user interface: Layout and components of the application's main screen.} 
  \label{fig:profy/interface}
\end{figure*}

\section{English Learning Application}
\label{sec:profy_application}

\subsection{Implementation}
\label{subsec:profy_implementation}

We developed a prototype Profy application to support Japanese people in learning to speak American English. The model speakers are native speakers of standard American English as defined in~\cite{Giegerich1992}. Two datasets were used to create this prototype.

One was LibriSpeech (960~h)~\cite{Vassil2015}, which was used as a large unlabeled dataset of English speech to train HuBERT. LibriSpeech consists of audiobook readings.

The other was UME-ERJ~\cite{Minematsu2001}, a database of English texts read by Japanese students and native speakers of standard American English, which was used as a small amount of labeled speech data. UME-ERJ contains $806$ sentences spoken by $202$ Japanese and $20$ American speakers. This dataset was used to train the classifier.

All audio files in the two datasets were standardized to a single-channel $16000$ $\mathrm{Hz}$ format and aligned to $4$ $\mathrm{s}$ in length.

We trained the model with HuBERT using the LibriSpeech data to obtain common speech features in English.

The attention layer used to visualize the differences was a $32$-dimensional hidden layer. We used LSTM with $128$-dimensional hidden layers and a linear layer with $512$ hidden layers for metric learning for distance visualization. We used Dropout~\cite{Nitish2014} to train these models with a probability of $0.5$ to suppress overfitting.

Since UME-ERJ is unbalanced, containing more speech data from Japanese than from American English speakers, we used focal loss~\cite{Lin2017} as an error function to reduce the effect of the imbalance.

The optimization algorithm used Adam~\cite{Diederik2015}, with a batch size of $32$ and a learning rate of $10^{-4}$.

We also augmented the data to make them more tolerant of noise and to compensate for the scarcity of data. The data augmentation included adding Gaussian noise in the background, randomly increasing or decreasing the audio volume, randomly shifting the pitch, and silencing randomly selected audio parts.

These deep models were implemented using PyTorch~\cite{Paszke2019} as the backend and an NVIDIA GeForce 1080Ti GPU. We then made it into a web application based on Flask, a microweb framework written in Python.

\subsection{Interface}
\label{subsec:profy_interface}

Figure~\ref{fig:profy/interface} shows the Profy interface. The system displays the target sentence that the user is about to learn to utter \ctext{1} and the model speech \ctext{2}. By playing back \ctext{2}, the user can listen to the correct pronunciation of the target sentence.

The user can input his/her speech into the system using the button \ctext{3}. The \textit{Pause} button can be used to interrupt the input. After the user has finished inputting his/her speech, the \textit{Analyze} button is used to analyze it.

When the \textit{Analyze} button is pressed, the degree of nativeness of the user's speech \ctext{4} and the difference \ctext{5} and distance \ctext{6} between the model speech and the user's speech are displayed. The difference is displayed in red on the waveform. The red area is the part that differs from the model speech, and the larger the difference, the darker the color.

The distance is displayed as two points on two-dimensional coordinates. The red point is the model's speech, and the blue point is the user's speech. The closer these points are, the closer the user's speech is to the model speech. This visual representation allows learners to intuitively understand the degree of their improvement.

\subsection{Learning Process}
\label{subsec:profy_learning_process}

\begin{figure*}[t]
  \centering
  \includegraphics[width=\textwidth]{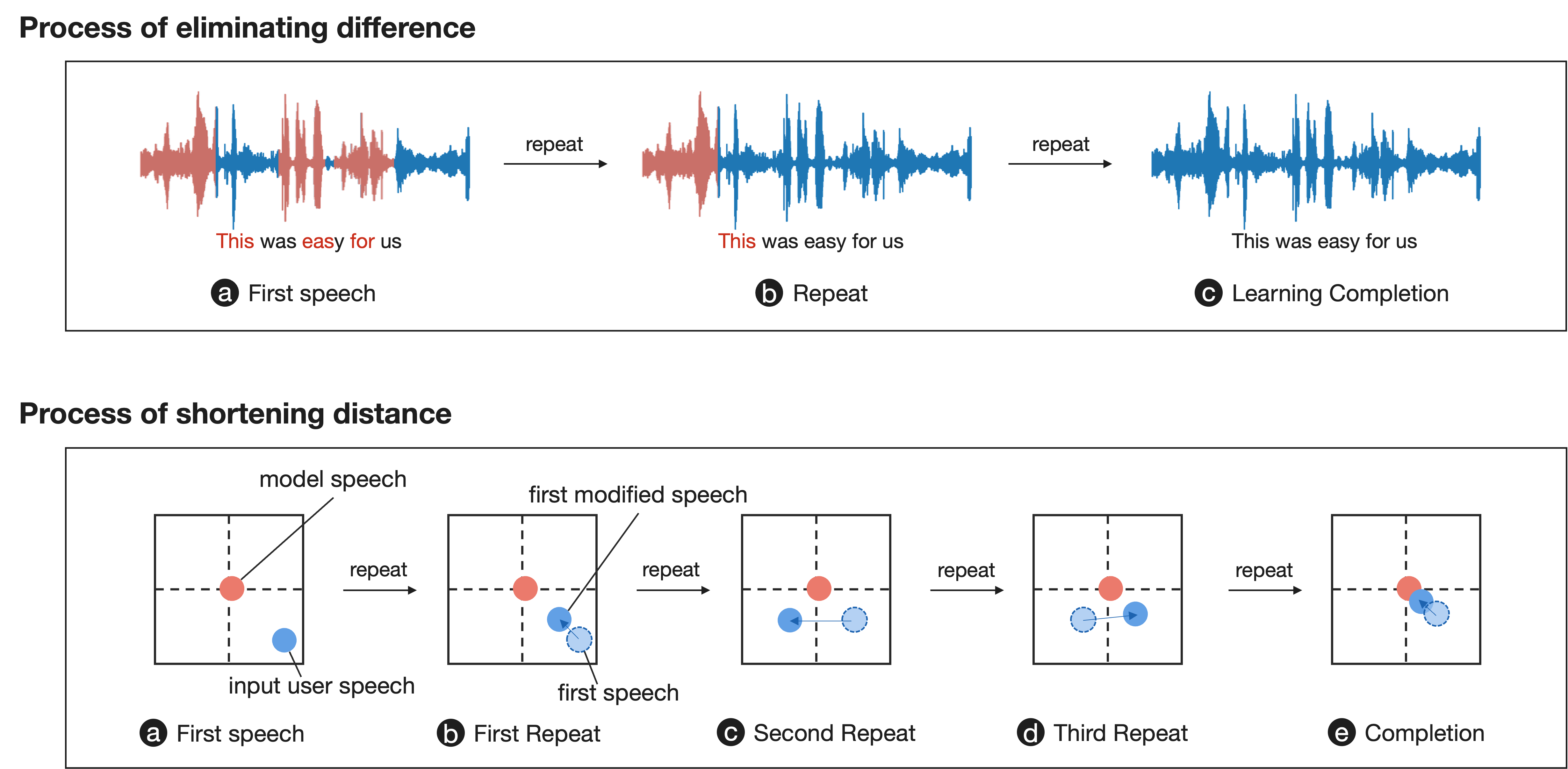}
  \caption[Examples of Profy interactions]{Examples of Profy interactions: Demonstration of eliminating differences and shortening distances in user speech.}
  \label{fig:profy/learning_process}
\end{figure*}

Here, we describe how a user can interact with the application to practice speaking the target language. The user's practice process is shown in Figure~\ref{fig:profy/learning_process}.

The user speaks given sentences into the system. Then, the system evaluates the user's utterance, gives a numerical value for the user's pronunciation score, and visually shows the difference and distance from the model utterance. The user repeatedly modifies his or her speech until the pronunciation score is satisfactory.

By repeating the sentence, the user can compare changes in the displayed difference and distance. The waveform highlights localize regions emphasized by the classifier, but do not by themselves identify the phonetic cause or prescribe a correction.

\subsubsection{Eliminating the difference}
\label{subsubsec:profy_eliminating_difference}

Figure~\ref{fig:profy/learning_process}~(upper panel) shows how the difference between the user's speech and the model's speech is eliminated as the user repeatedly modifies his or her speech. In the waveform of the user's speech shown in the figure, the parts that differ from the model speech are indicated in red.

When the user first speaks the sentence ``This was easy for us,'' the system highlights regions corresponding to ``this,'' ``eas,'' and ``for'' (\ctext{a}). After the user repeats the sentence, the highlights corresponding to ``eas'' and ``for'' disappear (\ctext{b}).

Finally, after another repetition, the remaining highlight corresponding to ``this'' also disappears, as shown in \ctext{c}.

\subsubsection{Shortening the distance}
\label{subsubsec:profy_shortening_distance}

Figure~\ref{fig:profy/learning_process}~(lower panel) shows how the distance between the user's speech and the model's speech decreases as the user repeatedly revises his or her speech. The model speech and the learner's speech are represented by dots in a two-dimensional coordinate space.

The red dots represent the model speech, and the blue dots represent the learner's speech. The greater the distance between the dots, the more the user's speech differs from the model speech. The user changes the pronunciation so that the distance between these points becomes shorter.

If the user's dot moves closer to the model-speech dot across repetitions, the displayed model-derived distance decreases (\ctext{a}$\rightarrow$\ctext{b} and \ctext{d}$\rightarrow$\ctext{e}). If it moves elsewhere, the distance representation changes in another direction (\ctext{b}$\rightarrow$\ctext{c}); the axes do not by themselves determine whether a phonetic change is correct.

The visualization lets the user observe whether successive attempts move closer to the model point, although the coordinate axes themselves do not identify which phonetic feature changed.

\section{Evaluation}
\label{sec:profy_evaluation}

We conducted a user study to evaluate the learning effectiveness of speaking practice using Profy. We examined whether the prototype application helped Japanese learners improve their English pronunciation. Ten individuals aged 24--35 years participated in the experiment.

\subsection{Procedure}
This experiment compared Profy with elicited imitation (EI), a method of learning speaking by having the user listen to a model speech and imitate it several times. On the other hand, practice using our system provides feedback on the user's pronunciation score and difference and distance from model speech to help the user modify his or her speech accordingly.

Each participant was presented with $10$ sentences from TIMIT~\cite{Garofolo1993}, a major English speech corpus, and learned to speak half of them using EI and the other half using Profy's method. To ensure that the effect estimation would not be affected by the level of difficulty, the sentences were divided into five-sentence sets A and B; half of the participants learned set A using Profy, and the other half learned set B using Profy.

During the experiment, all participants' utterances were recorded, and the first and last utterances of each sentence were used to estimate the learning effect.

Once the participants had completed the speaking practice with both methods, a brief questionnaire was administered regarding their user experiences.

\subsection{Results}
\label{subsec:profy_results}

\subsubsection{Learning Effect}
\label{subsubsec:profy_learning_effect}

To compare the effectiveness of Profy with that of the standard EI method, we presented five American raters with two sets of pre- and post-practice sentence pairs and asked them to judge which sentence was more intelligible. The raters were blinded to the pre- and post-practice sentences.

\begin{figure}
  \includegraphics[width=\textwidth]{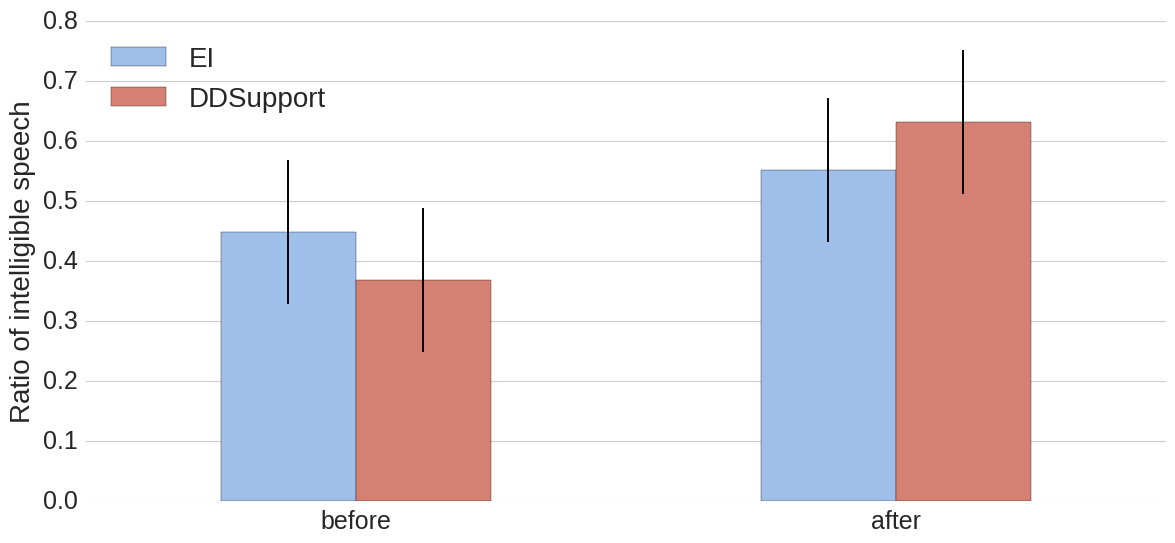}
  \caption[Comparison of post-practice intelligibility gains]{Comparison of post-practice intelligibility gains for users of Profy and standard EI method.}
  \label{fig:profy/user_study_1}
\end{figure}

The experimental results are shown in Figure~\ref{fig:profy/user_study_1}. The y-axis indicates the percentage of the user's speech judged to be more intelligible when compared before and after learning. The sentences practiced using both Profy and EI were found more intelligible than the pre-practice sentences.

However, the sentences practiced using Profy showed a larger observed improvement in intelligibility than those practiced using EI.

Furthermore, unlike EI, Profy had non-overlapping pre- and post-practice confidence intervals. The source paper did not report a direct paired significance test for this comparison.

\subsubsection{User Experience and Questionnaires}
\label{subsubsec:profy_user_experience}

After the experiment, we conducted questionnaires with the following Likert scale questions about the Profy:
\begin{description}
  \item[Q1] I found it easy to use Profy.
  \item[Q2] I found the display of pronunciation score helpful.
  \item[Q3] I found the display of difference helpful.
  \item[Q4] I found the display of distance helpful.
\end{description}

\begin{figure}
  \includegraphics[width=\textwidth]{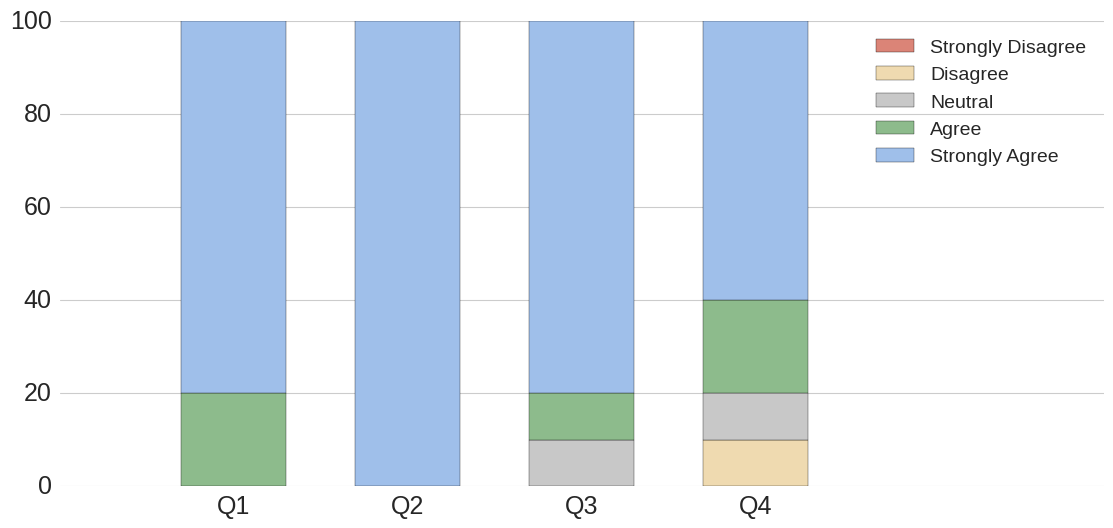}
  \caption[Questionnaire results on Profy usability]{Questionnaire results: User feedback on Profy's ease of use and helpfulness of features.}
  \label{fig:profy/user_study_2}
\end{figure}

For each question, we asked respondents to rate on a five-point scale of strongly agree, agree, neutral, disagree and strongly disagree. The questionnaire results are shown in Figure~\ref{fig:profy/user_study_2}.

All participants found the proposed system easy to use (``agree''/``strongly agree'').

Moreover, all participants found the presentation of the pronunciation score helpful.

On the other hand, while the majority found the presentation of differences and distances useful, some answered ``neutral'' or ``disagree'' (one and two, respectively).

In summary, useful features in Profy's functional assessment are pronunciation scores, differences, and distance, in that order. This result shows that learners evaluate feedback in order from the most intuitively understandable.

\section{Discussion}
\label{sec:profy_discussion}

\subsection{Effectiveness of Data-Driven Proficiency Learning}
\label{subsec:profy_effectiveness}

One of Profy's technical contributions is learning a proficiency classifier end-to-end. Unlike approaches that rely on phoneme-level annotations and explicit linguistic rules, Profy trains from unannotated utterance-level data.

This design may reduce annotation requirements, but the present study did not test adaptation to new languages, dialects, or non-speech imitation tasks. Such transfer requires new data, models, and domain-specific validation.

The visualization of classifier-emphasized regions through an attention mechanism also sets Profy apart from overall-score-only feedback. It shows which regions contributed to the model's judgment, while not identifying their phonetic cause.

\subsection{Limitations and Future Improvements}
\label{subsec:profy_limitations}

The current implementation has several limitations. First, distance visualization is difficult for some users to understand. Because the axes of the learned two-dimensional coordinates do not have directly interpretable acoustic meanings, the visualization does not by itself explain how to correct an error.

Second, the system is currently limited to feedback at the single sentence level. In actual conversation, prosodic features at larger units such as intonation, rhythm, and stress are also important. In the future, it will be necessary to extend to a feedback system that also considers these suprasegmental features.

Third, the current system only works with recorded audio and cannot provide real-time feedback. By utilizing edge computing on mobile devices, real-time feedback could be enabled, providing a more natural learning experience.

Fourthly, and perhaps most significantly from a pedagogical standpoint, Profy highlights classifier-relevant regions and displays an abstract acoustic distance, but it does not validate those regions as phonetic errors or provide guidance on \textit{how to correct} them. The learner is still responsible for interpreting the display and deducing any articulatory adjustment. Future work should critically explore validating the highlighted regions and integrating corrective actions, for example by retrieving relevant native-speaker examples, linking to targeted phonetic exercises, or generating minimal pairs.

\subsection{Relation to Existing Language Learning Tools and CAPT Systems}
\label{subsec:profy_comparison_existing}

When comparing Profy to established language learning applications like Duolingo or Babbel, or other research-based CAPT systems, several distinctions and potential synergies emerge. Many commercial platforms (e.g., Duolingo, Busuu) offer comprehensive, gamified curricula covering various language skills (vocabulary, grammar, listening, speaking). Their pronunciation feedback, if available, often relies on ASR to provide a general correctness score for an utterance or identifies very coarse-grained errors at the word level, rather than pinpointing specific phonetic deviations within a word. While effective for motivation and broad practice, this level of feedback may be insufficient for learners seeking to refine specific aspects of their pronunciation.

Specialized CAPT systems often provide more detailed phonetic analysis (e.g., based on Goodness of Pronunciation (GOP) scores derived from forced alignment, or analysis of specific acoustic features like formants or pitch contours)~\cite{witt2000phone, Li2016}. However, these systems frequently require: (a) accurately transcribed speech, (b) detailed annotations of target pronunciations, often at the phoneme level, (c) pre-defined linguistic rules for error detection, or (d) a single, reference native speaker model, which may not capture the natural variability in pronunciation.

Profy's approach offers several differentiating factors in this landscape:
\begin{itemize}
    \item \textbf{Data-Driven Proficiency Learning from Largely Unannotated Data:} Profy learns the characteristics of proficient speech directly from audio examples of native speakers and contrasts them with learner speech, minimizing the need for extensive manual annotation or explicit rule-setting for every possible mispronunciation. This is particularly advantageous for scalability and adaptation to different languages or dialects.
    \item \textbf{Localized Visualization of Classifier-Relevant Regions:} The attention-based waveform highlighting shows which segments of an utterance contributed most to its classification as non-native-like. This is more localized than an overall score, but is not a phonetic error diagnosis.
    \item \textbf{Learning from a Group of Native Speakers:} By training on data from multiple native speakers, Profy has the potential to learn a more robust and representative model of target pronunciation, implicitly accommodating natural variations (although the current ``Distance'' visualization uses an average native representation as a target).
\end{itemize}

It is important to acknowledge that Profy, in its current form, is a specialized tool focused on model-derived feedback for relatively short, isolated utterances. It does not offer a complete language learning curriculum, personalized learning paths, or the gamification elements found in many commercial applications. A promising avenue for future work lies in integrating Profy's feedback mechanisms into broader language learning platforms after further validation.

\subsection{Broader Applications Beyond Language Learning}
\label{subsec:profy_broader_applications}

The Profy framework suggests possible future research on other skills acquired by imitating expert performance. For example, related models might compare musical performances or movement data from experts and learners.

In sports or professional training, such systems would require domain-specific representations, labels or evaluation criteria, and user studies before claims about corrective feedback or learning effectiveness could be made.

The current pronunciation study supplies no evidence for effectiveness in music, sports, surgery, or other professional skills; these are examples of research directions only.

A transferable design principle is learning a performance classifier from examples rather than encoding every rule manually, while still requiring expert validation of what the learned representation means in each domain.

\section{Conclusion}
\label{sec:profy_conclusion}

This chapter has presented Profy, a language learning support system that addresses the challenge of providing scalable feedback for learner skill assessment in the Imitate stage of the AI-Guided Learning framework. Profy uses deep learning-based speech processing to provide model-derived visual feedback on pronunciation; whether this improves the efficiency of skill acquisition requires further evaluation.

The key contributions of Profy are threefold. Firstly, it introduces a technical framework for assessing learners' speech proficiency and visualizing classifier-emphasized regions using unannotated speech data without comparison to a specific native speaker. Particularly important is that it learns a proficiency classifier end-to-end and localizes regions that influence the classifier output.

Secondly, Profy presents a language learning support system for non-native English speakers that provides visual feedback on pronunciation. By highlighting classifier-emphasized waveform regions and visualizing distances in latent space, it provides more localized model-derived information than systems that present only an overall score.

Thirdly, the user study with Japanese learners of English showed a larger observed intelligibility improvement for Profy than for conventional Elicited Imitation (EI). Unlike EI, the pre- and post-practice confidence intervals did not overlap for Profy.

Despite these promising results, it is important to acknowledge the limitations of the current implementation of Profy. The system's performance may vary depending on factors such as the linguistic distance between the learner's native language and the target language, the diversity of accents represented in the model speech data, and the complexity of the practiced sentences.

Looking ahead, Profy holds considerable promise for applications beyond language learning. The core concept of learning proficiency from data and visualizing model-emphasized regions could be adapted to support skill acquisition in fields such as sports, music, and professional training, subject to domain-specific validation.

For instance, future systems could analyze an athlete's movements or compare musical performances with expert examples. These examples illustrate possible extensions; the present study does not provide evidence that Profy's framework generalizes beyond pronunciation.

Within the broader scope of this thesis, Profy complements the other two systems presented: AIxSpeed, which focuses on optimizing audio playback speed, and FastPerson, which addresses rapid comprehension of video content. While AIxSpeed and FastPerson enhance the efficiency of content consumption, Profy optimizes the process of skill acquisition and feedback. Together, these three systems form a comprehensive approach to enhancing time efficiency in learning through multimedia content, collectively contributing to a more effective and personalized learning experience across various stages of the learning process.

In conclusion, Profy represents a significant advancement in the field of computer-assisted language learning and, more broadly, in the development of efficient skill acquisition support systems. By enabling learners to receive immediate, visual feedback on their pronunciation without the need for constant expert supervision, it addresses a critical challenge in language education and opens new possibilities for self-directed learning.

As research in this area continues to evolve, future work should focus on expanding the system's capabilities to cover more aspects of language learning, such as grammar and vocabulary acquisition, and exploring its application in other skill domains.
Crucially, enhancing the feedback to include guidance on \textit{how to correct} identified errors, alongside exploring pathways for practical deployment in educational settings and addressing associated pedagogical and logistical considerations, will be vital steps towards realizing Profy's full potential.
Furthermore, integrating Profy with other learning technologies, such as virtual reality or augmented reality, could potentially create even more immersive and effective learning experiences.

The insights gained from this research have the potential to reshape how we approach skill acquisition in educational and professional settings, contributing to more efficient and effective learning methods in our increasingly digital and globally connected world. In particular, the approach of learning proficiency in a data-driven manner holds the potential to become a new design principle for imitation learning support systems in various fields.

\chapter{Discussion}

\section{Chapter Overview}
This chapter synthesizes how the three systems—AIxSpeed, FastPerson, and Profy—collectively instantiate the Consume–Understand–Imitate (CUI) framework for audio‑visual learning. We first provide a brief recap of CUI and the role of each system, then position the work within established learning theories. We next discuss how the systems complement one another, examine limitations and open questions, and conclude with a vision‑driven trajectory that extends beyond the three prototypes.

\section{Brief Recap: CUI and the Three Systems}
In this dissertation, CUI denotes a process model tailored to audio‑visual learning: (i) Consume—efficient access to content; (ii) Understand—deep comprehension of multimodal material; and (iii) Imitate—practice with feedback for skill acquisition.
AIxSpeed supports efficient consumption (and early understanding) by adaptively adjusting playback speed based on acoustic features. FastPerson supports understanding by integrating linguistic and visual cues and providing summaries that preserve the speaker's voice and context. Profy supports imitation by analyzing performance against model patterns and providing model-derived, localized feedback for practice.

\section{Positioning in the Learning Sciences}
We position the work within the learning sciences by relating CUI to established models and clarifying its scope for audio‑visual learning.

Relation to established models. CUI partially aligns with (a) Kolb's experiential cycle~\cite{kolb1984experiential} (Consume $\leftrightarrow$ concrete experience; Understand $\leftrightarrow$ reflection and conceptualization; Imitate $\leftrightarrow$ active experimentation), while recognizing Kolb's cyclic nature (learners may move back and forth) and individual differences; (b) Bloom's revised taxonomy~\cite{anderson2001revised}, which is two-dimensional (knowledge $\times$ cognitive process) rather than a strict linear hierarchy, consistent with CUI's cyclic use; and (c) ICAP~\cite{chi2014icap}, which classifies engagement by visible learner behaviors (e.g., notetaking, summarizing, self-explanation, interaction). Our systems may help shifts from Passive to Active, and in some uses toward Constructive, when such generative behaviors are present. Support for fully Interactive learning is limited and remains an avenue for future work.

This work reinterprets general learning theories for the characteristics of audio‑visual media and offers practical system design choices addressing challenges such as temporal inefficiency, multimodal integration, and feedback scarcity. Whereas prior theory emphasizes describing cognitive processes, this research materializes learning support as implementable systems.

\section{Synthesis: How the Systems Work Together}
The three systems are designed to be complementary along the Consume→Understand→Imitate flow. A typical workflow is to summarize a long lecture with FastPerson (with seamless access back to details), use AIxSpeed to increase listening efficiency, and then connect to practice with Profy using model-derived feedback. Together they support an end‑to‑end learner workflow; understanding potential synergies and trade‑offs (e.g., cognitive load, controllability) requires further empirical study.

Furthermore, this research can be situated with respect to Chi and Wylie's ICAP framework~\cite{chi2014icap} (Passive, Active, Constructive, Interactive). Because ICAP is defined by visible learner behaviors, AIxSpeed and FastPerson may support Passive→Active shifts when learners engage in generative actions (e.g., notetaking, summarizing, self‑explanation), and Profy may prompt Constructive activity via feedback on learner productions. Support for the Interactive level (collaborative knowledge construction) is limited and remains future work.

\section{Redefining Learning in the Digital Age: Learner Autonomy and the AI-Guided Consume–Understand–Imitate Framework}
The AI‑Guided Learning framework proposed in this research goes beyond system design considerations to discuss fundamental questions about the nature of learning. The long‑standing idea of widening learner autonomy and access (e.g., ``learning webs''~\cite{Illich1971}) gains new practical avenues through AI. This section describes how the Consume–Understand–Imitate (CUI) cycle and the three systems—AIxSpeed, FastPerson, and Profy—address challenges such as the inefficiency of long‑form content, the difficulty of understanding multimodal information, and the lack of feedback in imitation‑based learning, thereby supporting learner autonomy and broadening access to expert‑like guidance.

The traditional school system has been described as potentially commodifying knowledge and transforming learning into an act of consumption, thereby constraining spontaneous learning capacities. In contrast, concepts like ``learning webs'' proposed networks where learners could access knowledge resources based on their interests, meet peers, and learn from experts. The AI-Guided Learning and CUI framework in this research can be positioned as a reinterpretation of this vision for the digital age, implemented as a concrete set of technological systems aiming to solve challenges specific to audio-visual learning. AIxSpeed supports efficient information intake (Consume), FastPerson facilitates deep multimodal understanding (Understand), and Profy aids in the imitation and refinement of practical skills (Imitate), thereby strengthening the entire learning cycle.

This research aims not only to technically realize this vision but also to open new theoretical horizons. First, AI-Guided Learning introduces ``autonomous learning orchestration.'' This refers to AI's ability to understand learner needs, progress, and context, and dynamically adjust optimal learning paths or information presentation. In this work, orchestration is framed as supportive of self‑regulated learning (SRL): goal setting, monitoring, and evaluation remain learner‑led, with AI assisting rather than replacing metacognitive control. For instance, AIxSpeed adjusts playback speed based on estimated listening difficulty from acoustic features, attempting to reduce the cognitive load learners would otherwise expend on manual speed adjustments. This capability, viewed through the lens of distributed cognition theory and the extended mind thesis, suggests that learning processes unfold across the entire human-AI-environment interaction, extending beyond individual cognition~\cite{hutchins1995cognition, clark1998extended}.

Second, this research suggests the potential for ``predictive learning support.'' AI-Guided Learning, beyond responding to explicit learner requests, can predict implicit needs from behavioral patterns and content characteristics, offering proactive support. The automatic extraction of important segments in FastPerson, or the presentation of differences from expert models in Profy, can serve as support before learners even identify a problem themselves. This extends self-regulated learning theory, enabling AI to assist with parts of the metacognitive process, thereby allowing learners to focus on higher-order thinking~\cite{zimmerman2002becoming}.

Third, the three systems—AIxSpeed, FastPerson, and Profy—each aim to reduce common constraints in audio‑visual learning. AIxSpeed addresses time constraints by adjusting redundant or difficult‑to‑hear segments. FastPerson addresses information overload and comprehension constraints through efficient presentation of multimodal information. Profy addresses constraints of expert access and self‑assessment by providing automated, model-derived feedback. These are intended to function not merely as efficiency tools but as instruments that reduce common constraints on audio‑visual learning, enabling learners to proactively design their own learning processes.

The potential for these systems to reconstruct power relationships in learning is also noteworthy. In traditional education, teachers and institutions often functioned as gatekeepers of knowledge. AI-Guided Learning aims to alter this structure, supporting learners in becoming architects of their own learning. This resonates with the principles of critical pedagogy, which repositions learners not as passive recipients of knowledge but as active knowledge constructors~\cite{freire1970pedagogy}.

\section{The Synthesis of Human and Machine Intelligence in the CUI Cycle}
AI‑Guided Learning reframes the human–AI relationship from a unidirectional teacher–student framework to one of collaborative intelligence. In this paradigm, AI does not unilaterally transmit knowledge but functions as a cognitive partner that augments learners' abilities and helps optimize processes at each stage of the CUI cycle.

This concept can be understood in the context of post-humanist pedagogy, particularly in relation to the notion of ``entanglement''~\cite{barad2007meeting}. In AI-Guided Learning, humans and AI are not viewed as separate entities interacting, but as mutually influencing beings that jointly constitute the phenomenon of learning. Learning does not occur solely within the human or the AI but emerges from the dynamic relationship between them.

The three systems developed realize this collaborative intelligence within the CUI cycle in distinct ways. In the Consume and initial Understand stages, AIxSpeed utilizes ASR model confidence as an indicator of listening difficulty, technologically supplementing the limits of human auditory processing. The system does not replace human understanding but optimizes auditory information presentation so learners can process and comprehend information more efficiently. This can be seen as an application of the Zone of Proximal Development concept for the AI age, where AI functions as digital scaffolding, enabling learners to tackle tasks they might not manage alone~\cite{vygotsky1978mind}. However, its inability to discern semantic importance (e.g., intentionally slow speech by an instructor) based solely on acoustic features remains a challenge for future work.

In the Understand stage, FastPerson integrates and efficiently presents multimodal information. This is based on the ``economy of attention'' concept, where attention is a scarce resource in an information-rich world; it uses an LLM to summarize transcribed speech and detected visual information~\cite{simon1971designing}. Furthermore, generating summary videos in the original speaker's voice and enabling seamless switching between summary and full versions aligns with the Cognitive Theory of Multimedia Learning by retaining visual and auditory channels~\cite{mayer2005cognitive}. This video‑to‑video summarization differs from text‑based summarization by preserving the speaker’s voice and visual context. The current prototype does not implement citation-backed generation or a separate factual-verification stage; these are priorities for mitigating hallucination and fidelity loss in future deployment.

In the Imitate stage, Profy ventures into the realm of ``tacit knowledge'' using self-supervised learning. Skills like pronunciation are typical examples of tacit knowledge, a state of ``knowing more than we can tell''~\cite{polanyi1966tacit}. Profy learns norms of proficiency from large amounts of expert data in a data-driven manner and visualizes discrepancies with learner performance as waveform regions or distances in a learned two-dimensional space. This is an attempt to partially formalize tacit knowledge and make it learnable. The meaning of this learned distance and its axes requires further validation; the current representation does not map axis directions to specific phonetic features. It nevertheless provides model-derived feedback without one-on-one coaching.

\section{Limitations and Open Questions}
While each system was evaluated independently, combined workflows have not yet been studied; potential synergies and trade‑offs in cognitive load and usability remain to be examined. A factorial evaluation (e.g., summary × speed optimization × feedback) could isolate interaction effects on comprehension and transfer. The CUI model covers key phases of audio‑visual learning but does not directly address metacognitive regulation, affective factors, or fully interactive collaboration. Our evaluations primarily capture immediate or short‑term outcomes; long‑term retention, transfer, and behavioral change warrant longitudinal study. Efficiency gains must be balanced with pedagogical pacing (e.g., intentional pauses and repetitions). For Profy, explaining why specific errors occur and how to correct them remains challenging; establishing mappings from the learned representations to actionable articulatory guidance is a priority for future work. 

Finally, beyond the limitations of individual systems, we must recognize that AI-Guided Learning as a whole provides an incomplete answer to the fundamental question of ``what is learning''. While efficient information intake, deep understanding, and skill imitation are important aspects of learning, addressing other crucial elements that constitute the complete picture of human learning—such as creativity, critical thinking, and collaborative problem-solving—remains a future challenge.

\section{Broadened Access and Educational Justice}
One notable social implication of AI‑Guided Learning is broadened access to expert‑like feedback and resources. By this, we mean widening equitable access to top‑tier instruction, expert models, and high‑quality feedback regardless of geography, affiliation, or socioeconomic status. High‑quality education and expert guidance have traditionally been limited to a privileged few. The systems developed in this research, particularly Profy, demonstrate the potential to technologically alleviate this inequality. At the same time, fairness risks in speech technologies (e.g., accent‑ or race‑related performance gaps in ASR and pronunciation scoring) must be addressed. We plan to audit datasets and models (e.g., score validity across groups and environment effects) and to calibrate thresholds to reduce systematic disparities.

Profy addresses the challenge of expensive and inaccessible expert coaching by extracting patterns of proficiency from speech data via self-supervised learning, offering automated feedback in a widely accessible form. This can be interpreted as an attempt to embody the ``difference principle'' within educational technology, contributing to educational justice by maximizing benefits for the least advantaged~\cite{rawls1971theory}.

This democratization also prompts a reconsideration of concepts like standards and correct answers. As AI learns from diverse expert data, it may become possible to recognize and present not a single correct method but various patterns of excellence. This could challenge the reproduction mechanisms of cultural capital in educational systems that tend to favor those with specific cultural backgrounds~\cite{bourdieu1977outline}. AI-Guided Learning might contribute to creating environments where learners from diverse backgrounds can pursue excellence in their respective contexts.

Furthermore, AI-Guided Learning adds a new dimension to the ``politics of recognition''~\cite{taylor1994politics}. When AI systems consistently acknowledge learner effort and progress while making the basis and limitations of their feedback transparent, they may enhance learner self-efficacy and motivation. The question remains whether AI-based recognition is equivalent to human recognition or can embody true intersubjectivity~\cite{honneth1995struggle}, but it may play a complementary role alongside human instructors.

\section{Future Trajectories: The Co-evolution of Humans, AI, and Learning Ecosystems}
The future of AI-Guided Learning should be considered not only in terms of technological advancements but also from the perspective of human-AI co-evolution and the transformation of surrounding learning ecosystems. The development of generative AI and large language models, in particular, presents new possibilities and challenges.

Technologically, further evolution of multimodal AI is expected to lead to systems that can more sophisticatedly integrate text, audio, images, and video, dynamically selecting optimal media and linguistic expressions using evidence‑based signals (e.g., prior knowledge, task difficulty, learning history) rather than debated constructs such as learning styles. AI with extensive context windows will enable more continuous support based on long‑term learning histories and deep contextual understanding.

System improvements could include, for Profy, not only error indication but also suggestions for specific correction methods or step-by-step approaches to target pronunciations, plus stronger local explanations (e.g., projecting onto IPA/phonetic features or using SHAP‑style attributions). AIxSpeed might aim for more context-aware speed adjustments by incorporating semantic information or speaker intent, and provide spaced‑repetition prompts or retrieval cues to support long‑term retention. FastPerson could be enhanced with greater interactivity and fidelity safeguards for LLM outputs (e.g., extractive bias, citation‑backed responses, verification QA).

Socially and institutionally, new learning communities and certification systems supporting AI-Guided Learning will be crucial. AI-mediated skill exchange and peer matching could be fostered, and communities of practice might be reconstructed in digital spaces~\cite{lave1991situated}. Furthermore, to accommodate diverse individual learning paths, the development of new learning record and competency authentication systems (e.g., digital credentials) using technologies like blockchain is anticipated.

Culturally, a transformation in the societal perception of ``learning'' itself is required. In a society where lifelong learning becomes the norm, AI-Guided Learning provides the technological foundation, but fostering a culture that accepts and utilizes it is essential. This co-evolution of humans and AI also carries risks, such as the degeneration of human cognitive abilities, loss of critical thinking towards AI, and attenuation of human relationships. Therefore, in parallel with technological development, the formulation of ethical guidelines, revision of educational policies, and formation of social consensus are a priority for future work.

\section{Complementarity with LLM-based Systems}
The rapid development of Large Language Models (LLMs) and generative AI is fundamentally expanding the possibilities of educational technology. Advanced LLMs like GPT-4 enable answering complex questions, generating personalized explanations, and even creating educational content. Amid such technological advances, it is important to clarify what unique value the three systems proposed in this research—AIxSpeed, FastPerson, and Profy—possess.

First, it is important that these systems focus on enhancement rather than generation. While LLMs excel at generating new content, existing high-quality educational content (expert lectures, authoritative materials, practitioner demonstrations) contains value that is difficult to reproduce with generative AI. This includes deep domain expertise, structuring of explanations based on years of teaching experience, and subtle nuances in demonstrations.

\subsection{AIxSpeed: Beyond Text-Based Interaction}
The value of AIxSpeed lies in its direct action on the auditory characteristics of audio content, whereas LLMs primarily rely on text-based interaction. While state‑of‑the‑art multimodal LLMs can control players and assist navigation, phoneme‑level, signal‑aware speed optimization is a distinct function; AIxSpeed targets this perceptual stream specifically.

More importantly, AIxSpeed transforms passive consumption into active consumption. Adaptive speed control based on acoustic features maximizes information intake efficiency while minimizing learners' cognitive load. This is an optimization of the learning experience from continuous audio streams that is not directly addressed by LLM‑based dialogue systems. Prior work on time-compressed multimedia and instructional design has shown that playback rate and cognitive load can materially affect learning efficiency and comprehension~\cite{pastore2015using,sweller2019cognitive}, and AIxSpeed's contribution in this domain is difficult to replace with LLMs.

Additionally, many learners engage in concurrent learning (while commuting, exercising, doing housework), and in this context, optimized audio content is more practical than text-based LLM interfaces. AIxSpeed has unique value in technically supporting such learning styles.

\subsection{FastPerson: Preserving Multimodal Richness in the Age of Text Generation}
The uniqueness of FastPerson lies in the integrated processing and preservation of multimodal information. While current LLMs excel mainly at text generation and image/video generation models are developing, it remains difficult to integratively summarize and reconstruct complex multimodal content like lecture videos (speaker explanations, slides, board writing, gestures, demonstrations).

Particularly important is that FastPerson preserves the original speaker's voice and visual context. Mayer et al.~\cite{mayer2003social} found that learners performed better on a transfer test when multimedia narration used a human rather than a machine-synthesized voice. This finding supports the design choice to preserve the original voice instead of replacing it with generic synthesized narration.

Furthermore, FastPerson's interaction design—seamless switching between summary and detail—provides a learning experience that is difficult to achieve with LLM-based systems. Learners can dynamically adjust the granularity of information according to their understanding level, which can be considered a dynamic implementation of scaffolding~\cite{reiser2018scaffolding}. Even if video generation AI becomes capable of generating educational videos including lecture slides and board writing in the future, the value of this interaction design that provides multiple levels of abstraction while maintaining relationships with the original content will be maintained.

Additionally, context preservation of educational content is an important issue. In specialized field lectures, educators' considerations are reflected in terminology usage, concept introduction order, and example selection. FastPerson seeks to retain more of this context than a text-only summary by linking each generated summary back to the original video segments, although preservation of pedagogical intent was not directly evaluated.

\subsection{Profy: Tacit Knowledge and Embodied Learning Beyond Language Models}
The domain of skill acquisition that Profy targets remains challenging for current LLMs. Physical skills like pronunciation belong to the realm of tacit knowledge that is difficult to verbalize, and text-based explanations or instructions yield only limited learning effects. While LLM‑based dialogue can explain pronunciation rules, it does not directly evaluate learners' actual performance or pinpoint specific improvements; robust automatic evaluation remains challenging.

Profy's self-supervised learning approach extracts implicit patterns of proficiency from large amounts of expert data. This is learning support at the perceptual-motor level that goes beyond the linguistic descriptions that LLMs rely on. Recent findings in embodied cognitive science emphasize the importance of perception-action coupling in skill acquisition~\cite{pacheco2019search}, and Profy realizes AI-based support in this domain.

Particularly important is the visual feedback that Profy provides. Highlighting waveform regions emphasized by the classifier and visualizing distances in latent space gives learners more localized information than an overall score. The current system does not identify the phonetic cause or prescribe how to correct it, and it was not compared with LLM-generated text feedback. The immediacy and concreteness of visual feedback can play an important role in motor learning~\cite{sigrist2013augmented}.

Profy is trained on speech from multiple native speakers rather than a single reference speaker. However, the present study did not quantify how well the training data cover acceptable pronunciation variation or evaluate fairness across accents and learner groups; these require explicit bias and coverage audits.

\subsection{Integration and Complementarity with LLM-Based Systems}
Importantly, these systems do not compete with LLMs but have a complementary relationship. Recent multimodal LLMs (e.g., with real‑time audio I/O) can support conversational navigation of content; our systems remain focused on perceptual‑signal‑level optimization (AIxSpeed), source‑grounded multimodal summarization (FastPerson), and embodied skill feedback (Profy). For example, one could interactively explore content efficiently listened to with AIxSpeed through dialogue with LLMs, have LLMs provide additional context or related knowledge about content summarized by FastPerson, or have LLMs linguistically interpret Profy's analysis results to generate personalized practice plans.

Such integration points to the potential of AI‑Guided Learning. By combining the domains where each technology excels—LLMs' linguistic reasoning and generation, and our systems' perceptual processing and enhancement of existing content—more comprehensive and effective learning support can be realized, with uncertainty display and robust error‑handling following human‑AI interaction guidelines~\cite{amershi2019guidelines}. This suggests a future vision of learning support through coordination of specialized AI systems rather than a single omnipotent AI.

\section{Vision and Concluding Synthesis: An AI‑Orchestrated Open Learning Ecosystem Beyond Media Boundaries}
This section integrates the dissertation’s discussion into a single conclusion about the vision of AI‑Guided Learning. Crucially, the scope of AI‑Guided Learning extends far beyond audio and video. AI is conceived as transforming \emph{any} medium or environment—including tools, spaces, events, and everyday practices that are not ordinarily treated as instructional materials—into learnable substrates. It does so by orchestrating contexts and pathways that account for each learner’s interests, goals, current state, and constraints, thereby enabling self‑directed learning. The emphasis is not merely “humans learning from AI,” but AI \emph{shaping the learning ecosystem} so that humans can learn with agency through an optimized, adaptive environment.

\textbf{Vision.} Our long‑term vision is an \emph{open learning ecosystem} in which AI provides \emph{autonomous learning orchestration} and \emph{predictive support} that generalize and continuously optimize the Consume–Understand–Imitate (CUI) cycle across diverse media and environments. This ecosystem democratizes access to expert‑level models, guidance, and objective feedback while preserving learner agency and advancing equity.

\textbf{Generalizing CUI.} We reinterpret \emph{Consume} as access and sensing, \emph{Understand} as multimodal integration and contextualization, and \emph{Imitate} as acting, practicing, and receiving feedback. This generalization applies across knowledge, skills, and dispositions, and reaches beyond audio‑video to physical spaces, devices, social interaction, and data streams—enhancing existing resources while connecting them to the learner’s personal context.

\textbf{Design commitments.} (1) \emph{Agency first:} learners retain control over goals, monitoring, and evaluation; AI assists as coach and orchestrator. (2) \emph{Predictive and adaptive support:} anticipate needs from behavioral, content, and environmental signals with transparency and user control. (3) \emph{Fidelity and provenance:} ground summaries and guidance in sources with uncertainty display to prevent mislearning. (4) \emph{Equity and access:} audit cross‑group validity, calibrate thresholds, and follow universal design to make democratization substantive. (5) \emph{Interoperability and openness:} APIs and standards, auditable learning records, and privacy protection open the ecosystem.

\textbf{From prototypes to an ecosystem.} AIxSpeed, FastPerson, and Profy address time efficiency, multimodal content access, and model-derived feedback, respectively. The next step is integration: standardizing the conversion of resources and environments into learnable form while coordinating with general‑purpose reasoning (e.g., LLMs). Rather than a single omnipotent AI, a federation of specialized AIs could compose learner‑centered pathways.

\textbf{Toward responsible deployment.} We treat ethics, governance, and evaluation—provenance, fairness audits, learning validity, privacy—as first‑class design requirements. Through collaboration among learners, educators, researchers, developers, and policymakers, we can transition toward a human‑centered ecosystem that \emph{democratizes excellence} and substantially reduces time‑to‑proficiency without trading off agency.

\textbf{Closing.} The three implementations lay foundations for AI‑Guided Learning that turns \emph{everything}—across media and environments—into learnable opportunities. Going forward, by centering orchestration and predictive support, extending CUI to the lived world, and coupling enhancement of existing resources with interoperability and equitable access, we move toward a world in which learners design their own learning with AI‑supported, context‑aware pathways.

\chapter{Conclusion}

\section{Summary of Research Contributions}
This doctoral thesis presents a learning support framework, AI-Guided Learning, together with three implementations (AIxSpeed, FastPerson, and Profy) tailored for audio-visual learning. These offer an integrated approach to fundamental challenges—efficient access to long-form content (Consume), understanding of multimodal information (Understand), and acquisition and refinement of practical skills based on observation (Imitate)—by supporting the Consume-Understand-Imitate (CUI) learning cycle.

The academic contributions of this research are organized along four dimensions: theoretical, technical, empirical, and societal.

\subsection{Theoretical Contributions}
First, it presented the three-stage learning model of Consume-Understand-Imitate (CUI), organizing audio‑visual learning processes in a cross‑cutting way and offering a perspective that positions the potential roles of AI support at each stage.

Second, through the concept of AI-Guided Learning, it situated a view of human–AI collaboration in education. Rather than a tool or teacher replacement, AI can be viewed as a cognitive partner that supports learners and helps optimize processes. This provides a basis for connecting distributed cognition and the extended mind with modern AI technology in educational contexts~\cite{hutchins1995cognition, clark1998extended}.

Third, it demonstrated the potential to reinterpret and extend ideals of liberation from institutionalized education and autonomous learning, as raised by thinkers like Ivan Illich, for the contemporary era by leveraging AI technology~\cite{Illich1971}. Concepts such as ``autonomous learning orchestration" and ``predictive learning support" are attempts to concretize this direction.

\subsection{Technical Contributions}
Technically, we designed and implemented three systems corresponding to each stage of the CUI cycle:
(1) AIxSpeed uses ASR confidence as a proxy for listening difficulty and implements phoneme‑level speed adaptation;
(2) FastPerson integrates audio and visual information to summarize lecture videos, preserves the original speaker’s voice in short clips, and allows chapter‑level back‑and‑forth navigation (with consent and clear labelling when synthetic voice is used);
(3) Profy leverages unannotated expert audio and small learner datasets to provide model-derived feedback that highlights classifier-emphasized regions and visualizes distance from model speech.
Together, these implementations strengthen existing content and represent practical contributions in speech processing, multimedia analysis, and educational AI.

\subsection{Empirical Contributions}
Empirically, we evaluated each system via user studies.
AIxSpeed achieved average playback factors of 1.30x for LibriSpeech and 1.29x for UME-ERJ, with higher mean opinion scores than constant-speed playback at the same average factors.
FastPerson reduced viewing time by an average of 53\%, with no statistically significant difference in quiz-measured comprehension from conventional playback.
Profy showed a larger observed improvement in pronunciation intelligibility than elicited imitation, and its pre- and post-practice confidence intervals did not overlap.

\subsection{Societal Contributions}
Societal contributions include: (i) highlighting transparency principles for summarization fidelity and explicit labelling/consent when synthetic voice is used; (ii) emphasizing the need for bias auditing in pronunciation assessment; and (iii) articulating design considerations that lower temporal and cognitive barriers to high‑quality existing content for diverse learners. These provide concrete guidance to advance equitable access to learning opportunities.

\section{Addressing the Research Questions}
This research empirically addressed the Central Research Question (CRQ) and three Specific Research Questions (RQ1-3) presented in the introduction.

\textbf{CRQ: To what extent can adaptive support at each stage of consuming, understanding, and imitating audio-visual content through deep learning technologies achieve reductions in learning time and improvements in learning outcomes?}
This research demonstrated, through systems tailored to each CUI stage, a 53\% average viewing-time reduction with FastPerson, adaptive playback at average factors of approximately 1.3x with AIxSpeed, and an observed improvement in pronunciation intelligibility with Profy.

\textbf{RQ1 (AIxSpeed: Balancing Listening Efficiency and Listenability in Audio Content)}:
Phoneme-level adaptive speed control achieved average playback factors of approximately 1.3x and higher mean opinion scores than constant-speed playback at the same average factors. This supports efficient information consumption (Consume).

\textbf{RQ2 (FastPerson: Efficient Summarization Integrating Visual and Audio Information in Lecture Videos)}:
Video summarization preserving multimodal information reduced viewing time by an average of 53\%, with no statistically significant difference in quiz-measured comprehension from conventional playback; 78\% of FastPerson users rated the ability to switch between summary and detailed content as useful or very useful. This supports content understanding (Understand).

\textbf{RQ3 (Profy: Model-Derived Localized Feedback in Imitation Learning)}:
The feedback system based on self-supervised learning showed a larger observed improvement in pronunciation intelligibility than elicited imitation, and the Profy condition had non-overlapping pre- and post-practice confidence intervals. This supports practical skill acquisition (Imitate) through scalable feedback.

\section{Implications for Theory and Practice}
The theoretical implications of this research extend to learning sciences, cognitive science, and HCI. They include the extension of multimedia learning theory for the AI era, the presentation of a collaborative cognitive model for humans and AI, and new implementation examples of learner-centered design for adaptive interfaces.

Practical implications include concrete examples and design considerations for AI‑leveraged personalized learning support, effective AI‑assisted content design (e.g., fidelity in summarization, explicit labelling/consent for synthetic voice), and operational guidance (e.g., bias audits across diverse learner groups). It also illustrates applications of user‑centered design principles~\cite{norman2013design}; and for policy (e.g., teacher training extending TPACK~\cite{mishra2006technological}).

\section{Limitations and Future Research Directions}
Limitations of this research include: (1) limited diversity and scale of experimental participants, and unverified long-term effects; (2) lack of support for collaborative learning or more complex social learning contexts; (3) limited evaluation of long-term knowledge retention and transfer effects (cf. levels of processing theory~\cite{craik1972levels}), where longitudinal designs incorporating spaced repetition and retrieval practice would be valuable; (4) insufficient consideration of new possibilities (e.g., more interactive feedback) and challenges (e.g., misinformation risks~\cite{bender2021dangers}) arising from the rapid development of generative AI.

Future research should primarily address these limitations. Specifically:
(1) Conducting large-scale, long-term empirical studies with more diverse learner populations.
(2) Extending AI-Guided Learning to collaborative learning environments (e.g., AI-facilitated peer learning).
(3) Developing features for Profy to provide specific corrective guidance, and enhancing semantic understanding and contextual awareness in AIxSpeed and FastPerson.
(4) Advancing personalized and interactive learning support through deeper integration with generative AI, alongside developing risk management methods.
(5) Establishing design and operational guidelines based on ethical AI principles, and researching societal acceptance.
(6) Expanding Profy's framework to support the acquisition of diverse skills beyond pronunciation (e.g., musical instruments, sports, presentations).

\section{Future Vision: The Evolution of AI-Guided Learning Ecosystems}
We articulate a future in which AI transforms any medium or environment into a learnable substrate and orchestrates individualized pathways that respect each learner’s interests, goals, and state. Extending beyond audio‑visual resources, AI reframes tools, spaces, events, and everyday practices as contexts for learning while continuously optimizing the Consume–Understand–Imitate cycle—reinterpreted as access and sensing, multimodal integration and contextualization, and acting, practicing, and receiving feedback. Rather than a single omnipotent system, this ecosystem is composed by coordinating specialized AIs—exemplified by AIxSpeed, FastPerson, and Profy—with general‑purpose reasoning to assemble learner‑centered routes.

The evolution of this ecosystem rests on treating provenance and fidelity, auditable fairness, interoperability, and privacy as first‑class design requirements. Learners traverse from summaries to sources with minimal friction, understand the uncertainty of feedback, and modulate the degree of AI intervention to fit their aims and constraints. Educators and developers align on demonstrable outcomes such as reductions in time‑to‑proficiency and improvements in retention and transfer on real‑world tasks, while communities connect through portable learning records enriched with context. In this future, AI is not primarily a lecturer but an orchestrator of environments, uniting enhancement of existing resources with broadened and equitable access.

\section{Final Reflections}
This dissertation has shown, through AIxSpeed, FastPerson, and Profy, that we can advance CUI in implementable ways that enhance existing resources and reconcile time efficiency with comprehension and skill acquisition without eroding learner agency. It also reframed AI as a cognitive partner and orchestrator, offering a view in which learning emerges across human–AI–environment couplings. The vision of AI‑Guided Learning articulated here moves beyond audio and video to extend CUI to the lived world, building an open ecosystem that renders diverse media and environments learnable.

Realizing this vision requires not only technical progress but also embedding principles of ethics, governance, and evaluation into design and advancing deployment through collaboration among learners, educators, researchers, developers, and policymakers. The aim is to democratize access to excellence and materially shorten the time to proficiency without trading off agency or equity. We hope that the contributions and limitations identified here, together with the consolidated vision, serve as a starting point for the next wave of research and practice toward making learning a more autonomous and flourishing endeavor for everyone.

\bibliographystyle{plain}
\bibliography{myref_intro, myref_related_work, myref_methodology, myref_discussion, myref_conclusion}

\begin{thebibliography}{100}

\bibitem{almusharraf2024capt}
Asma Almusharraf, Hassan~Saleh Mahdi, Haifa Al-Nofaie, and Amal Aljasser.
\newblock The effect of settings, educational level and tools on
  computer-assisted pronunciation training: A meta-analysis.
\newblock {\em Journal of Computer Assisted Learning}, 40(4):1605--1615, 2024.

\bibitem{alpert1970advances}
Daniel Alpert and Donald~Lester Bitzer.
\newblock Advances in computer-based education: The {Plato} program will
  provide a major test of the educational and economic feasibility of this
  medium.
\newblock {\em Science}, 167(3925):1582--1590, 1970.

\bibitem{amershi2019guidelines}
Saleema Amershi, Dan Weld, Mihaela Vorvoreanu, Adam Fourney, Besmira Nushi,
  Penny Collisson, Jina Suh, Shamsi Iqbal, Paul~N. Bennett, Kori Inkpen, Jaime
  Teevan, Ruth Kikin-Gil, and Eric Horvitz.
\newblock Guidelines for human-{AI} interaction.
\newblock In {\em Proceedings of the 2019 CHI Conference on Human Factors in
  Computing Systems}, pages 1--13, 2019.

\bibitem{amrate2024capt}
Moustafa Amrate and Pi-hua Tsai.
\newblock Computer-assisted pronunciation training: A systematic review.
\newblock {\em ReCALL}, 37(1):22--42, 2025.
\newblock Published online 19 September 2024.

\bibitem{Anderson1985LISP}
John~R. Anderson and Brian~J. Reiser.
\newblock The {LISP} tutor.
\newblock {\em BYTE}, 10(4):159--175, 1985.

\bibitem{anderson2001revised}
Lorin~W. Anderson and David~R. Krathwohl, editors.
\newblock {\em A Taxonomy for Learning, Teaching, and Assessing: A Revision of
  Bloom's Taxonomy of Educational Objectives}.
\newblock Longman, New York, 2001.

\bibitem{Apostolidis2021}
Evlampios Apostolidis, Eleni Adamantidou, Alexandros~I. Metsai, Vasileios
  Mezaris, and Ioannis Patras.
\newblock Video summarization using deep neural networks: A survey.
\newblock {\em Proceedings of the IEEE}, 109(11):1838--1863, 2021.

\bibitem{Ardila2019}
Rosana Ardila, Megan Branson, Kelly Davis, Michael Kohler, Josh Meyer, Michael
  Henretty, Reuben Morais, Lindsay Saunders, Francis Tyers, and Gregor Weber.
\newblock Common voice: {A} massively-multilingual speech corpus.
\newblock In {\em Proceedings of the Twelfth Language Resources and Evaluation
  Conference}, pages 4218--4222. European Language Resources Association, 2020.

\bibitem{Arik2017}
Sercan~\"{O}. Arik, Mike Chrzanowski, Adam Coates, Gregory Diamos, Andrew
  Gibiansky, Yongguo Kang, Xian Li, John Miller, Andrew Ng, Jonathan Raiman,
  Shubho Sengupta, and Mohammad Shoeybi.
\newblock Deep voice: Real-time neural text-to-speech.
\newblock In {\em Proceedings of the International Conference on Machine
  Learning}, volume~70, pages 195--204. PMLR, 2017.

\bibitem{arnab2021vivit}
Anurag Arnab, Mostafa Dehghani, Georg Heigold, Chen Sun, Mario Lu{\v{c}}i{\'c},
  and Cordelia Schmid.
\newblock {ViViT}: A video vision transformer.
\newblock In {\em Proceedings of the IEEE/CVF International Conference on
  Computer Vision}, pages 6816--6826, 2021.

\bibitem{Barry1997}
Barry Arons.
\newblock {SpeechSkimmer}: A system for interactively skimming recorded speech.
\newblock {\em ACM Transactions on Computer-Human Interaction}, 4(1):3--38,
  1997.

\bibitem{Azevedo2004AdaptiveScaffolding}
Roger Azevedo, Jennifer~G. Cromley, and Diane Seibert.
\newblock Does adaptive scaffolding facilitate students' ability to regulate
  their learning with hypermedia?
\newblock {\em Contemporary Educational Psychology}, 29(3):344--370, 2004.

\bibitem{baevski2020wav2vec}
Alexei Baevski, Yuhao Zhou, Abdelrahman Mohamed, and Michael Auli.
\newblock wav2vec 2.0: A framework for self-supervised learning of speech
  representations.
\newblock In {\em Advances in Neural Information Processing Systems},
  volume~33, pages 12449--12460. Curran Associates, Inc., 2020.

\bibitem{Bahdanau2014}
Dzmitry Bahdanau, Kyunghyun Cho, and Yoshua Bengio.
\newblock Neural machine translation by jointly learning to align and
  translate.
\newblock In {\em Proceedings of the International Conference on Learning
  Representations}, 2015.

\bibitem{Zx21}
Zhongxin Bai and Xiao-Lei Zhang.
\newblock Speaker recognition based on deep learning: An overview.
\newblock {\em Neural Networks}, 140:65--99, 2021.

\bibitem{bandura1977social}
Albert Bandura.
\newblock {\em Social learning theory}.
\newblock Prentice Hall, 1977.

\bibitem{bandura1997self}
Albert Bandura.
\newblock {\em Self-efficacy: The exercise of control}.
\newblock W. H. Freeman, 1997.

\bibitem{barad2007meeting}
Karen Barad.
\newblock {\em Meeting the universe halfway: Quantum physics and the
  entanglement of matter and meaning}.
\newblock Duke University Press, 2007.

\bibitem{basyal2023}
Lochan Basyal and Mihir Sanghvi.
\newblock Text summarization using large language models: A comparative study
  of {MPT-7b-instruct}, {Falcon-7b-instruct}, and {OpenAI Chat-GPT} models.
\newblock {\em arXiv}, 2023.

\bibitem{beauchemin2024enhancing}
No{\'e}mie Beauchemin, Patrick Charland, Alexander Karran, Jared Boasen, Bella
  Tadson, Sylvain S{\'e}n{\'e}cal, and Pierre-Majorique L{\'e}ger.
\newblock Enhancing learning experiences: {EEG}-based passive {BCI} system
  adapts learning speed to cognitive load in real-time, with motivation as
  catalyst.
\newblock {\em Frontiers in Human Neuroscience}, 18:1416683, 2024.

\bibitem{bender2021dangers}
Emily~M. Bender, Timnit Gebru, Angelina McMillan-Major, and Margaret Mitchell.
\newblock On the dangers of stochastic parrots: Can language models be too big?
\newblock In {\em Proceedings of the ACM Conference on Fairness,
  Accountability, and Transparency}, pages 610--623. Association for Computing
  Machinery, 2021.

\bibitem{benedetto2024abstractive}
Irene Benedetto, Moreno La~Quatra, Luca Cagliero, Lorenzo Canale, and Laura
  Farinetti.
\newblock Abstractive video lecture summarization: Applications and future
  prospects.
\newblock {\em Education \& Information Technologies}, 29(3):2951--2971, 2024.

\bibitem{Betihavas2016}
Vasiliki Betihavas, Heather Bridgman, Rachel Kornhaber, and Merylin Cross.
\newblock The evidence for 'flipping out': A systematic review of the flipped
  classroom in nursing education.
\newblock {\em Nurse Education Today}, 38:15--21, 2016.

\bibitem{Bitzer1961PLATO}
Donald~L. Bitzer, Peter~G. Braunfeld, and Wayne~W. Lichtenberger.
\newblock {PLATO}: An automatic teaching device.
\newblock {\em IRE Transactions on Education}, 4(4):157--161, 1961.

\bibitem{Duolingo2024}
Cindy Blanco.
\newblock {Duolingo} language report 2024, 2024.
\newblock Retrieved May 25, 2025, from
  https://blog.duolingo.com/2024-duolingo-language-report/.

\bibitem{bloom19842}
Benjamin~S. Bloom.
\newblock The 2 sigma problem: The search for methods of group instruction as
  effective as one-to-one tutoring.
\newblock {\em Educational Researcher}, 13(6):4--16, 1984.

\bibitem{bourdieu1977outline}
Pierre Bourdieu.
\newblock {\em Outline of a theory of practice}.
\newblock Cambridge University Press, 1977.

\bibitem{brame2016effective}
Cynthia~J. Brame.
\newblock Effective educational videos: Principles and guidelines for
  maximizing student learning from video content.
\newblock {\em CBE—Life Sciences Education}, 15(4):es6, 2016.

\bibitem{bu2021pteacher}
Yaohua Bu, Tianyi Ma, Weijun Li, Hang Zhou, Jia Jia, Shengqi Chen, Kaiyuan Xu,
  Dachuan Shi, Haozhe Wu, Zhihan Yang, Kun Li, Zhiyong Wu, Yuanchun Shi, Xiaobo
  Lu, and Ziwei Liu.
\newblock {PTeacher}: A computer-aided personalized pronunciation training
  system with exaggerated audio-visual corrective feedback.
\newblock In {\em Proceedings of the CHI Conference on Human Factors in
  Computing Systems}, pages 1--14, 2021.

\bibitem{xtts2024}
Edresson Casanova, Kelly Davis, Eren G{\"o}lge, G{\"o}rkem G{\"o}knar, Iulian
  Gulea, Logan Hart, Aya Aljafari, Joshua Meyer, Reuben Morais, Samuel Olayemi,
  and Julian Weber.
\newblock {XTTS}: A massively multilingual zero-shot text-to-speech model.
\newblock In {\em Proceedings of INTERSPEECH 2024}, pages 4978--4982, 2024.

\bibitem{Chalmers2003Seamful}
Matthew Chalmers and Ian MacColl.
\newblock Seamful and seamless design in ubiquitous computing.
\newblock In {\em Workshop on At the Crossroads: The Interaction of HCI and
  Systems Issues in UbiComp (UbiComp 2003 Workshop)}, 2003.
\newblock Position paper.

\bibitem{chen2024effect}
Ashley Chen, Suchita~E. Kumar, Rhea Varkhedi, and Dillon~H. Murphy.
\newblock The effect of playback speed and distractions on the comprehension of
  audio and audio-visual materials.
\newblock {\em Educational Psychology Review}, 36:79, 2024.

\bibitem{Chen2013}
Liang Chen, Yipeng Zhou, and Dah~Ming Chiu.
\newblock Video browse - a study of user behavior in online {VoD} services.
\newblock In {\em Proceedings of the International Conference on Computer
  Communication and Networks}, pages 1--7, 2013.

\bibitem{ec2022systematic}
Xu~Chen, Jie Mu, and Tingting Zhang.
\newblock Computer assisted pronunciation training {(CAPT):} {A} systematic
  review of studies from 2012 to 2021.
\newblock In {\em Proceedings of the International Conference on Computers in
  Education}, pages 575--580. Asia-Pacific Society for Computers in Education,
  2022.

\bibitem{chen2024multipa}
Yu-Wen Chen, Zhou Yu, and Julia Hirschberg.
\newblock {MultiPA}: A multi-task speech pronunciation assessment model for
  open response scenarios.
\newblock In {\em Proceedings of INTERSPEECH}, pages 297--301, 2024.

\bibitem{chi2014icap}
Michelene~TH Chi and Ruth Wylie.
\newblock The {ICAP} framework: Linking cognitive engagement to active learning
  outcomes.
\newblock {\em Educational Psychologist}, 49(4):219--243, 2014.

\bibitem{chi2012}
Pei-Yu Chi, Sally Ahn, Amanda Ren, Mira Dontcheva, Wilmot Li, and Bj{\"o}rn
  Hartmann.
\newblock {MixT}: Automatic generation of step-by-step mixed media tutorials.
\newblock In {\em Proceedings of the Annual ACM Symposium on User Interface
  Software and Technology}, pages 93--102, 2012.

\bibitem{Choi2024VIVID}
Seulgi Choi, Hyewon Lee, Yoonjoo Lee, and Juho Kim.
\newblock {VIVID}: Human-{AI} collaborative authoring of vicarious dialogues
  from lecture videos.
\newblock In {\em Proceedings of the CHI Conference on Human Factors in
  Computing Systems}, pages 1--26. Association for Computing Machinery, 2024.

\bibitem{Chorowski2015}
Jan Chorowski, Dzmitry Bahdanau, Dmitriy Serdyuk, KyungHyun Cho, and Yoshua
  Bengio.
\newblock Attention-based models for speech recognition.
\newblock In {\em Proceedings of the International Conference on Neural
  Information Processing Systems}, pages 577--585, 2015.

\bibitem{clark1998extended}
Andy Clark and David Chalmers.
\newblock The extended mind.
\newblock {\em Analysis}, 58(1):7--19, 1998.

\bibitem{Clarke2020ReactiveVideo}
Christopher Clarke, Larissa Pschetz, Mor Trope, and Dave Murray-Rust.
\newblock Reactive video: Adaptive video playback based on user motion for
  supporting physical activity.
\newblock In {\em Proceedings of the International Conference on Intelligent
  User Interfaces}, pages 196--208, 2020.

\bibitem{conrad2021measuring}
Colin Conrad and Aaron Newman.
\newblock Measuring mind wandering during online lectures assessed with {EEG}.
\newblock {\em Frontiers in Human Neuroscience}, 15:697532, 2021.

\bibitem{conrad1989timecompressed}
Linda Conrad.
\newblock The effects of time-compressed speech on native and {EFL} listening
  comprehension.
\newblock {\em Studies in Second Language Acquisition}, 11(1):1--16, 1989.

\bibitem{craik1972levels}
Fergus I.~M. Craik and Robert~S. Lockhart.
\newblock Levels of processing: A framework for memory research.
\newblock {\em Journal of Verbal Learning and Verbal Behavior}, 11(6):671--684,
  1972.

\bibitem{crompton2023artificial}
Helen Crompton and Diane Burke.
\newblock Artificial intelligence in higher education: The state of the field.
\newblock {\em International Journal of Educational Technology in Higher
  Education}, 20(1):22, 2023.

\bibitem{Dear2017PLATO}
Brian Dear.
\newblock {\em The friendly orange glow: The untold story of the {PLATO} system
  and the dawn of cyberculture}.
\newblock Pantheon Books, 2017.

\bibitem{devlin-etal-2019-bert}
Jacob Devlin, Ming-Wei Chang, Kenton Lee, and Kristina Toutanova.
\newblock {BERT}: Pre-training of deep bidirectional transformers for language
  understanding.
\newblock In {\em Proceedings of the Conference of the North {A}merican Chapter
  of the Association for Computational Linguistics: Human Language
  Technologies, Volume 1 (Long and Short Papers)}, pages 4171--4186.
  Association for Computational Linguistics, 2019.

\bibitem{Dillahunt2014}
Tawanna~R. Dillahunt, Zengguang Wang, and Stephanie~D. Teasley.
\newblock Democratizing higher education: Exploring {MOOC} use among those who
  cannot afford a formal education.
\newblock {\em The International Review of Research in Open and Distributed
  Learning}, 15(5):177--196, 2014.

\bibitem{Dragicevic2008VideoManipulation}
Pierre Dragicevic, Gonzalo Ramos, Jacobo Bibliowitcz, Derek Nowrouzezahrai,
  Ravin Balakrishnan, and Karan Singh.
\newblock Video browse by direct manipulation.
\newblock In {\em Proceedings of the CHI Conference on Human Factors in
  Computing Systems}, pages 237--246, 2008.

\bibitem{Drucker2002SmartSkip}
Steven~M. Drucker, Asta Glatzer, Steven De~Mar, and Curtis Wong.
\newblock {SmartSkip}: Consumer level browse and skipping of digital video
  content.
\newblock In {\em Proceedings of the CHI Conference on Human Factors in
  Computing Systems}, pages 219--226, 2002.

\bibitem{Duan2019}
Songshuang Duan and Xiaoqian Chen.
\newblock Why college students watch streaming drama at higher playback speed:
  The uses and gratifications perspective.
\newblock In {\em Proceedings of the International Joint Conference on
  Information, Media and Engineering}, 2019.

\bibitem{ericsson1993role}
K.~Anders Ericsson, Ralf~T. Krampe, and Clemens Tesch-R{\"o}mer.
\newblock The role of deliberate practice in the acquisition of expert
  performance.
\newblock {\em Psychological Review}, 100(3):363--406, 1993.

\bibitem{evidenceforlearning2021metacognition}
{Evidence for Learning}.
\newblock Metacognition and self-regulation.
\newblock Teaching and Learning Toolkit, 2021.
\newblock Review last updated July 2021.

\bibitem{feng2025aial}
Lijuan Feng.
\newblock Investigating the effects of artificial intelligence-assisted
  language learning strategies on cognitive load and learning outcomes: A
  comparative study.
\newblock {\em Journal of Educational Computing Research}, 62(8):1741--1774,
  2025.
\newblock First published online 31 August 2024; issue published January 2025.

\bibitem{flege1997experience}
James~Emil Flege, Ocke-Schwen Bohn, and Sunyoung Jang.
\newblock Effects of experience on non-native speakers' production and
  perception of {English} vowels.
\newblock {\em Journal of Phonetics}, 25(4):437--470, 1997.

\bibitem{Lionel2017}
Lionel Fontan, Isabelle Ferran{\'e}, J{\'e}r{\^o}me Farinas, Julien Pinquier,
  Julien Tardieu, Cynthia Magnen, Pascal Gaillard, Xavier Aumont, and Christian
  F{\"u}llgrabe.
\newblock Automatic speech recognition predicts speech intelligibility and
  comprehension for listeners with simulated age-related hearing loss.
\newblock {\em Journal of Speech, Language, and Hearing Research},
  60(9):2394--2405, 2017.

\bibitem{Fouse2011ChronoViz}
Adam Fouse, Nadir Weibel, Edwin Hutchins, and James~D. Hollan.
\newblock {ChronoViz}: A system for supporting navigation of time-coded data.
\newblock In {\em Proceedings of the CHI Conference on Human Factors in
  Computing Systems Extended Abstracts}, pages 299--304, 2011.

\bibitem{freire1970pedagogy}
Paulo Freire.
\newblock {\em Pedagogy of the oppressed}.
\newblock Seabury Press, 1970.

\bibitem{gabay2017perceptual}
Yafit Gabay, Avi Karni, and Karen Banai.
\newblock The perceptual learning of time-compressed speech: A comparison of
  training protocols with different levels of difficulty.
\newblock {\em {PLoS ONE}}, 12(5):e0176488, 2017.

\bibitem{Garofolo1993}
J.~S. Garofolo, L.~F. Lamel, W.~M. Fisher, J.~G. Fiscus, D.~S. Pallett, and
  N.~L. Dahlgren.
\newblock {DARPA TIMIT} acoustic phonetic continuous speech corpus {CDROM}.
\newblock Technical Report NISTIR-4930, National Institute of Standards and
  Technology (NIST), 1993.

\bibitem{Giegerich1992}
Heinz~J. Giegerich.
\newblock {\em {English} phonology: An introduction}.
\newblock Cambridge Textbooks in Linguistics. Cambridge University Press, 1992.

\bibitem{Goodman1987HyperCard}
Danny Goodman.
\newblock {\em The complete {HyperCard} handbook}.
\newblock Bantam Books, 1987.

\bibitem{Graves:06icml}
A.~Graves, S.~Fernandez, F.~Gomez, and J.~Schmidhuber.
\newblock Connectionist temporal classification: Labelling unsegmented sequence
  data with recurrent neural nets.
\newblock In {\em Proceedings of the International Conference on Machine
  Learning}, 2006.

\bibitem{Graves2013}
Alex Graves, Abdel-rahman Mohamed, and Geoffrey Hinton.
\newblock Speech recognition with deep recurrent neural networks.
\newblock In {\em Proceedings of the International Conference on Acoustics,
  Speech and Signal Processing}, pages 6645--6649, 2013.

\bibitem{gui2023survey}
Jie Gui, Tuo Chen, Jing Zhang, Qiong Cao, Zhenan Sun, Hao Luo, and Dacheng Tao.
\newblock A survey on self-supervised learning: Algorithms, applications, and
  future trends.
\newblock {\em IEEE Transactions on Pattern Analysis and Machine Intelligence},
  46(12):9052--9071, 2024.

\bibitem{Guo2014}
Philip~J. Guo, Juho Kim, and Rob Rubin.
\newblock How video production affects student engagement: An empirical study
  of {MOOC} videos.
\newblock In {\em Proceedings of the ACM Conference on Learning at Scale},
  pages 41--50. ACM, 2014.

\bibitem{Habgood2011Intrinsic}
M.~P.~J. Habgood and S.~E. Ainsworth.
\newblock Motivating children to learn effectively: Exploring the value of
  intrinsic integration in educational games.
\newblock {\em The Journal of the Learning Sciences}, 20(2):169--206, 2011.

\bibitem{hannun2014}
Awni Hannun, Carl Case, Jared Casper, Bryan Catanzaro, Greg Diamos, Erich
  Elsen, Ryan Prenger, Sanjeev Satheesh, Shubho Sengupta, Adam Coates, and
  Andrew~Y. Ng.
\newblock Deep speech: Scaling up end-to-end speech recognition.
\newblock {\em arXiv}, 2014.

\bibitem{hao2024collaborative}
Xinyue Hao, Emrah Demir, and Daniel Eyers.
\newblock Exploring collaborative decision-making: A quasi-experimental study
  of human and generative {AI} interaction.
\newblock {\em Technology in Society}, 78:102662, 2024.

\bibitem{Hardison2004VisualFeedback}
Debra~M. Hardison.
\newblock Generalization of computer-assisted prosody training: Quantitative
  and qualitative findings.
\newblock {\em Language Learning \& Technology}, 8(1):34--52, 2004.

\bibitem{hartshorne2018critical}
Joshua~K. Hartshorne, Joshua~B. Tenenbaum, and Steven Pinker.
\newblock A critical period for second language acquisition: Evidence from 2/3
  million {English} speakers.
\newblock {\em Cognition}, 177:263--277, 2018.

\bibitem{Hermann2015}
Karl~Moritz Hermann, Tom{\'a}{\v s} Ko{\v c}isk{\'y}, Edward Grefenstette,
  Lasse Espeholt, Will Kay, Mustafa Suleyman, and Phil Blunsom.
\newblock Teaching machines to read and comprehend.
\newblock In {\em Proceedings of the International Conference on Neural
  Information Processing Systems}, pages 1693--1701, 2015.

\bibitem{Hew2014}
Khe~Foon Hew and Wing~Sum Cheung.
\newblock Students' and instructors' use of massive open online courses
  ({MOOCs}): Motivations and challenges.
\newblock {\em Educational Research Review}, 12:45--58, 2014.

\bibitem{Higuchi2017}
Keita Higuchi, Ryo Yonetani, and Yoichi Sato.
\newblock {EgoScanning}: Quickly scanning first-person videos with egocentric
  elastic timelines.
\newblock In {\em Proceedings of the CHI Conference on Human Factors in
  Computing Systems}, pages 6536--6546. Association for Computing Machinery,
  2017.

\bibitem{Elad2015}
Elad Hoffer and Nir Ailon.
\newblock Deep metric learning using triplet network.
\newblock {\em Similarity-Based Pattern Recognition}, pages 84--92, 2015.

\bibitem{holstein2022designing}
Kenneth Holstein and Vincent Aleven.
\newblock Designing for human-{AI} complementarity in {K-12} education.
\newblock {\em AI Magazine}, 43(2):239--248, 2022.

\bibitem{honneth1995struggle}
Axel Honneth.
\newblock {\em The struggle for recognition: The moral grammar of social
  conflicts}.
\newblock MIT Press, 1995.

\bibitem{hori2017attention}
Chiori Hori, Takaaki Hori, Teng-Yok Lee, Ziming Zhang, Bret Harsham, John~R.
  Hershey, Tim~K. Marks, and Kazuhiko Sumi.
\newblock Attention-based multimodal fusion for video description.
\newblock In {\em Proceedings of the IEEE International Conference on Computer
  Vision}, pages 4193--4202, 2017.

\bibitem{Horvitz1999Principles}
Eric Horvitz.
\newblock Principles of mixed-initiative user interfaces.
\newblock In {\em Proceedings of the CHI Conference on Human Factors in
  Computing Systems}, pages 159--166, 1999.

\bibitem{Hsu2021a}
Wei-Ning Hsu, Benjamin Bolte, Yao-Hung~Hubert Tsai, Kushal Lakhotia, Ruslan
  Salakhutdinov, and Abdelrahman Mohamed.
\newblock {HuBERT}: Self-supervised speech representation learning by masked
  prediction of hidden units.
\newblock {\em IEEE/ACM Transactions on Audio, Speech and Language Processing},
  29:3451--3460, 2021.

\bibitem{hutchins1995cognition}
Edwin Hutchins.
\newblock {\em Cognition in the {W}ild}.
\newblock MIT Press, 1995.

\bibitem{Illich1971}
Ivan Illich.
\newblock {\em Deschooling society}.
\newblock Harper \& Row, 1971.

\bibitem{inie2023designing}
Nanna Inie, Jeanette Falk, and Steve Tanimoto.
\newblock Designing participatory {AI}: Creative professionals’ worries and
  expectations about generative {AI}.
\newblock In {\em Proceedings of the CHI Conference on Human Factors in
  Computing Systems Extended Abstracts}, pages 1--8, 2023.

\bibitem{Ishii1997TangibleBits}
Hiroshi Ishii and Brygg Ullmer.
\newblock Tangible bits: Towards seamless interfaces between people, bits and
  atoms.
\newblock In {\em Proceedings of the CHI Conference on Human Factors in
  Computing Systems}, pages 234--241, 1997.

\bibitem{janse2004word}
Esther Janse.
\newblock Word perception in fast speech: Artificially time-compressed vs.
  naturally produced fast speech.
\newblock {\em Speech Communication}, 42:155--173, 2004.

\bibitem{Jiang2002}
Wenyu Jiang and Henning Schulzrinne.
\newblock Speech recognition performance as an effective perceived quality
  predictor.
\newblock In {\em Proceedings of the Tenth IEEE International Workshop on
  Quality of Service}, pages 269--275. IEEE, 2002.

\bibitem{jorgens2024curriculum}
Michelle~Celine J{"o}rgens, Florian Beier, Sebastian Kreibich, and Dirk Werth.
\newblock Using {ChatGPT} for course curriculum design: A systematic review.
\newblock In {\em The Paris Conference on Education 2024: Official Conference
  Proceedings}, pages 549--561, 2024.

\bibitem{Kabir2021ASO}
Muhammad~Mohsin Kabir, Muhammad~Firoz Mridha, Jungpil Shin, Israt Jahan, and
  Abu~Quwsar Ohi.
\newblock A survey of speaker recognition: Fundamental theories, recognition
  methods and opportunities.
\newblock {\em IEEE Access}, 9:79236--79263, 2021.

\bibitem{kalyuga2003expertise}
Slava Kalyuga, Paul Ayres, Paul Chandler, and John Sweller.
\newblock The expertise reversal effect.
\newblock {\em Educational Psychologist}, 38(1):23--31, 2003.

\bibitem{kasch2021educational}
Julia Kasch, Peter Van~Rosmalen, and Marco Kalz.
\newblock Educational scalability in {MOOCs}: Analysing instructional designs
  to find best practices.
\newblock {\em Computers \& Education}, 161:104054, 2021.

\bibitem{kasneci2023chatgpt}
Enkelejda Kasneci, Kathrin Se{\ss}ler, Stefan K{\"u}chemann, Maria Bannert,
  Daryna Dementieva, Frank Fischer, Urs Gasser, Georg Groh, Stephan
  G{\"u}nnemann, Eyke H{\"u}llermeier, Stephan Krusche, Gitta Kutyniok, Tilman
  Michaeli, Claudia Nerdel, J{\"u}rgen Pfeffer, Oleksandra Poquet, Michael
  Sailer, Albrecht Schmidt, Tina Seidel, Matthias Stadler, Jochen Weller,
  Jochen Kuhn, and Gjergji Kasneci.
\newblock {ChatGPT} for good? on opportunities and challenges of large language
  models for education.
\newblock {\em Learning and Individual Differences}, 103:102274, 2023.

\bibitem{Kawamura2014}
Shunya Kawamura, Tsukasa Fukusato, Tatsunori Hirai, and Shigeo Morishima.
\newblock Efficient video viewing system for racquet sports with automatic
  summarization focusing on rally scenes.
\newblock In {\em Proceedings of the ACM SIGGRAPH Posters}, 2014.

\bibitem{Kay2012}
Robin~H. Kay.
\newblock Exploring the use of video podcasts in education: A comprehensive
  review of the literature.
\newblock {\em Computers in Human Behavior}, 28(3):820--831, 2012.

\bibitem{Khosravi2022Explainable}
Hassan Khosravi, Simon~Buckingham Shum, Guanliang Chen, Cristina Conati,
  Yi-Shan Tsai, Judy Kay, Simon Knight, Roberto Martinez-Maldonado, Shazia
  Sadiq, and Dragan Ga{\v{s}}evi{\'c}.
\newblock Explainable artificial intelligence in education: A comprehensive
  review.
\newblock {\em Computers and Education: Artificial Intelligence}, 3:100074,
  2022.

\bibitem{Kiili2005Flow}
Kristian Kiili.
\newblock Digital game-based learning: Towards an experiential gaming model.
\newblock {\em The Internet and Higher Education}, 8(1):13--24, 2005.

\bibitem{kim2022automatic}
Eesung Kim, Jae-Jin Jeon, Hyeji Seo, and Hoon Kim.
\newblock Automatic pronunciation assessment using self-supervised speech
  representation learning.
\newblock In {\em Proceedings of INTERSPEECH}, pages 1411--1415. ISCA, 2022.

\bibitem{Kim2016}
Hyun~Hee Kim and Yong~Ho Kim.
\newblock Generic speech summarization of transcribed lecture videos: Using
  tags and their semantic relations.
\newblock {\em Journal of the Association for Information Science and
  Technology}, 67(2):366--379, 2016.

\bibitem{kim2014understanding}
Juho Kim, Philip~J. Guo, Daniel~T. Seaton, Piotr Mitros, Krzysztof~Z. Gajos,
  and Robert~C. Miller.
\newblock Understanding in-video dropouts and interaction peaks in online
  lecture videos.
\newblock In {\em Proceedings of the First ACM Conference on Learning at
  Scale}, pages 31--40. ACM, 2014.

\bibitem{Diederik2015}
Diederik~P. Kingma and Jimmy Ba.
\newblock Adam: A method for stochastic optimization.
\newblock In {\em Proceedings of the International Conference on Learning
  Representations}, 2015.

\bibitem{kolb1984experiential}
David~A. Kolb.
\newblock {\em Experiential Learning: Experience as the Source of Learning and
  Development}.
\newblock Prentice-Hall, Englewood Cliffs, NJ, 1984.

\bibitem{Kulik2016Effectiveness}
James~A. Kulik and J.~D. Fletcher.
\newblock Effectiveness of intelligent tutoring systems: A meta-analytic
  review.
\newblock {\em Review of Educational Research}, 86(1):42--78, 2016.

\bibitem{Kurihara2011}
Kazutaka Kurihara.
\newblock {CinemaGazer}: A system for watching video at very high speed.
\newblock In {\em Proceedings of the Workshop on Advanced Visual Interfaces},
  2012.

\bibitem{Lang2020}
David Lang, Guanliang Chen, Kathy Mirzaei, and Andreas Paepcke.
\newblock Is faster better? a study of video playback speed.
\newblock In {\em Proceedings of the International Conference on Learning
  Analytics \& Knowledge}, pages 260--269. Association for Computing Machinery,
  2020.

\bibitem{lave1991situated}
Jean Lave and Etienne Wenger.
\newblock {\em Situated learning: Legitimate peripheral participation}.
\newblock Cambridge University Press, 1991.

\bibitem{lei2021moment}
Jie Lei, Tamara~L. Berg, and Mohit Bansal.
\newblock Detecting moments and highlights in videos via natural language
  queries.
\newblock In {\em Advances in Neural Information Processing Systems},
  volume~34, pages 11846--11858, 2021.

\bibitem{Li2016}
Wei Li, Kehuang Li, Sabato~Marco Siniscalchi, Nancy~F. Chen, and Chin-Hui Lee.
\newblock Detecting mispronunciations of {L2} learners and providing corrective
  feedback using knowledge-guided and data-driven decision trees.
\newblock In {\em Proceedings of INTERSPEECH}, 2016.

\bibitem{li2024enhancing}
Xueyi Li, Youheng Bai, Teng Guo, Zitao Liu, Yaying Huang, Xiangyu Zhao, Feng
  Xia, Weiqi Luo, and Jian Weng.
\newblock Enhancing length generalization for attention based knowledge tracing
  models with linear biases.
\newblock In {\em Proceedings of the Thirty-Third International Joint
  Conference on Artificial Intelligence}, pages 5918--5926. International Joint
  Conferences on Artificial Intelligence Organization, 2024.
\newblock Official IJCAI article 654; the DOI registry currently returns
  metadata for a different paper for 10.24963/ijcai.2024/654.

\bibitem{Lin2017}
Tsung-Yi Lin, Priya Goyal, Ross Girshick, Kaiming He, and Piotr Doll{\'a}r.
\newblock Focal loss for dense object detection.
\newblock {\em Proceedings of the IEEE International Conference on Computer
  Vision}, pages 2980--2988, 2017.

\bibitem{Shuo2022}
Shuo Liu, Adria Mallol-Ragolta, Emilia Parada-Cabaleiro, Kun Qian, Xin Jing,
  Alexander Kathan, Bin Hu, and Bjoern~W Schuller.
\newblock Audio self-supervised learning: A survey.
\newblock {\em Patterns}, 3(12):100616, 2022.

\bibitem{loshchilov2018decoupled}
Ilya Loshchilov and Frank Hutter.
\newblock Decoupled weight decay regularization.
\newblock In {\em Proceedings of the 7th International Conference on Learning
  Representations}, 2019.

\bibitem{Low2008Modality}
Renae Low and John Sweller.
\newblock The modality principle in multimedia learning.
\newblock In Richard~E. Mayer, editor, {\em The Cambridge Handbook of
  Multimedia Learning (2nd ed.)}, pages 227--246. Cambridge University Press,
  2014.

\bibitem{Mishaim21}
Mishaim Malik, Muhammad Malik, Khawar Mehmood, and Imran Makhdoom.
\newblock Automatic speech recognition: A survey.
\newblock {\em Multimedia Tools and Applications}, 80:9411--9457, 2021.

\bibitem{Martin2020}
Florence Martin, Ting Sun, and Carl~D. Westine.
\newblock A systematic review of research on online teaching and learning from
  2009 to 2018.
\newblock {\em Computers \& Education}, 159:104009, 2020.

\bibitem{martini2020wmc}
Markus Martini, Robert Marhenke, Caroline Martini, Sonja Rossi, and Pierre
  Sachse.
\newblock Individual differences in working memory capacity moderate effects of
  post-learning activity on memory consolidation over the long term.
\newblock {\em Scientific Reports}, 10(1):17976, 2020.

\bibitem{Justin2012}
Justin Matejka, Tovi Grossman, and George Fitzmaurice.
\newblock Swift: Reducing the effects of latency in online video scrubbing.
\newblock In {\em Proceedings of the CHI Conference on Human Factors in
  Computing Systems}, pages 637--646. Association for Computing Machinery,
  2012.

\bibitem{mayer2005cambridge}
Richard~E. Mayer, editor.
\newblock {\em The {Cambridge} handbook of multimedia learning}.
\newblock Cambridge University Press, 2005.

\bibitem{mayer2005cognitive}
Richard~E. Mayer.
\newblock Cognitive theory of multimedia learning.
\newblock In Richard~E. Mayer, editor, {\em The Cambridge handbook of
  multimedia learning}, pages 31--48. Cambridge University Press, 2005.

\bibitem{mayer2020multimedia}
Richard~E. Mayer.
\newblock {\em Multimedia learning}.
\newblock Cambridge University Press, 3rd ed. edition, 2020.

\bibitem{mayer2003social}
Richard~E. Mayer, Kristina Sobko, and Patricia~D. Mautone.
\newblock Social cues in multimedia learning: Role of speaker's voice.
\newblock {\em Journal of Educational Psychology}, 95(2):419--425, 2003.

\bibitem{mcauliffe2017_interspeech}
Michael McAuliffe, Michaela Socolof, Sarah Mihuc, Michael Wagner, and Morgan
  Sonderegger.
\newblock {Montreal Forced Aligner}: Trainable text-speech alignment using
  {Kaldi}.
\newblock In {\em Proceedings of INTERSPEECH}, pages 498--502, 2017.

\bibitem{mcfee2015librosa}
Brian McFee, Colin Raffel, Dawen Liang, Daniel~P. Ellis, Matt McVicar, Eric
  Battenberg, and Oriol Nieto.
\newblock {Librosa}: Audio and music signal analysis in {Python}.
\newblock In {\em Proceedings of the Python in Science Conference}, pages
  18--24, 2015.

\bibitem{Minematsu2001}
Nobuaki Minematsu, Yoshihiro Tomiyama, Kei Yoshimoto, Katsumasa Shimizu,
  Seiichi Nakagawa, Masatake Dantsuji, and Shozo Makino.
\newblock Development of {English} speech database spoken by {Japanese}
  learners.
\newblock In {\em Proceedings of the COCOSDA Workshop 2001}, pages 76--81,
  2001.

\bibitem{Nobuaki2002}
Nobuaki Minematsu, Yoshihiro Tomiyama, Kei Yoshimoto, Katsumasa Shimizu,
  Seiichi Nakagawa, Masatake Dantsuji, and Shozo Makino.
\newblock {English} speech database read by {Japanese} learners for {CALL}
  system development.
\newblock In {\em Proceedings of the International Conference on Language
  Resources and Evaluation}, pages 896--903. European Language Resources
  Association, 2002.

\bibitem{mishra2006technological}
Punya Mishra and Matthew~J. Koehler.
\newblock Technological pedagogical content knowledge: A framework for teacher
  knowledge.
\newblock {\em Teachers College Record}, 108(6):1017--1054, 2006.

\bibitem{Moreno2000Multimedia}
Roxana Moreno and Richard~E. Mayer.
\newblock Cognitive principles of multimedia learning: The role of modality and
  contiguity.
\newblock {\em Journal of Educational Psychology}, 91(2):358--368, 1999.

\bibitem{Mu2020MultimodalAnalytics}
Su~Mu, Meng Cui, and Xiaodi Huang.
\newblock Multimodal data fusion in learning analytics: A systematic review.
\newblock {\em Sensors}, 20(23):6856, 2020.

\bibitem{Murphy2022LectureSpeed}
Dillon~H. Murphy, Kara~M. Hoover, Karina Agadzhanyan, Jesse~C. Kuehn, and
  Alan~D. Castel.
\newblock Learning in double time: The effect of lecture video speed on
  immediate and delayed comprehension.
\newblock {\em Applied Cognitive Psychology}, 36(1):69--82, 2022.

\bibitem{neri2002pedagogy}
Ambra Neri, Catia Cucchiarini, Helmer Strik, and Lou Boves.
\newblock The pedagogy--technology interface in computer assisted pronunciation
  training.
\newblock {\em Computer Assisted Language Learning}, 15(5):441--467, 2002.

\bibitem{ngo2024effectiveness}
Thuy Thi-Nhu Ngo, Howard Hao-Jan Chen, and Kyle Kuo-Wei Lai.
\newblock The effectiveness of automatic speech recognition in {ESL}/{EFL}
  pronunciation: A meta-analysis.
\newblock {\em ReCALL}, 36(1):4--21, 2024.

\bibitem{Nixon2019}
Jeremy Nixon, Michael~W. Dusenberry, Linchuan Zhang, Ghassen Jerfel, and Dustin
  Tran.
\newblock Measuring calibration in deep learning.
\newblock In {\em Proceedings of the IEEE/CVF Conference on Computer Vision and
  Pattern Recognition Workshops}, pages 38--41, 2019.

\bibitem{norman2013design}
Donald~A. Norman.
\newblock {\em The design of everyday things: Revised and expanded edition}.
\newblock Basic Books, 2013.

\bibitem{OpenAI2023GPT4V}
{OpenAI}.
\newblock {GPT-4V(ision)} system card.
\newblock Technical report, OpenAI, 2023.

\bibitem{pacheco2019search}
Matheus~M. Pacheco, Charley~W. Lafe, and Karl~M. Newell.
\newblock Search strategies in the perceptual-motor workspace and the
  acquisition of coordination, control, and skill.
\newblock {\em Frontiers in Psychology}, 10:1874, 2019.

\bibitem{Paganus2006VowelGame}
Annu Paganus, Vesa-Petteri Mikkonen, Tomi M{\"a}ntyl{\"a}, Sami Nuuttila, Jouni
  Isoaho, Olli Aaltonen, and Tapio Salakoski.
\newblock The vowel game: Continuous real-time visualization for pronunciation
  learning with vowel charts.
\newblock In {\em Advances in Natural Language Processing}, pages 696--703.
  Springer, 2006.

\bibitem{Palaskar2019How2Summ}
Shruti Palaskar, Jind{\v{r}}ich Libovick{\'y}, Spandana Gella, and Florian
  Metze.
\newblock Multimodal abstractive summarization for {How2} videos.
\newblock In Anna Korhonen, David Traum, and Llu{\'i}s M{\`a}rquez, editors,
  {\em Proceedings of the Annual Meeting of the Association for Computational
  Linguistics}, pages 6587--6596, July 2019.

\bibitem{Vassil2015}
Vassil Panayotov, Guoguo Chen, Daniel Povey, and Sanjeev Khudanpur.
\newblock {Librispeech}: An {ASR} corpus based on public domain audio books.
\newblock In {\em Proceedings of the IEEE International Conference on
  Acoustics, Speech and Signal Processing}, pages 5206--5210, 2015.

\bibitem{Papert1980Mindstorms}
Seymour Papert.
\newblock {\em Mindstorms: Children, computers, and powerful ideas}.
\newblock Basic Books, 1980.

\bibitem{pastore2015using}
Raymond~S. Pastore and Albert~D. Ritzhaupt.
\newblock Using time-compression to make multimedia learning more efficient:
  Current research and practice.
\newblock {\em TechTrends}, 59:66--74, 2015.

\bibitem{Paszke2019}
Adam Paszke, Sam Gross, Francisco Massa, Adam Lerer, James Bradbury, Gregory
  Chanan, Trevor Killeen, Zeming Lin, Natalia Gimelshein, Luca Antiga, Alban
  Desmaison, Andreas K{\"o}pf, Edward Yang, Zach DeVito, Martin Raison, Alykhan
  Tejani, Sasank Chilamkurthy, Benoit Steiner, Lu~Fang, Junjie Bai, and Soumith
  Chintala.
\newblock {PyTorch}: An imperative style, high-performance deep learning
  library.
\newblock In {\em Proceedings of the Advances in Neural Information Processing
  Systems}, pages 8024--8035. Curran Associates, Inc., 2019.

\bibitem{pavel2014}
Amy Pavel, Colorado Reed, Bj{\"o}rn Hartmann, and Maneesh Agrawala.
\newblock Video digests: A browsable, skimmable format for informational
  lecture videos.
\newblock In {\em Proceedings of the 27th Annual ACM Symposium on User
  Interface Software and Technology}, pages 573--582. Association for Computing
  Machinery, 2014.

\bibitem{Piech2013Peer}
Chris Piech, Jonathan Huang, Zhenghao Chen, Chuong Do, Andrew Ng, and Daphne
  Koller.
\newblock Tuned models of peer assessment in {MOOCs}.
\newblock In {\em Proceedings of the Educational Data Mining}, pages 153--160,
  2013.

\bibitem{pilicita2025llms}
Anabel Pilicita and Enrique Barra.
\newblock {LLMs} in education: Evaluation {GPT} and {BERT} models in student
  comment classification.
\newblock {\em Multimodal Technologies and Interaction}, 9(5):44, 2025.

\bibitem{Plass2015Gaming}
Jan~L. Plass, Bruce~D. Homer, and Charles~K. Kinzer.
\newblock Foundations of game-based learning.
\newblock {\em Educational Psychologist}, 50(4):258--283, 2015.

\bibitem{polanyi1966tacit}
Michael Polanyi.
\newblock {\em The tacit dimension}.
\newblock Doubleday, 1966.

\bibitem{Pongnumkul2010Timeline}
Suporn Pongnumkul, Jue Wang, Gonzalo Ramos, and Michael Cohen.
\newblock Content-aware dynamic timeline for video browse.
\newblock In {\em Proceedings of the annual ACM symposium on User interface
  software and technology}, pages 139--142, 2010.

\bibitem{Qian2010}
Xiaojun Qian, Frank Soong, and Helen Meng.
\newblock Discriminatively trained acoustic models for improving
  mispronunciation detection and diagnosis in computer aided pronunciation
  training ({CAPT}).
\newblock In {\em Proceedings of INTERSPEECH}, 2010.

\bibitem{radford2023}
Alec Radford, Jong~Wook Kim, Tao Xu, Greg Brockman, Christine McLeavey, and
  Ilya Sutskever.
\newblock Robust speech recognition via large-scale weak supervision.
\newblock In {\em Proceedings of the International Conference on Machine
  Learning}, volume 202 of {\em Proceedings of Machine Learning Research},
  pages 28492--28518. PMLR, 2023.

\bibitem{radford2018}
Alec Radford, Karthik Narasimhan, Tim Salimans, and Ilya Sutskever.
\newblock Improving language understanding by generative pre-training.
\newblock Technical report, OpenAI, 2018.

\bibitem{radford2019}
Alec Radford, Jeffrey Wu, Rewon Child, David Luan, Dario Amodei, and Ilya
  Sutskever.
\newblock Language models are unsupervised multitask learners.
\newblock Technical Report TR-2019-1, OpenAI, 2019.

\bibitem{Nisreen2012}
Nisreen~I. Radwan, Nancy~M. Salem, and Mohamed~I. El~Adawy.
\newblock Histogram correlation for video scene change detection.
\newblock In {\em Proceedings of the Second International Conference on
  Computer Science, Engineering and Applications}, pages 765--773, 2012.

\bibitem{raffel2020}
Colin Raffel, Noam Shazeer, Adam Roberts, Katherine Lee, Sharan Narang, Michael
  Matena, Yanqi Zhou, Wei Li, and Peter~J. Liu.
\newblock Exploring the limits of transfer learning with a unified text-to-text
  transformer.
\newblock {\em Journal of Machine Learning Research}, 21(140):1--67, 2020.

\bibitem{ravenor2024adequacy}
R.~Yamamoto Ravenor.
\newblock On the adequacy of {Elsa Speak} in formal education: A survey of
  teacher-users.
\newblock {\em Advances in Artificial Intelligence and Machine Learning},
  04:2387--2394, 2024.

\bibitem{rawls1971theory}
John Rawls.
\newblock {\em A theory of justice}.
\newblock Harvard University Press, 1971.

\bibitem{ray2023}
Partha~Pratim Ray.
\newblock {ChatGPT}: A comprehensive review on background, applications, key
  challenges, bias, ethics, limitations and future scope.
\newblock {\em Internet of Things and Cyber-Physical Systems}, 3:121--154,
  2023.

\bibitem{reiser2018scaffolding}
Brian~J. Reiser.
\newblock Scaffolding complex learning: The mechanisms of structuring and
  problematizing student work.
\newblock {\em The Journal of the Learning Sciences}, 13(3):273--304, 2004.

\bibitem{Cortes15}
Shaoqing Ren, Kaiming He, Ross Girshick, and Jian Sun.
\newblock Faster {R-CNN}: Towards real-time object detection with region
  proposal networks.
\newblock In C.~Cortes, N.~Lawrence, D.~Lee, M.~Sugiyama, and R.~Garnett,
  editors, {\em Proceedings of the Advances in Neural Information Processing
  Systems}, volume~28, pages 91--99, 2015.

\bibitem{Resnick2009Scratch}
Mitchel Resnick, John Maloney, Andr{\'e}s Monroy-Hern{\'a}ndez, Natalie Rusk,
  Evelyn Eastmond, Karen Brennan, Amon Millner, Eric Rosenbaum, Jay Silver,
  Brian Silverman, and Yasmin Kafai.
\newblock Scratch: Programming for all.
\newblock {\em Communications of the ACM}, 52(11):60--67, 2009.

\bibitem{Risko2012Wandering}
Evan~F. Risko, Nicola Anderson, Amrita Sarwal, Megan Engelhardt, and Alan
  Kingstone.
\newblock Everyday attention: Variation in mind wandering and memory in a
  lecture.
\newblock {\em Applied Cognitive Psychology}, 26(2):234--242, 2012.

\bibitem{RogersonRevell2021}
Pamela~M. Rogerson-Revell.
\newblock Computer-assisted pronunciation training ({CAPT}): Current issues and
  future directions.
\newblock {\em RELC Journal}, 52(1):189--205, 2021.

\bibitem{rousso2024tradition}
Rotem Rousso, Eyal Cohen, Joseph Keshet, and Eleanor Chodroff.
\newblock Tradition or innovation: A comparison of modern {ASR} methods for
  forced alignment.
\newblock In {\em Proceedings of INTERSPEECH 2024}, pages 1525--1529, 2024.

\bibitem{roy2024lesson}
Palak Roy, Helen Poet, Ruth Staunton, Katherine Aston, and David Thomas.
\newblock {ChatGPT} in lesson preparation: A teacher choices trial.
\newblock Technical report, National Foundation for Educational Research,
  December 2024.

\bibitem{saito2019effects}
Kazuya Saito and Luke Plonsky.
\newblock Effects of second language pronunciation teaching revisited: A
  proposed measurement framework and meta-analysis.
\newblock {\em Language Learning}, 69(3):652--708, 2019.

\bibitem{Scheirer1997}
E.~Scheirer and M.~Slaney.
\newblock Construction and evaluation of a robust multifeature speech/music
  discriminator.
\newblock In {\em Proceedings of the IEEE International Conference on
  Acoustics, Speech and Signal Processing}, volume~2, pages 1331--1334, 1997.

\bibitem{Schneider2019}
Steffen Schneider, Alexei Baevski, Ronan Collobert, and Michael Auli.
\newblock wav2vec: Unsupervised pre-training for speech recognition.
\newblock {\em Proceedings of INTERSPEECH}, pages 3465--3469, 2019.

\bibitem{zhang2023generative}
Tony Sheehan.
\newblock Generative {AI} in education: Past, present, and future.
\newblock {\em EDUCAUSE Review}, 2023.

\bibitem{Shen2018}
Jonathan Shen, Ruoming Pang, Ron~J. Weiss, Mike Schuster, Navdeep Jaitly,
  Zongheng Yang, Zhifeng Chen, Yu~Zhang, Yuxuan Wang, R.~J. Skerrv-Ryan, Rif~A.
  Saurous, Yannis Agiomyrgiannakis, and Yonghui Wu.
\newblock Natural {TTS} synthesis by conditioning {Wavenet} on {MEL}
  spectrogram predictions.
\newblock In {\em Proceedings of the IEEE International Conference on
  Acoustics, Speech and Signal Processing}, pages 4779--4783, 2018.

\bibitem{Sherwood1979ComputerSpeaks}
Bruce~A. Sherwood.
\newblock The computer speaks.
\newblock {\em IEEE Spectrum}, 16(8):18--25, 1979.

\bibitem{Shneiderman1983Direct}
Ben Shneiderman.
\newblock Direct manipulation: A step beyond programming languages.
\newblock {\em Computer}, 16(8):57--69, 1983.

\bibitem{Shneiderman1997Direct}
Ben Shneiderman.
\newblock Direct manipulation for comprehensible, predictable and controllable
  user interfaces.
\newblock In {\em Proceedings of the International Conference on Intelligent
  User Interfaces}, pages 33--39, 1997.

\bibitem{Shneiderman2022HumanCenteredAI}
Ben Shneiderman.
\newblock {\em Human-centered {AI}}.
\newblock Oxford University Press, 2022.

\bibitem{Shneiderman1997DebateMaes}
Ben Shneiderman and Pattie Maes.
\newblock Direct manipulation versus interface agents.
\newblock {\em Interactions}, 4(6):42--61, 1997.

\bibitem{shortt2023gamification}
Mitchell Shortt, Shantanu Tilak, Irina Kuznetcova, Bethany Martens, and
  Babatunde Akinkuolie.
\newblock Gamification in mobile-assisted language learning: A systematic
  review of {Duolingo} literature from public release of 2012 to early 2020.
\newblock {\em Computer Assisted Language Learning}, 36(3):517--554, 2023.

\bibitem{sigrist2013augmented}
Roland Sigrist, Georg Rauter, Robert Riener, and Peter Wolf.
\newblock Augmented visual, auditory, haptic and multimodal feedback in motor
  learning: A review.
\newblock {\em Psychonomic Bulletin \& Review}, 20:21--53, 2013.

\bibitem{simon1971designing}
Herbert~A. Simon.
\newblock Designing organizations for an information-rich world.
\newblock In Martin Greenberger, editor, {\em Computers, Communications, and
  the Public Interest}, pages 37--72. Johns Hopkins Press, 1971.

\bibitem{singleton2021cph}
David Singleton and Justyna Leśniewska.
\newblock The critical period hypothesis for {L2} acquisition: An unfalsifiable
  embarrassment?
\newblock {\em Languages}, 6(3):149, 2021.

\bibitem{Smith07}
R.~Smith.
\newblock An overview of the {Tesseract OCR Engine}.
\newblock In {\em Proceedings of the International Conference on Document
  Analysis and Recognition}, volume~2, pages 629--633, 2007.

\bibitem{Nitish2014}
Nitish Srivastava, Geoffrey Hinton, Alex Krizhevsky, Ilya Sutskever, and Ruslan
  Salakhutdinov.
\newblock Dropout: A simple way to prevent neural networks from overfitting.
\newblock {\em Journal of Machine Learning Research}, 15(56):1929--1958, 2014.

\bibitem{sullivan2008motor}
Katherine~J. Sullivan, Shailesh~S. Kantak, and Patricia~A. Burtner.
\newblock Motor learning in children: Feedback effects on skill acquisition.
\newblock {\em Physical Therapy}, 88(6):720--732, 2008.

\bibitem{Sutherland1963Sketchpad}
Ivan~E. Sutherland.
\newblock Sketchpad: a man-machine graphical communication system.
\newblock In {\em Proceedings of the May 21-23, 1963, Spring Joint Computer
  Conference}, pages 329--346. ACM, 1963.

\bibitem{sweller1988cognitive}
John Sweller.
\newblock Cognitive load during problem solving: Effects on learning.
\newblock {\em Cognitive Science}, 12(2):257--285, 1988.

\bibitem{sweller2011cognitive}
John Sweller.
\newblock Cognitive load theory.
\newblock In {\em Psychology of learning and motivation}, volume~55, pages
  37--76. Elsevier, 2011.

\bibitem{sweller2019cognitive}
John Sweller, Jeroen J.~G. van Merri{\"e}nboer, and Fred Paas.
\newblock Cognitive architecture and instructional design: 20 years later.
\newblock {\em Educational Psychology Review}, 31:261--292, 2019.

\bibitem{taylor1994politics}
Charles Taylor.
\newblock The politics of recognition.
\newblock In Amy Gutmann, editor, {\em Multiculturalism}, pages 25--74.
  Princeton University Press, 1994.

\bibitem{Cristian2021}
Cristian Tejedor-Garc\'{\i}a, Valent\'{\i}n Carde\~{n}oso Payo, and David
  Escudero-Mancebo.
\newblock Automatic speech recognition ({ASR}) systems applied to pronunciation
  assessment of {L2} {Spanish} for {Japanese} speakers.
\newblock {\em Applied Sciences}, 11(15), 2021.

\bibitem{tharumalingam2025increasing}
Theepan Tharumalingam, Brady R.~T. Roberts, Jonathan~M. Fawcett, and Evan~F.
  Risko.
\newblock Increasing video lecture playback speed can impair test
  performance---a meta-analysis.
\newblock {\em Educational Psychology Review}, 37, 2025.

\bibitem{thompson2011khan}
Clive Thompson.
\newblock How {Khan Academy} is changing the rules of education.
\newblock Wired, July 2011.

\bibitem{thomson2018measurement}
Ron~I. Thomson.
\newblock Measurement of accentedness, intelligibility and comprehensibility.
\newblock In Okim Kang and April Ginther, editors, {\em Assessment in second
  language pronunciation}, pages 11--29. Routledge, 2018.

\bibitem{tio2024eduqate}
Sidney Tio, Dexun Li, and Pradeep Varakantham.
\newblock {EduQate}: Generating adaptive curricula through {RMABs} in education
  settings.
\newblock In {\em Proceedings of the 24th International Conference on
  Autonomous Agents and Multiagent Systems}, pages 2042--2050, 2025.
\newblock ACM Digital Library identifier 10.5555/3709347.3743842; the DOI
  resolver does not currently resolve this identifier.

\bibitem{truong2007video}
Ba~Tu Truong and Svetha Venkatesh.
\newblock Video abstraction: A systematic review and classification.
\newblock {\em ACM Transactions on Multimedia Computing, Communications, and
  Applications}, 3(1), 2007.

\bibitem{Vandenoord16}
Aaron van~den Oord, Sander Dieleman, Heiga Zen, Karen Simonyan, Oriol Vinyals,
  Alex Graves, Nal Kalchbrenner, Andrew Senior, and Koray Kavukcuoglu.
\newblock {WaveNet}: A generative model for raw audio.
\newblock In {\em Proceedings of the ISCA Workshop on Speech Synthesis
  Workshop}, page 125, 2016.

\bibitem{vartakavi2021audio}
Aneesh Vartakavi, Amanmeet Garg, and Zafar Rafii.
\newblock Audio summarization for podcasts.
\newblock In {\em Proceedings of the European Signal Processing Conference},
  pages 431--435. IEEE, 2021.

\bibitem{Vaswani2017}
Ashish Vaswani, Noam Shazeer, Niki Parmar, Jakob Uszkoreit, Llion Jones,
  Aidan~N. Gomez, \L~ukasz Kaiser, and Illia Polosukhin.
\newblock Attention is all you need.
\newblock In {\em Proceedings of the Advances in Neural Information Processing
  Systems}, volume~30, pages 5998--6008, 2017.

\bibitem{vygotsky1978mind}
Lev~S. Vygotsky.
\newblock {\em Mind in society: The development of higher psychological
  processes}.
\newblock Harvard University Press, 1978.

\bibitem{w3c2021webaudio}
{W3C Audio Working Group}.
\newblock {Web Audio API}.
\newblock W3C Recommendation, 2021.
\newblock 17 June 2021.

\bibitem{w3c2021webrtc}
{W3C Web Real-Time Communications Working Group}.
\newblock {WebRTC} 1.0: Real-time communication between browsers.
\newblock W3C Recommendation, 2021.
\newblock 26 January 2021.

\bibitem{wang2020}
Changhan Wang, Yun Tang, Xutai Ma, Anne Wu, Dmytro Okhonko, and Juan Pino.
\newblock Fairseq {S2T}: Fast speech-to-text modeling with {Fairseq}.
\newblock In {\em Proceedings of the Conference of the Asia-Pacific Chapter of
  the Association for Computational Linguistics and the 10th International
  Joint Conference on Natural Language Processing: System Demonstrations},
  pages 33--39, 2020.

\bibitem{wang2024diffusion}
Xiaolong Wang, Zhijian He, and Xiaojiang Peng.
\newblock Artificial-intelligence-generated content with diffusion models: A
  literature review.
\newblock {\em Mathematics}, 12(7):977, 2024.

\bibitem{Yuxuan2017}
Yuxuan Wang, R.~J. Skerry-Ryan, Daisy Stanton, Yonghui Wu, Ron~J. Weiss,
  Navdeep Jaitly, Zongheng Yang, Ying Xiao, Zhifeng Chen, Samy Bengio, Quoc~V.
  Le, Yannis Agiomyrgiannakis, Rob Clark, and Rif~A. Saurous.
\newblock {Tacotron}: Towards end-to-end speech synthesis.
\newblock In {\em Proceedings of INTERSPEECH}, pages 4006--4010, 2017.

\bibitem{wilson2002six}
Margaret Wilson.
\newblock Six views of embodied cognition.
\newblock {\em Psychonomic Bulletin \& Review}, 9:625--636, 2002.

\bibitem{witt2000phone}
Silke~M. Witt and Steve~J. Young.
\newblock Phone-level pronunciation scoring and assessment for interactive
  language learning.
\newblock {\em Speech Communication}, 30(2):95--108, 2000.

\bibitem{wood1976role}
David Wood, Jerome~S. Bruner, and Gail Ross.
\newblock The role of tutoring in problem solving.
\newblock {\em Journal of Child Psychology and Psychiatry}, 17(2):89--100,
  1976.

\bibitem{Woolley1994PLATO}
David~R. Woolley.
\newblock {PLATO}: The emergence of on-line community.
\newblock {\em Computer-Mediated Communication Magazine}, 1(3):5, 1994.

\bibitem{xiao2025pedagogical}
Ruiwei Xiao, Xinying Hou, Runlong Ye, Majeed Kazemitabaar, Nicholas Diana,
  Michael Liut, and John Stamper.
\newblock Improving student-{AI} interaction through pedagogical prompting: An
  example in computer science education.
\newblock arXiv preprint arXiv:2506.19107v1, 2025.
\newblock Submitted 23 June 2025.

\bibitem{Xiong2016}
Wayne Xiong, Jasha Droppo, Xuedong Huang, Frank Seide, Michael~L. Seltzer,
  Andreas Stolcke, Dong Yu, and Geoffrey Zweig.
\newblock Toward human parity in conversational speech recognition.
\newblock {\em IEEE/ACM Transactions on Audio, Speech, and Language
  Processing}, pages 2410--2423, 2017.

\bibitem{xu-etal-2024-conformer}
Mingbin Xu, Alex Jin, Sicheng Wang, Mu~Su, Tim Ng, Henry Mason, Shiyi Han,
  Zhihong Lei, Yaqiao Deng, Zhen Huang, and Mahesh Krishnamoorthy.
\newblock Conformer-based speech recognition on extreme edge-computing devices.
\newblock In {\em Proceedings of the Conference of the North American Chapter
  of the Association for Computational Linguistics: Human Language Technologies
  (Industry Track)}, pages 131--139. Association for Computational Linguistics,
  2024.

\bibitem{Yang2023}
Antoine Yang, Arsha Nagrani, Ivan Laptev, Josef Sivic, and Cordelia Schmid.
\newblock {VidChapters-7M}: Video chapters at scale.
\newblock {\em arXiv}, 2023.
\newblock arXiv:2309.13952. NeurIPS 2023 Datasets and Benchmarks Track.

\bibitem{Yang2020UnremarkableAI}
Qian Yang, Aaron Steinfeld, Carolyn Ros{\'e}, and John Zimmerman.
\newblock Re-examining whether, why, and how human-{AI} interaction is uniquely
  difficult to design.
\newblock In {\em Proceedings of the CHI Conference on Human Factors in
  Computing Systems}, pages 1--13, 2020.

\bibitem{Saelyne2022SoftVideo}
Saelyne Yang, Jisu Yim, Aitolkyn Baigutanova, Seoyoung Kim, Minsuk Chang, and
  Juho Kim.
\newblock {SoftVideo}: Improving the learning experience of software tutorial
  videos with collective interaction data.
\newblock In {\em Proceedings of the International Conference on Intelligent
  User Interfaces}, pages 646--660. Association for Computing Machinery, 2022.

\bibitem{Shu-wen2021}
Shu-wen Yang, Po-Han Chi, Yung-Sung Chuang, Cheng-I~Jeff Lai, Kushal Lakhotia,
  Yist~Y. Lin, Andy~T. Liu, Jiatong Shi, Xuankai Chang, Guan-Ting Lin,
  Tzu-Hsien Huang, Wei-Cheng Tseng, Ko-tik Lee, Da-Rong Liu, Zili Huang, Shuyan
  Dong, Shang-Wen Li, Shinji Watanabe, Abdelrahman Mohamed, and Hung-yi Lee.
\newblock {SUPERB}: Speech processing {Universal} {PERformance} {Benchmark}.
\newblock In {\em Proceedings of INTERSPEECH 2021}, pages 1194--1198, 2021.

\bibitem{yoshitaka2012personalized}
Atsuo Yoshitaka and Kazuya Sawada.
\newblock Personalized video summarization based on behavior of viewer.
\newblock In {\em Proceedings of the International Conference on Signal Image
  Technology and Internet Based Systems}, pages 661--667, 2012.

\bibitem{yurum2024predictive}
Ozan~Ra{\c s}it Y{\"u}r{\"u}m, Tu{\u g}ba Ta{\c s}kaya-Temizel, and Soner
  Y{\i}ld{\i}r{\i}m.
\newblock Predictive video analytics in online courses: A systematic literature
  review.
\newblock {\em Technology, Knowledge and Learning}, 29:1907--1937, 2024.

\bibitem{zekveld2011cognitive}
Adriana~A. Zekveld, Sophia~E. Kramer, and Joost~M. Festen.
\newblock Cognitive load during speech perception in noise: The influence of
  age, hearing loss, and cognition on the pupil response.
\newblock {\em Ear and Hearing}, 32(4):498--510, 2011.

\bibitem{Zen2019}
Heiga Zen, Viet Dang, Rob Clark, Yu~Zhang, Ron~J. Weiss, Ye~Jia, Zhifeng Chen,
  and Yonghui Wu.
\newblock {LibriTTS}: {A} corpus derived from {LibriSpeech} for text-to-speech.
\newblock In {\em Proceedings of INTERSPEECH}, pages 1526--1530, 2019.

\bibitem{Zhang2006}
Dongsong Zhang, Lina Zhou, Robert~O Briggs, and Jay~F Nunamaker~Jr.
\newblock Instructional video in e-learning: Assessing the impact of
  interactive video on learning effectiveness.
\newblock {\em Information \& Management}, 43(1):15--27, 2006.

\bibitem{zhang2022collective}
Jingjing Zhang, Yicheng Huang, and Ming Gao.
\newblock Video features, engagement, and patterns of collective attention
  allocation: An open flow network perspective.
\newblock {\em Journal of Learning Analytics}, 9(1):32--52, 2022.

\bibitem{Zhang2020}
Xinlei Zhang, Takashi Miyaki, and Jun Rekimoto.
\newblock {WithYou}: Automated adaptive speech tutoring with context-dependent
  speech recognition.
\newblock In {\em Proceedings of the CHI Conference on Human Factors in
  Computing Systems}, pages 1--2, 2020.

\bibitem{zhang2024embodied}
Yuan Zhang, Florence Baills, and Pilar Prieto.
\newblock Embodied music training can help improve speech imitation and
  pronunciation skills.
\newblock {\em Language Teaching}, 2024.
\newblock Published online 18 December 2024.

\bibitem{zhao2018l2}
Guanlong Zhao, Evgeny Chukharev-Hudilainen, Sinem Sonsaat, Alif Silpachai,
  Ivana Lucic, Ricardo Gutierrez-Osuna, and John Levis.
\newblock {L2-Arctic}: A non-native {English} speech corpus.
\newblock In {\em Proceedings of Interspeech 2018}, pages 2783--2787, 2018.

\bibitem{zhou2025crowd}
Xin Zhou, Aixin Sun, Jie Zhang, and Donghui Lin.
\newblock The crowd in {MOOCs}: A study of learning patterns at scale.
\newblock {\em Interactive Learning Environments}, 33(3):2136--2150, 2025.

\bibitem{zimmerman2002becoming}
Barry~J. Zimmerman.
\newblock Becoming a self-regulated learner: An overview.
\newblock {\em Theory into Practice}, 41(2):64--70, 2002.

\bibitem{zong2024playback}
Jiaxuan Zong, Nihat Polat, and Laura Mahalingappa.
\newblock Effects of playback speed and language proficiency on listening
  comprehension of multilingual {English} learners.
\newblock {\em International Multilingual Research Journal}, 2024.
\newblock Published online 8 June 2024; later assigned to volume 19, issue 3.

\end{thebibliography}

\chapter*{Publications}
\addcontentsline{toc}{chapter}{Publications}

\section*{Peer-Reviewed International Conference Proceedings}
\begin{enumerate}
    \item Kazuki Kawamura and Jun Rekimoto, ``DDSupport: Language Learning Support System that Displays Differences and Distances from Model Speech,'' 21st IEEE International Conference on Machine Learning and Applications (ICMLA), Nassau, Bahamas, 2022, pp. 313-320, \\doi: 10.1109/ICMLA55696.2022.00051

    \item Kazuki Kawamura and Jun Rekimoto, ``AIxSpeed: Playback Speed Optimization Using Listening Comprehension of Speech Recognition Models,'' Augmented Humans International Conference 2023 (AHs), Glasgow, United Kingdom, 2023, pp. 200-208, doi: 10.1145/3582700.3582722

    \item Kazuki Kawamura and Jun Rekimoto, ``FastPerson: Enhancing Video Learning through Effective Video Summarization that Preserves Linguistic and Visual Contexts,'' Augmented Humans International Conference 2024 (AHs), Melbourne, Australia, 2024, pp. 205-216, doi: 10.1145/3652920.3652922
\end{enumerate}

\section*{Other Academic Achievements}
\subsection*{Posters/Workshops}
\begin{enumerate}
    \item Kazuki Kawamura and Jun Rekimoto, ``A Language Acquisition Support System that Presents Differences and Distances from Model Speech,'' 34th Annual ACM Symposium on User Interface Software and Technology (UIST), Virtual Event, USA, 2021, pp. 44-46, doi: 10.1145/3474349.3480225

    \item Kazuki Kawamura and Jun Rekimoto, ``Visualization of Speech Differences for Dialect Speech Training,'' 2021 CHI Conference on Human Factors in Computing Systems Workshop on Human Augmentation for Skill Acquisition and Skill Transfer (CHI), Virtual Event, Japan, 2021.

    \item Kazuki Kawamura and Jun Rekimoto, ``AIxSpeed: Playback Speed Optimization Using Listening Comprehension of Speech Recognition Models,'' 35th Annual ACM Symposium on User Interface Software and Technology (UIST), Bend, USA, 2022, pp. 1-3, doi: 10.1145/3526114.3558727

    \item Kazuki Kawamura and Jun Rekimoto, ``QA-FastPerson: Extending Video Platform Search Capabilities by Creating Summary Videos in Response to User Queries,'' Augmented Humans International Conference 2024 (AHs), Melbourne, Australia, 2024, pp. 290-293, doi: 10.1145/3652920.3653052

    \item Kazuki Kawamura and Jun Rekimoto, ``Generating Summary Videos from User Questions to Support Video-Based
    Learning,'' 2024 CHI Conference on Human Factors in Computing Systems Workshop on Generative AI and HCI
    (CHI), Honolulu, USA, 2024.

\end{enumerate}

\subsection*{Peer-Reviewed Domestic Conference Proceedings}
\begin{enumerate}
    \item Kazuki Kawamura and Jun Rekimoto, ``DDSupport: A Language Learning Support System that Presents Differences and Distances from Model Pronunciation,'' Interaction, 2022, pp. 77-86 (in Japanese)

    \item Kazuki Kawamura and Jun Rekimoto, ``FastPerson: Lecture Video Summarization Based on Visual and Audio Information for User-Centered Learning Experience,'' Interaction, 2024, pp. 11-20 (in Japanese)
\end{enumerate}

\end{document}